# Gauss-Seed Nets of Sturm-Liouville Problems
# With Energy-Independent Characteristic Exponents
# and
# Related Sequences of Exceptional Orthogonal Polynomials
# I. Canonical Darboux Transformations
# Using Almost-Everywhere Holomorphic Factorization Functions

**G. Natanson**

ai-solutions Inc.
2232 Blue Valley Dr.
Silver Spring MD 20904
U.S.A.
greg_natanson@yahoo.com

The paper applies the so-called 'Canonical-Darboux-Transformation' (CDT) method to reproduce general expressions for rational potentials (RPs) quantized in terms of exceptional orthogonal polynomial systems (X-OPSs). The benchmark of the new method recently developed by the author for implicit potentials solvable via hypergeometric functions is that rationally-extended SUSY partners of the original potential are quantized in terms of sequences of the so-called 'Gauss-seed' (GS) Heine polynomials starting from a polynomial of non-zero order. The common mark of the Darboux-Poschl-Teller (DPT) potential and isotonic oscillator discussed in this paper is that the appropriate rational Sturm-Liouville (RSL) equations have energy-independent characteristic exponents at both singular end points and as a result the appropriate sequences of GS Heine polynomials turn into X-OPSs with infinitely many members. Some features of our approach are closely related to Quesne's studies on rational extensions (REs) of shape-invariant potentials using the Schrödinger equation expressed in terms of an intermediate variable – which is nothing but the 'Liouville transform' of the rational Sturm-Liouville problem of our interest. Whenever it is applicable, our results were verified versus particular cases discussed in her papers.

To construct nodeless factorization functions (FFs) regular at one of the end points we require that the appropriate Almost-Everywhere-Holomorphic (AEH) solutions (also referred in the literature as 'quasi-algebraic' or 'quasi-rational') lie below the ground energy level. Contrary to the technique using the Hilbert-Klein formulas and Kienast-Lawton-Hahn (KLH) theorem for locations of zeros of Jacobi and generalized Laguerre (GL) polynomials, the conventional unbroken-symmetry approach allows one to easily extend the analysis to multi-step Darboux transformations (DTs). On other hand, the Hilbert-Klein formulas were used to construct of broken-symmetry SUSY partners of the DPT potential, similarly to the so-called 'Case-III' REs of the isotonic oscillator found by Grandati by means of the KLH theorem. It is stressed that REs of both DPT potential and isotonic oscillator can be used to generate $X_m$-Jacobi and $X_m$-Laguerre polynomials only within the validity range of SUSY quantum mechanics and should not be considered as an alternative to Gomez-Ullate, Kamran, and Milson' theory of exceptional orthogonal polynomials which covers a larger range of polynomial indexes.

In addition, as a side result, we refine Odake and Sasaki's analysis of rationally extended hyperbolic Poschl-Teller (h-PT) potentials. It is shown that the Schrödinger equation with both DPT and h-PT potential can be converted into the same RSL equation, with the only difference that one has to change the quantization interval from $(0,+1)$ to $(-\infty,0)$ and alter the energy sign. As a result, Darboux transforms of two problems share the same 16 sets of GS Heine polynomials. Depending on the choice of the quantization interval, one comes to two different subsets of orthogonal Heine polynomials: exceptional Jacobi polynomials for $(0, +1)$ and 'abnormally orthogonal' Heine polynomials -- Darboux transforms of Romanovsky-Jacobi polynomials -- for $(-\infty, 0)$.

## 1. Introduction

Several years ago Gomez-Ullate, Kamran, and Milson [1, 2] discovered two complete orthogonal sets of Heun [3, 4] and 'confluent Heun' (*c*-Heun) [4, 5] polynomials commonly referred to as '$X_1$-Jacobi' and '$X_1$-Laugerre' exceptional orthogonal polynomials, respectively. Their discovery was almost immediately utilized by Quesne [6] who constructed two rational potentials (RPs) quantized in terms of these polynomials. One of these potentials – the deformed isotonic oscillator - has been earlier constructed by Levai and Roy [7] who, in following Junker and Roy [8], referred to it as 'conditionally exactly solvable' (CES) potential since one has to impose additional restrictions on the coefficients of the appropriate polynomial fraction (PF).

Within the framework of this paper, the most significant result of Quesne's pioneering work [6] is the prove that the constructed potentials are quantized in terms of non-classical polynomials -- the particular cases of Heine polynomials [9, 10] of a very special type recently elicited by the author [11] in his analysis of SUSY partners of Gauss-reference (GRef) potentials solvable in terms of hypergeometric functions [12]..

By analyzing the associate superpotentials Quesne later proved [13] that the constructed potentials are nothing but isospectral SUSY partners of the Darboux-Poschl-Teller (DPT) potential [14, 15][x)] and isotonic oscillator, though in the latter case Quesne's results can be more directly obtained from Levai and Roy's analysis [7]. In the same work [13] Quesne also found new rational extensions (RE) of the DPT potential and isotonic oscillator exactly quantized via $X_2$-Jacobi and $X_2$-Laugerre polynomials which satisfy the Fuschian equation with five regular singular points (including infinity) and its confluent counter-part, respectively. In this connection it seems appropriate to mention again Levai and Roy's work [7] – we point the reader to their Eq. (7) which explicitly specifies the structure of the given RE of the isotonic

---

[x)] In following Matveev's suggestion [16], we use this term, instead of the conventional name 'trigonometric Poschl-Teller' (t-PT) potential, to give a proper credit to Darboux [14], who was the first to obtain eigenfunctions and eigenvalues for the appropriate Sturm-Liouville (SL) problem.

oscillator assuming that corresponding generalized Laguerre (GL) polynomials have only real roots. From our view the most important element of Levai and Roy's analysis is polynomial truncation of factorization functions (FFs) introduced by Junker and Roy [8]. This truncation leads to RPs which are the main subject of our current studies [11, 17-19].

Independently from Quesne's second work, Odake and Sasaki [20] came up with a more universal procedure for generating series of RPs quantized via the so-called '$X_m$-Jacobi' and '$X_m$-Laugerre' orthogonal polynomials. As explained in next Section, their approach reproduces many features of the polynomial truncation suggested by Levai and Roy [7]. A more direct link between the potentials quantized via $X_m$-Laugerre orthogonal polynomials [13, 20] and Junker and Roy's CES potentials [8] was revealed by Dutta and Roy [21]. Surprisingly they made no reference to Roy's own unpublished paper with Levai [7]. It is especially disappointing since the latter paper explicitly restricted the discussion of FFs regular either at the origin or at infinity to solutions lying below the ground energy level which covers both Case-I [20] and Case-II [22] $X_m$-Laugerre polynomials, in Quesne's [13] terms. In both papers [20] and [22] Odake and Sasaki simply mentioned without a proof that the appropriate FFs do not have nodes inside the quantization interval. The proof for Cases I GL polynomials was then given in a separate publication [23] based on the observation that coefficients of GL polynomials $L_m^{(\lambda_0)}(-\varsigma)$ expressed in terms of $\varsigma$ are all positive for $\lambda_0 > 0$.

The new findings were quickly incorporated by Gomez-Ullate, Kamran, and Milson [24] into a rigorous theory of $X_m$-Laugerre polynomials. In particular, they pointed to the fact that FFs used in [20] and [22] for DTs of the isotonic oscillator must be nodeless in both Cases I and II as a direct consequence of the analysis performed in [24] for locations of zeros of GL polynomials [26] -- the so-called 'Kienast-Lawton-Hahn' (KLH) theorem [27-30] in Grandati's [31, 32] terms.

In this connection we would like also to mention the *rational* factorization (RF) performed by Ho, Odake, and Sasaki [33] for rational Sturm-Liouville (RSL) problems quantized via $X_m$-Jacobi and Case-I $X_m$-Laugerre orthogonal polynomials in terms of forward and backward shift operators, though (except the simplest case of the 'algebraic ground state condition' [33]) it is unclear whether their technique is equivalent to the 'covariant rational factorization' scheme utilized in Gomez-Ullate, Kamran, and Milson's theory of exceptional orthogonal polynomials [24, 35].

Some interesting properties of called $X_m$-Jacobi and $X_m$-Laugerre orthogonal polynomials have been also recovered in more recent works of Sasaki, Ho et al [36-38], Tanaka [39], Takemura [40], and Gomez-Ullate, Milson et al [41, 42].

New *regular* sequences of multi-index orthogonal polynomials were more recently introduced by Quesne et al [43-49] by analyzing multi-step Darboux transforms of the isotonic oscillator. We shall come back to a more detailed discussion of these papers in Part II where we outline the general formalism for constructing multi-index RPs quantized via orthogonal Heine polynomials.

Some preliminary attempts to extend the theory of exceptional orthogonal polynomials to two-step Darboux transformations of GL polynomials has been also made by Gomez-Ullate, Kamran, and Milson [50]. As explained below, use of REs of the DPT potential and isotonic oscillator does not allow one to construct all the sequences of $X_m$-Jacobi and $X_m$-Laugerre orthogonal polynomials and therefore cannot be used as a substitute for the accurate theory developed by Gomez-Ullate, Kamran, and Milson [1, 2, 24, 35] (see also [42] for more details).

The purpose of this paper is to demonstrate that the RPs quantized via 'exceptional orthogonal polynomial systems' (X-OPSs) represent some particular cases of a more general family of CES potentials [11, 17, 19] which are obtained from the exactly solvable GRef potentials using SUSY transformations with rational superpotentials. The benchmark of the new approach is that that eigenfunctions of bound states in the rational CES potential constructed in such a way are expressible in terms of polynomial solutions of an energy-dependent manifold of Heine equations [9, 10]. We refer to these solutions as 'Gauss-seed' (GS) Heine polynomials. In the limiting cases of RSL problems with energy-independent characteristic exponents at both end points the coefficient of the first derivatives becomes energy-independent and as a result all GS Heine polynomials from the given sequence turn into solutions of the same RSL equation. One of these sequences is formed by orthogonal Heine polynomials and thereby coincides with the appropriate X-OPS.

Our original interest in this problem was stimulated by the pioneering work of Cooper, Ginoccio, and Khare [47] who opened a new direction in the theory of exactly-quantized RPs by applying the state-erasing DT to the generic GRef' potential quantized via Jacobi polynomials [12]. While analyzing their work, the author [11] made several important observations briefly summarized below and then elaborated in more details in Section 2.

First, we realized that eigenfunctions for the Coop er-Ginoccio-Khare (CGK) potential [47] have the so-called 'quasi-algebraic' [48] form in the sense that they turn into algebraic functions if all (generally irrational) power exponents happened to be rational numbers. In following [49], Gomez-Ullate, Kamran and Milson [35] (as well as Grandati in [31]) refer to these solutions as 'quasi-rational' to stress that their logarithmic derivatives are rational functions. However, for the purpose of this paper it seems more essential to stress another important feature of these solutions, namely, as pointed in [4] with regard to the Heun equation [3], they are 'valid in the entire plane, except of course, at the singularities and, in most cases, with various cuts made to ensure single-valueness'. For this reason we prefer to refer to them as 'almost-everywhere holomorphic' (AEH) solutions. When necessary, the AEH solutions can be converted into polynomials by making the appropriate gauge transformations. In other word, in following Bose [50] we move in the direction opposite to conventional one [26, 51] utilized in works of Levai [52, 53] and (in following him) by Quesne [6] who started from the Fuschian second-order equation with polynomial coefficients and then converted it to the 'normal' [50] (self-adjoin 'canonical [54]) to construct RPs solvable in polynomials.

Secondly, to preserve the rational form of the canonical SL problem, one should use the so-called 'generalized Darboux transformation' (GDT), originally introduced by Rudyak and Zakhariev [55] in the scattering theory and then studied more cautiously by Leib and of Schnizer [56-58] and independently by Suzko [59-61] in the same context. Keeping in mind that the factorization of the SL equation with no first derivative in terms of GDTs is nothing but a particular case of the factorization of the generic Fuschian equation suggested by Gomez-Ullate, Kamran, and Milson [34], it seems more appropriate to use the term 'canonical Darboux transformations' (CDTs), instead of 'GDTs' at least for applications of our interest since.

Our initial studies of RSL equations with at least two AEH solution [11] revealed the following important feature repeatedly exploited in this series of publications [11, 17-19]:

*i*) *any CDT using one of AEH solutions as the FF leads to the partner RSL equation sharing the density function with the original one;*

*ii) the canonical Darboux (CD) transform of any other AEH solution is itself an AEH solution of the partner equation.*

The direct consequence of the cited result is that the partner equation is quantized via GS Heine polynomials if this true for the original RSL problem. Disappointedly, our draft [11] overlooked Levai and Roy's study [7] on REs of the isotonic oscillator. It directly followed from our analysis that the resultant potentials can be quantized via confluent Heine (*c*-Heine) polynomials.

Another important work overlooked in [11] is Gomez-Ullate, Kamran, and Milson's break-through analysis [25] of positions of zeros of 'quasi-algebraic' solutions (in terms of [11, 17]) for the Morse and symmetric h-PT potentials. As a result, we focused solely on 'basic' AEH solutions ('basic quasi--algebraic kernels' in terms of [11, 17]) associated with polynomials of zeroth-order and thereby nodeless by definition. The author ran into these solutions many years ago [65] while analyzing zeros of the Jost function [67] for Darboux transforms of the generic centrifugal-barrier (CB) potential on the half-line ($0 < r < \infty$) and using the hyperbolic Poschl-Teller (h-PT) potential [15] as an example. (Some elements of the author's approach can be traced in later works of Sukumar [68] and Bauer [69, 70].) In particular, it was proven that DTs erasing any solution regular either at 0 or ∞ (types **a** and **b**) lead to isospectral potentials, nearly 10 years before similar results were derived in a more general way using the SUSY factorization of the Schrödinger equation on the line [71-69]. It was also shown that one could add the new bound state below the ground energy level of the original potential using a nodeless solution irregular at both end points (type **d**), approximately at the same time as Deift and Trubowitz [74, 75] came up with similar results in their fundamental studies on mathematical aspects of the inverse scattering.

A remarkable feature of the CKG potential (like any other potential generated by means of basic FFs) is that it is exactly solvable, i.e., its parameters can be varied without changing locations of singular points. On the contrary, all REs of shape-invariant potentials can be obtained by polynomial truncation [7] of Junker and Roy's FFs [8] and therefore are CES in terms of [8].

A fundamentally important distinction between different families of RPs comes from the fact that we use the density function of a very special structure – the so-called 'tangent polynomial' (TP) of the order K ≤ 2 divided by $z^2(1-z)^2$ or by $\zeta^2$ for the *r*-GRef or *c*-GRef potentials (exactly solvable via hypergeometric and confluent hypergeometric functions [12] , respectively). In the generic case of the second-order TP with only 'outer' roots (i.e., either with real roots outside of the quantization interval or with negative discriminant), CDTs erasing basic solutions of the *r*-GRef SL equation lead to RPs quantized in terms of GS Heine polynomials which satisfy the Fuschian equation with 5 regular singular points (including infinity). In case of the TP with a single outer root the latter equation turns into the Heun equation and the appropriate potentials are quantized in terms of Heun polynomials [3, 4]. It was also proven [11] that SL equations utilizing TPs with positive discriminant have exactly four basic solutions, each of its own $\mathsf{a}$, $\mathsf{b}$, $\mathsf{c}$, and $\mathsf{d}$ so that the appropriate *r*-GRef potential has three other SUSY partners, in addition to the CGK potential. If the TP with positive discriminant has no outer roots (which is true for both *t*- and *h*-versions of the PT potential), then all four SUSY partners collapse into the appropriate shape-invariant potential. In [19] most of the results proven in [11, 17] were extended to *c*-GRef potentials quantized via GL polynomials.

As discussed in Section 2, there are two types of commonly-applicable Heine-polynomial generators (HPGs). The HPGs of the first type referred to as 'general' ($g$−HPGs) are applicable to an arbitrary GRef potential on the line. In particular, as demonstrated in [11], they can be used to generate 'quantized' Heine polynomials describing bound states in the CKG potential. Each *r*- or *c*-manifold of GRef potentials on the line has only one shape-invariant limit: the Rosen-Morse [76] or Morse [77] potential, respectively. Their REs have been recently constructed by Quesne [78, 79] using the first-order differential operators labeled as types I, II and III which correspond to CDTs with FFs of type a, b and d in our terms. (Grandati's prescription [80] for rational extensions of the Rosen-Morse potential requires a more sophisticated analysis which will be presented in a separate publication [18].) Selecting regular AEH solution below the ground energy allowed Quesne [78] to select nodeless AEH solutions and thus to explicitly assure that the resultant REs of the Rosen-Morse potential do not have singularities within the quantization interval. For solutions of type d [18] one can make use of the Hubert-Klein [81, 82] formula in [10], similarly to the discussion presented Appendix C below for two versions of the PT potential.

The $\mathcal{g}$-HPGs are also applicable to CDTs with FFs of type b and d for GRef potentials on the half-line (including the isotonic oscillator) as well as to the Darboux transforms of the DPT potential using FFs of type d. To treat CDTs with FFs of type a for GRef potentials on the half-line one has to use polynomial generators of a different kind referred to here as 'abridged HPGs' ($\mathcal{a}$-HPGs). Since the DPT potential is the only GRef potential quantized with boundary conditions at the ends of a finite interval, CDTs with regular FFs for require a special consideration. One thus has to re-define HPGs for CDTs with FFs of both types a and b, which is done in subsection 4.1.

As already mentioned in connection of Quesne's study of REs of the Rosen-Morse potential in [79], the important feature of regular solutions (types a and b) below the ground energy level of any potential on the line is that they are necessarily nodeless. However, when dealing with radial potentials one has to define a 'regular' solution in more precise terms. As clarified in Appendix A, in case of a CB potential on the half-line there is always a range of characteristic exponents at the origin where conventional rules of SUSY quantum mechanics become invalid. Since successful DTs of type b inévitably bring potential parameters into this region, the number of DTs with factorization functions of this type is also limited. In case of the DPT potential the same is true for factorization functions of type a, keeping in mind that the potential has the CBs at both end points.

As far as potential parameters stay within the validity range of SUSY quantum mechanics, Darboux transforms of solutions of type a and b below the ground energy level remain nodeless. We will repeatedly use this observation both in Part II and in the following papers [18, 19] to construct multi-step SUSY partners of *r*-GRef and *c*-GRef potentials.

Similarly to solutions of the Schrödinger equation with the generic *r*-GRef potential [12], RSL problems for most shape-invariant potentials generally have energy-dependent characteristic exponents at least at one of the end points. As a result, Heine polynomials describing bound energy states in those potentials do not form orthogonal sequences and will be covered in a separate publication [18]. Two potentials introduced in Section 3: the DPT potential and isotonic oscillator quantized in terms of Jacobi and GL polynomials with energy-independent indices represent two remarkable exclusions from the general rule. The direct consequence of this observation is that their RE are quantized in terms of infinite orthogonal sets of Heine polynomials commonly referred to as X-OPSs.

It should be stressed that our definition of the *r*-GRef potential automatically assumes that its argument lies between 0 and 1. Since one can always convert the hypergeometric equation from the finite interval (0, 1) onto the infinite intervals $(-\infty, 0)$ or $(1, +\infty)$ by linear fractional transformations, it was assumed that such a choice can be done without loss of generality. Our original perception [11, 12] changed due to Odake and Sasaki'a discovery [20] that the Schrödinger equation with the h-PT potential converted into the hypergeometric equation on the infinite interval $(1, +\infty)$ can be quantized in terms of an orthogonal subset of polynomials. It could be easily shown that these are nothing but Romavsky-Jacobi polynomials [83] as they are referred to by Lesky [84] (see also [85, 86]). It turns out that their Darboux transforms also form finite sets of orthogonal polynomials which makes the infinite quantization interval preferable in this particular case.

In Section 4 we present explicit expressions for characteristic exponents of AEH solutions of type $\mathfrak{a}$ and $\mathfrak{b}$ at the singular end points for RSL problems associated with both versions of the PT potential and then select the ones lying below the appropriate ground energy level. In following [42], we also make use of the Hilbert-Klein formulas [10, 81, 82] for distributions of zeros of Jacobi polynomials with indexes of opposite sign to confirm that factorization functions of our interest for the DPT potential do not have nodes inside the quantization interval. It is verified that appropriately scaled orthogonal '*J*-Heine' polynomials coincide with $X_m$-Jacobi polynomials in Gomez-Ullate, Kamran, and Milson's theory of exceptional orthogonal polynomials [1, 2, 24, 35] (see also [42] for more details).

We also demonstrated that there are many so-called 'parse' sequences of orthogonal Heine polynomials generated using DTs with FFs of type **d**. The first representative of these sequences has been already discovered by Quesne [13] using a second-order nodeless polynomial to generate a SUSY partner of the DPT potential with an inserted ground state (case III). The common puzzling feature of these sequences is that they all consist of a constant and a regular sub-sequence starting from a GS Heine polynomial of the order larger than 1.

For completeness we also included the isotonic oscillator though in the latter case we mostly confirm that the constructed SUSY partners match Grandati's cases I, II, and III [31]. It has been also explicitly demonstrated that appropriately scaled quantized Heine polynomials coincide with $X_m$-Laguerre orthogonal as defined in [35, 42].

In Section 5 we briefly outline some remarkable features of the Heine equation associated with Quesne's linear case [13] for the DPT potential, namely, four CDTs of the hypergeometric equation with FFs of types **a**, **b**, **c**, and **d** result in the same Heun equation which is CES via four infinite sequences of Heun polynonials.

Finally, Section 6 summarizes main conclusions of the paper and outlines directions for future development.

The paper is accompanied by three Appendices. In Appendix A we illuminate some limitations of SUSY quantum mechanics for radial potentials. Appendix B outlines general features of the Leeb-Schnizer-Susko theory of GDTs (CDTs in our current terms). Appendix C relates the Hilbert-Klein formulas for positions of zeros of Jacobi polynomials to nodeless AEH solutions of the RSL equation converted into the Schrödinger equation with either t- or h-PT potential by the appropriate Liouville transformation.

## 2. General Recipe for Constructing REs of Shape-Invariant GRef Potentials

Let us start our analysis from the RSL problem

$$\left\{\frac{d^2}{d\xi^2} + I[\xi;\varepsilon \mid {}_\iota\mathcal{G}]\right\}\Phi[\xi;\varepsilon \mid {}_\iota\mathcal{G}] = 0 \qquad (2.1)$$

exactly quantized via Jacobi ($\iota = 1$) and GL ($\iota = 0$) polynomials. In following Milson's suggestion [54], we refer to the energy-dependent PF

$$I[\xi;\varepsilon \mid {}_\iota\mathcal{G}] \equiv {}_\iota I^o[\xi;\varepsilon \mid {}_\iota Q^o; \mathcal{T}_K] = {}_\iota I^o[\xi \mid {}_\iota Q^o] + \varepsilon\, {}_\iota\wp[\xi;\, {}_\iota\mathcal{T}_K], \qquad (2.2)$$

as the Bose invariant [50]. The density function of our current interest thus has the form

$${}_\iota\wp[\xi;\, {}_\iota\mathcal{T}_K] = \frac{{}_\iota T_K[\xi]}{4\xi^2(1-\iota\xi)^2}, \qquad (2.3)$$

where K is the order of the TP mentioned in Introduction:

$${}_1T_K[z] = c_0(1-z)^2 + c_1 z^2 + {}_1d\, z(1-z) \qquad (2.4)$$

($\xi$ =z) or

$${}_0T_K[\zeta] = {}_0a\zeta^2 + {}_0d\,\zeta + c_0 \qquad (2.4')$$

($\xi$ =$\zeta$) for $\iota$ = 1or 0, respectively. We say that each Bose invariant (2.2) describes a PF beam $\mathsf{B}$. In particular, Gauss PF (GPF) beams studied in [12] are associated with the reference PFs

$${}_\iota I^o[\xi;\, {}_\iota Q^o] = -\frac{h_{o;0}}{4\xi^2} - \frac{h_{o;1}}{4(1-\iota\xi)^2} + (-1)^\iota \frac{{}_\iota O_0^o}{4\xi(1-\iota\xi)}. \qquad (2.5)$$

Note that we slightly changed the parameterization compared with our earlier works [12, 95, 96], with $h_{o;0}$ and $h_{o;1}$ standing for $h_0$ and $h_1$. As for the coefficient ${}_1O_0^o$, it is the combination:

$${}_1O_0^o = h_{o;0} + h_{o;1} - f_o \qquad (\iota = 1), \qquad (2.6)$$

with $f_0$ used instead of f in [12]. For the *c*-GRef potential the coefficients $h_{0;1}$ and ${}_0O_0^o$ stand respectively for $g_2$ and $-g_1$ in terms of [12]. We refer to the appropriate RPs as the *r*- and *c*-GRef potentials

$$V[\xi(x) \mid {}_\iota\mathcal{G}] = -{}_\iota\wp^{-1}[\xi(x); {}_\iota T_K]\ {}_\iota I^o[\xi(x) \mid {}_\iota Q^o] - \tfrac{1}{2}\{\xi, x\} \tag{2.7}$$

to distinguish them from the one ('*i*-GRef') introduced by Milson [54] when the appropriate Bose invariant has complex-conjugated singular points -*i* and *i* ($\iota = i$) so that

$${}_i\wp[\eta; {}_i T_2] = \frac{{}_i T_2[\eta]}{4(\eta^2 + 1)^2}. \tag{2.8}$$

As pointed by Roychoudhury et al [88] , one needs to differentiate between parameters ${}_\iota Q^o$, defining the reference PF ${}_\iota I^o[\xi \mid {}_\iota Q^o]$ and the coefficients of the TP in the numerator of density function (2.3), which specifies the change of variable z(x) used to convert the RSL problem with Bose invariant (2.2) into the 1D Schrödinger equation. In our definition of the PF beam the density function and therefore the appropriate change of variable are always fixed. A set of parameters ${}_\iota Q^o$ defines individual PF rays in the given beam and for this reason are referred to below as 'ray identifiers'. The term 'PF ray' is thus an alternative name for Bose invariant (2.1) with fixed values of all the parameters except than the energy $\varepsilon$. Each PF ray corresponds to a single potential curve when the RSL problem in question is converted into the Schrödinger equation via the appropriate Liouville transformation [89-94].

The type of the RSL problem for the Gauss PF (GPF) beam can be specified by three indexes: the TP order K, the number $\Im$ of outer roots, and the maximum degeneracy $\aleph$ of both zero and nonzero roots which, when necessary, are added as superscripts to the PF beam notation: ${}_\iota\mathcal{G}^{K\Im\aleph}$. Any shape-invariant potential is thus represented by one of the GPF beams ${}_{\iota}\mathcal{G}^{K0\aleph}$, namely, Rosen-Morse [76] and Manning-Rosen[x)] [97] potentials and both *h*- and *t*-versions of the PT potentials are represented by the *r*-GPF beams ${}_{1}\mathcal{G}^{000}$, ${}_{1}\mathcal{G}^{202}$, ${}_{1}\mathcal{G}^{101}$, and ${}_{1}\mathcal{G}^{201}$, respectively, whereas three confluent potentials: the Morse oscillator, the effective Coulomb potential, and the isotonic oscillator correspond to the *c*-GPF beams ${}_{0}\mathcal{G}^{000}$, ${}_{0}\mathcal{G}^{202}$, and ${}_{0}\mathcal{G}^{101}$. Note that K = ι+1 in the cases of our current interest: the DPT potential and isotonic oscillator.

For the generic *r*-GRef potential mapped onto the quantization interval 0 < z < 1 one can always choose the energy reading point in such a way ($h_{o;1} = -1$) that the indicial equation for the upper end has real or complex-conjugated roots for negative and positive energies, respectively. As a result, the appropriate potential $V[\xi(x);{}_{1}\mathcal{G}]$ approaches zero as $x \to +\infty$. The DPT potential as well as its confluent limiting case: isotonic oscillator represent the only exclusion from this rule because the indicial equations for both singular end points become energy-independent.

The *c*-GRef potentials represent a more complicated case [19] because the reference PF generally tends to a constant as $\xi \to \infty$. The requirement for *c*-GRef potentials to have nonnegative values at both ends leads to two separate families of potentials depending on the location of the potential barrier [19]. Two families intersect on the so-called 'asymptotically-leveled' potential curves approaching zero at both ends.

It has been already shown in [12] that the bound eigenfunctions associated with the $v^{th}$ energy levels $\varepsilon_v$ in the *r*-GRef and *c*-GRef potentials have the form:

$$\phi_v[\xi\,|\,{}_{\iota}\mathcal{G}_{\downarrow v}] = {}_{\iota}\Theta[\xi;\lambda_v,\nu_v]\;{}_{\iota}P_v[\zeta;\lambda_v,\nu_v] \qquad (\lambda_v,\,\nu_v > 0), \tag{2.9}$$

where

---

[x)] Both Quesne [79] and Grandati [80] refer to this potential as the Eckart potential [98]; however, we prefer to keep Infeld and Hull's terminology [99] since it was Manning and Rosen who first derived the explicit formula for bound energy levels in this potential whereas Eckart's work was devoted exclusively to electron scattering.

$$ {}_1\Theta[z;\lambda,\nu] \equiv \sqrt{|z(1-z)|}\,|z|^{\lambda/2}|1-z|^{\nu/2}, \qquad (2.10) $$

and

$$ {}_0\Theta[\zeta;\lambda,\nu] \equiv \sqrt{\zeta}\,\zeta^{\lambda/2}\exp[-\nu\zeta/2]; \qquad (2.10') $$

It has been already shown in [12] that the bound eigenfunctions associated with the $v^{th}$ energy levels $\varepsilon_v$ in the *r*-GRef and *c*-GRef potentials have the form:

$$ {}_1P_v[z;\lambda_v,\nu_v] \equiv P_v^{(\lambda_v,\nu_v)}(2z-1), \qquad (2.11) $$

$$ {}_0P_v[\zeta;\lambda_v,\nu_v] \equiv L_v^{(\lambda_v)}(\nu_v\zeta). \qquad (2.11') $$

(By definition the sub-beam ${}_\iota G_{\downarrow\tau}$ is formed by PF rays with the ray identifiers restricted to a sub-domain $\tau$. In particular, the notation ${}_\iota G_{\downarrow v}$ means that we deal with the region where the potential in question has at least v bound energy states.) For the reason explained in next Section, the so-called 'Gauss AEH weights' are defined via (2.10) in such a way that they can be applied to any of the quantization intervals: $z < 0$, $0 < z < 1$, or $z > 1$.

It has been proven in [17] that the discrete energy spectrum for the *r*-GRef potential generated by means of the TP with a *positive* discriminant is always accompanied by three basic solutions of type $\mathbf{t}$ = a, b, and d:

$$ {}_1\phi_{\mathbf{t}}[z\,|\,{}_1G_{\downarrow\mathbf{t},m}] = {}_1\Theta[z;\lambda_{\mathbf{t}},\nu_{\mathbf{t}}] \qquad (2.12) $$

where

$$ \lambda_{\mathbf{t}} \equiv \sigma_{0;\mathbf{t}}|\lambda_{\mathbf{t}}| > 0,\ \nu_{\mathbf{t}} \equiv \sigma_{1;\mathbf{t}}|\nu_{\mathbf{t}}| < 0 \quad \text{for } \mathbf{t} = \mathrm{a}; \qquad (2.13a) $$

$$ \lambda_{\mathbf{t}} \equiv \sigma_{0;\mathbf{t}}|\lambda_{\mathbf{t}}| < 0,\ \nu_{\mathbf{t}} \equiv \sigma_{1;\mathbf{t}}|\nu_{\mathbf{t}}| > 0 \quad \text{for } \mathbf{t} = \mathrm{b}; \qquad (2.13b) $$

$$ \lambda_{\mathbf{t}} \equiv \sigma_{0;\mathbf{t}}|\lambda_{\mathbf{t}}| > 0,\ \nu_{\mathbf{t}} \equiv \sigma_{1;\mathbf{t}}|\nu_{\mathbf{t}}| > 0 \quad \text{for } \mathbf{t} = \mathrm{c}; \qquad (2.13c) $$

$$ \lambda_{\mathbf{t}} \equiv \sigma_{0;\mathbf{t}}|\lambda_{\mathbf{t}}| < 0,\ \nu_{\mathbf{t}} \equiv \sigma_{1;\mathbf{t}}|\nu_{\mathbf{t}}| < 0 \quad \text{for } \mathbf{t} = \mathrm{d}. \qquad (2.13d) $$

The author has used both ground eigenfunction (in following Cooper, Ginoccio, and Khare [47]) and three other solutions irregular at least at one of the end points to generate rational potentials quantized in terms of GS Heine polynomials. Similarly, basic solutions ${}_{0}\Theta[\xi;\lambda_{\dagger},\nu_{\dagger}]$ are used in [19] to construct exactly quantized SUSY partners of the *c*-GRef potential. For both *r*-GRef and *c*-GRef potentials there are three cases of the TP with a single outer root ($\Im$=1);

i) TP with zero root (ZRtTP) associated with the radial potential [95, 96];

ii) linear TP (LTP);

iii) TP with a double root (DRtTP)

so that the appropriate SUSY partners generated using basic FFs is exactly quantized via Heun [3, 4] and *c*-Heun [4, 5] polynomials for $\iota$ = 1or 0, respectively. This thus adds 3 new members to the families of rational potentials [100] exactly solvable via *c*-Heun functions. On other hand, the author is unaware of any other example of rational potentials which can be exactly solvable via Heun functions both below and above zero energy. In this connection with recent works of Sasaki, Takemura, and Ho [101, 102] on second-order higher-degree apparent singularities, it is worth pointing out that the outer singularity in the exactly-solvable partner RSL equation generated using the TP with a double root also has the lower characteristic exponent of –2).

In [17] we have extended quantization condition [12] for energies of bound states in the *r*-GRef potential to an arbitrary AEH solution:

$$\phi_{\dagger,m}[z\,|\,{}_{1}G_{\downarrow\dagger,m}]=\sqrt{|\,z(1-z)\,|}\,|\,z\,|^{\lambda_{0;\dagger,m}/2}\,|1-z\,|^{\lambda_{1;\dagger,m}/2}\,\Pi_{m}[z;\bar{z}_{\dagger,m}]; \quad (2.14)$$

namely, it was proven that the appropriate factorization energies satisfy the set of algebraic equations

$$[\sigma_{0;\dagger}\lambda_{0}({}_{1}\varepsilon_{\dagger,m})+\sigma_{1;\dagger}\lambda_{1}({}_{1}\varepsilon_{\dagger,m})+2m+1]^{2}=f_{o}+1-{}_{1}a\;{}_{1}\varepsilon_{\dagger,m}, \quad (2.15)$$

where

$$\lambda_{r}(\varepsilon)\equiv\sqrt{h_{o;r}+1-c_{r}\varepsilon}\,. \quad (2.16)$$

Except the DPT potential discussed in subsection 4.1 below, the parameter $h_{o;1}$ is set to –1 so that the given *r*-GRef potential vanishes as $x\to\infty$. As pointed by Grosche [103], the quantization condition in question ($\dagger$ = c) can be reduced to sequences of quartic equations with respect the bound energies $\varepsilon_{v}$ (v=0,1,...) for both *r*- and *c*-GR potentials.

By setting $h_{o;1}$ to –1, one can easily convert the set of algebraic equations (2.15) and (2.16) into a similar quartic equation with respect to $\lambda_{1;\mathbf{t},m} \equiv \sigma_{1;\mathbf{t}}\lambda_1({}_1\varepsilon_{\mathbf{t},m})$ for any $\mathbf{t}$ = a, b, c, or d. It can be proven [17, 19] that algebraic equation (2.15) has a single solution for each real root of this quartic equation which implies that the number of AEH solutions of the given order m may not exceed 4 for any *r*-GRef potential. (AEH solutions for *c*-GRef potential represent a more challenging problem which will be discussed in detail in [19].) If the TP has a double root then the leading coefficient in the mentioned quartic equation vanishes and the number of AEH solutions of the given order m does not exceed 3. REs of both Manning-Rosen potential and effective Coulomb potential constructed by Quesne [78] and Grandati [104] obey this rule.

While focusing on basic solutions of a different type, the author [11, 17] wrongly assumed that nodeless AEH solutions (2.14) formed by higher-order polynomials

$$\Pi_m[z;\bar{z}_{\mathbf{t},m}] \equiv \Pi[z;\bar{z}_{\mathbf{t},m}] \equiv \prod_{r=1}^{m}(z - z_{\mathbf{t},m;r}) \qquad (2.17)$$

represent some exotic cases rather than the general rule. A turn-around in our way of thinking took place under influence of Gomez-Ullate, Kamran, and Milson’s discovery [25] of rationally-extended Morse oscillators (especially thank to an excellent exposition of the KLH theorem presented by Grandati [31, 32]). The recent work of Gomez-Ullate, Marcellan, and Milson [42] drew our attention to the Hubert-Klein [81, 82] formulas in [10] making it possible to analyze *r*- and *c*-GRef potentials in parallel.

The common remarkable feature of both versions of the PT potentials as well as isotonic oscillator is that the m$^{th}$ bound eigenfunction is accompanied by three AEH of order m:

$$\phi_{\mathbf{t},m}[\xi \mid {}_\iota G_{\downarrow\mathbf{t},m}] = {}_\iota\Theta[\xi; {}_\iota\lambda_{\mathbf{t},m}, {}_\iota\nu_{\mathbf{t},m}]\, \Pi_m[\xi; {}_\iota\bar{\xi}_{\mathbf{t},m}] \qquad (2.18)$$

with $\mathbf{t}$ = a, b, and d. Co-existence of four AEH soulutions of the same order m is closely related to the fact that the TP used to generate the appropriate GPF beam has positive discriminant. For any other shape-invariant potential the polynomial discriminant is equal to zero [18] and the appropriate RSL equation has only 3 AEH solutions of the same order m (see [78, 79] for particular examples).

Using generalized expression (2.18) for an AEH solution of order m one can represent the appropriate superpotentials as

$$W_{\dagger;m}[z \mid {}_1G] = \tfrac{1}{2} z' \,[(1+ {}_\iota\lambda_{\dagger,m})/z - ([1+ {}_\iota\lambda_{\dagger,m})/(1-z)] \\ + \tfrac{1}{4} ld\, {}_1\wp[z; {}_1T_K] + ld\, \Pi_m[z; \bar{z}_{\dagger,m}] \} \tag{2.19}$$

and

$$W_{\dagger,m}[\zeta \mid {}_0G] = \zeta' \{ \tfrac{1}{2}(1+ {}_0\lambda_{\dagger,m})/\zeta - \tfrac{1}{2}\nu_{\dagger,m} \\ + \tfrac{1}{4} ld\, {}_0\wp[\zeta; {}_0T_K] + ld\Pi_m[\zeta; \bar{\zeta}_{\dagger,m}] \} \tag{2.19′}$$

where prime and dot denote derivatives with respect to x and $\xi$= z or $\zeta$, respectively, and the symbol *ld* stands for the logarithmic derivative. It crucial that each term in figure brackets represents a PF. For the isotonic oscillator

$${}_0\wp[\zeta; {}_0T_1] = {}_0d\,\zeta^{-1},\ \zeta' = 2\sqrt{\zeta / {}_0\zeta},\ \ \zeta = {}_0dx^2 \tag{2.20}$$

(${}_0d$=C/4 in Quesne's terms [6]) so that (2.19') turns into superpotential (7) in Levai and Roy's paper [7 under assumption that the appropriate GL polynomial has only real roots.

The polynomials $\Pi_m[\xi; {}_\iota\bar{\xi}_{\dagger,m}]$ are solutions of the following Fuschian equation with zero characteristic exponents at any finite singular point:

$$[{}_\iota\hat{D}({}_\iota\lambda_{\dagger,m}, {}_\iota\nu_{\dagger,m}) + (-1)^\iota\ {}_\iota C_0(\dagger, m) + \tfrac{1}{4}{}_\iota d\ {}_\iota\varepsilon_{\dagger,m}]\Pi_m[\xi; {}_\iota\bar{\xi}_{\dagger,m}] = 0, \tag{2.21}$$

where

$${}_\iota\hat{D}(\lambda,\nu) \equiv \xi(1-\iota\xi)\frac{d^2}{d\xi^2} + 2(-1)^\iota\ {}_\iota B_1[\xi;\lambda,\nu]\frac{d}{d\xi}, \tag{2.22}$$

$${}_\iota B_1[\xi;\lambda,\nu] = (-1)^\iota\ \tfrac{1}{2}[(\lambda+1)(1-\iota\xi) - (\nu+\iota)\xi], \tag{2.22*}$$

$${}_\iota C_0(\dagger, m) = {}_\iota C_0^0(\lambda_{\dagger,m}, \nu_{\dagger,m}) + \tfrac{1}{4}\,{}_\iota O_0^o, \tag{2.23}$$

and

$${}_\iota C_0^0(\lambda,\nu) = \tfrac{1}{2}(\lambda+1)(\nu+\iota). \tag{2.23*}$$

Since any regular solution lying below the ground energy level ${}_\iota\varepsilon_0$ may not have nodes inside the quantization interval, polynomials $\Pi_m[\xi; {}_\iota\bar{\xi}_{\dagger,m}]$ have only outer roots if ${}_\iota\varepsilon_{\dagger,m} < {}_\iota\varepsilon_0$ for $\dagger = \mathrm{a}$ or $\mathrm{b}$. As discussed below, DTs do not change the type of the AEH solution [11, 19] as far as the latter retains within the validity range of SUSY quantum mechanics. In particular, this implies that Darboux transforms of AEH solutions (2.18) of types a and b with energies ${}_\iota\varepsilon_{\dagger,m} < {}_\iota\varepsilon_0$ do not have nodes inside the quantization interval. In Part II we use this general feature of DTs with regular FFs to construct multi-step sequences of orthogonal Heine polynomials.

By applying the Leeb-Schnizer-Suzko reciprocal formula (B6) in Appendix B to the *r*- and *c*-GRef potentials generated by means of the TP

$$ {}_\iota T_K[\xi;\aleph] = {}_\iota d\, \xi^{\aleph}(1-\iota\xi)^{K-\aleph} \tag{2.24}$$

(and commonly referred to as 'shape-invariant' potentials in following English translation[x)] of Gendenshtein's visionary paper [105]) one can represent the FF for the reverse CDT as

$$ {}^*\phi_{\dagger,m}[\xi \mid {}_\iota G^{K0\aleph}_{\downarrow\dagger,m}] = \Theta^{(\iota)}_{\dagger,m}[\xi; {}^1_\iota\lambda_{\dagger,m}, {}^1_\iota\nu_{\dagger,m}], \tag{2.25}$$

where the parameters of the AEH weight

$$ \Theta^{(\iota)}_{\dagger,m}[\xi;\lambda,\nu] \equiv \frac{{}_\iota\Theta[\xi;\lambda,\nu]}{\Pi_m[\xi; {}_\iota\bar{\xi}_{\dagger,m}]} \tag{2.26}$$

are defined as follows

$$ {}^1_\iota\lambda_{\dagger,m} = -\lambda_{\dagger,m} - \aleph, \tag{2.27a}$$

and

$$ {}^1_\iota\nu_{\dagger,m} = -\nu_{\dagger,m} - K + \aleph. \tag{2.27b}$$

---

*) It should be mentioned that in the original (Russian) versions of Gendenshtein's papers [105, 106], he uses the Russian equivalent of the English term 'form-invariance', retained in the translation of his joint paper with Krive [106]. The broadly accepted term: 'shape-invariance' first appearing in the English translation of [105] is nothing but a translation slip.

Note that $\aleph$ coincides with the TP order K for any shape-invariant potential except the DPT potential (assuming that the h-PT potential is mapped onto the interval (0,1) like any other r-GRef potential). This implies that

$$ {}_{\iota}^{1}\nu_{\dagger,m} = -\nu_{\dagger,m} \quad (\aleph = K). \tag{2.27*}$$

Since $\aleph = K = 0$ for both Rosen-Morse and Morse potentials the appropriate CDTs do not change exponent differences at ξ=0 -- a typical feature of *r*- and -GRef potentials on the line.

As outlined in Appendix A, DTs of CB potentials necessarily change the characteristic exponents at the origin. A good news is that the new characteristic exponents remain to be energy-independent. Taking into account that the TP for a radial GRef potential (except the Manning-Rosen and effective Coulomb potentials with $\aleph$ =2) has the form:

$$ {}_{\iota}T_K[\xi] = {}_{\iota}a\,\xi(\xi - \xi_{T;1}), \tag{2.28}$$

we find that

$$ {}^{*}\phi_{\dagger,m}[\xi \mid {}_{\iota}G^{K\Im 1}] \sim \xi^{-\sigma_{0;\dagger}\lambda_o/2} \quad \text{near } \xi = 0. \tag{2.29}$$

so that the exponent difference at the singular point ξ=0 in the partner RSL equation is shifted from $\lambda_o$ by 1:

$$ \lambda_{o;\dagger} = |\,\sigma_{0;\dagger}\lambda_o + 1\,|. \tag{2.30}$$

The CB parameter $s_o$ is related to the exponent difference $\lambda_o$ in a trivial fashion

$$ \lambda_o = |\,s_o - \tfrac{1}{2}\,| \tag{2.31}$$

Taking into account that [95, 96]

$$ s_o = \lambda_o + \tfrac{1}{2} \quad \text{for} \quad s_o > \tfrac{1}{2} \tag{2.31*}$$

we conclude that the derived expression for changes in the exponent difference at the origin is just an alternative representation for the general rule [65] prescribing the change in CB parameter (2.31*) under action of DTs of different types $\dagger$ = a, b, c, or d. As a direct consequence of the discussion presented in Appendix A, parameter (2.31) is required to be larger than 1 to stay within the validity range of the SUSY quantum mechanics.

The superpotentials associated with AEH solutions (2.22) has the form

$$
\begin{aligned}
{}^{*}W_{\dagger;m}[z\,|\,{}^{1}_{1}G^{K0\aleph}_{\dagger,m}] = &-\tfrac{1}{2}z'[(1+{}^{1}_{1}\lambda_{\dagger,m})/z-(1+{}^{1}_{1}\nu_{\dagger,m})/(1-z)] \\
&+\tfrac{1}{4}\mathrm{ld}\,{}_{1}\wp[z;{}_{1}T_{K}^{(\aleph)}]-\dot{\Pi}_{m}[z;\bar{z}_{\dagger,m}]/\Pi^{2}_{m}[z;\bar{z}_{\dagger,m}]\}
\end{aligned}
\tag{2.32}
$$

and

$$
\begin{aligned}
{}^{*}W_{\dagger;m}[\zeta\,|\,{}^{1}_{0}G^{K0\aleph}_{\dagger,m}] = \zeta'\{&\tfrac{1}{4}ld\,{}_{0}\wp[\zeta;{}_{0}T_{K}]+\tfrac{1}{2}(1+{}^{1}_{0}\lambda_{\dagger,m})/\zeta-\tfrac{1}{2}\,{}^{1}_{0}\nu_{\dagger,m} \\
&+\tfrac{1}{4}\mathrm{ld}\,{}_{1}\wp[\zeta;{}_{0}T_{K}^{(\aleph)}]-\dot{\Pi}_{m}[\zeta;\bar{\zeta}_{\dagger,m}]/\Pi^{2}_{m}[\zeta;\bar{\zeta}_{\dagger,m}]\}
\end{aligned}
\tag{2.32′}
$$

for $\iota$ = 1 and 0, respectively. Note that Quesne [13] in her original analysis of RPs quantized via exceptional orthogonal polynomials started from superpotentials (2.32) and (2.32′) with K= $\iota$ + 1, $\aleph$= 1 and then looked for all possible combinations of parameters which turn $\Pi_1[\xi;\,{}_{\iota}\bar{\xi}_{\dagger,1}]$ and $\Pi_2[\xi;\,{}_{\iota}\bar{\xi}_{\dagger,2}]$ into either Jacobi ($\iota$ = 1) or GL ($\iota$ = 0) polynomials.

The reference PF in the partner RSL equation

$$
\left\{\frac{d^2}{d\xi^2}+I^{o}[\xi\,|\,{}^{1}_{\iota}G^{K0\aleph}_{\dagger,m}]+\frac{{}_{\iota}d\varepsilon}{4\xi^{2-\aleph}(1-\iota\xi)^{2+\aleph-K}}\right\}\Phi[\xi;\varepsilon\,|\,{}^{1}_{\iota}G^{K0\aleph}_{\dagger,m}]=0 \tag{2.33}
$$

can be thus represented as

$$
I^{o}[z\,|\,{}^{1}_{1}G^{K0\aleph}_{\dagger,m}]=\frac{1-{}^{1}_{1}\lambda^{2}_{\dagger,m}}{4z^2}+\frac{1-{}^{1}_{1}\nu^{2}_{\dagger,m}}{4(1-z)^2}-2\sum_{r=1}^{m}\frac{1}{(z-z_{\dagger,m;r})^2}+\frac{O^{o}_{m}[z\,|\,{}^{1}_{1}G^{K0\aleph}_{\dagger,m}]}{4z(z-1)\Pi_{m}[z;\bar{z}_{\dagger,m}]}
\tag{2.34}
$$

and

$$
I^{o}[\zeta\,|\,{}^{1}_{0}G^{K0\aleph}_{\dagger,m}]=\frac{1-{}^{1}_{0}\lambda^{2}_{\dagger,m}}{4\zeta^2}-2\sum_{r=1}^{m}\frac{1}{(\zeta-\zeta_{\dagger,m;r})^2}+\frac{O^{o}_{m}[\zeta\,|\,{}^{1}_{0}G^{K0\aleph}_{\dagger,m}]}{4\zeta\Pi_{m}[\zeta;\bar{\zeta}_{\dagger,m}]}-\tfrac{1}{4}\,{}^{1}_{0}\nu^{2}_{\dagger,m}.
\tag{2.34′}
$$

In addition to decompositions (2.26) and (2.26′) explicitly specifying coefficients of all the second-order poles, we also introduce two alternative 'partial decompositions' (PD) of the reference PFs. As explained below, the PDs

$$I^{o}[z\,|\,{}^{1}_{1}G^{K0\aleph}_{\dagger,m}] = \frac{1-{}^{1}_{1}\lambda^{2}_{\dagger,m}}{4z^{2}} + \frac{1-{}^{1}_{1}\nu^{2}_{\dagger,m}}{4(1-z)^{2}} + \frac{\breve{O}^{\downarrow}_{m}[z\,|\,{}^{1}_{1}G^{K0\aleph}_{\dagger,m}]}{4z(z-1)\Pi_{m}[z;\bar{z}_{\dagger,m}]} + 2\breve{Q}[z;\bar{z}_{\dagger,m}] \tag{2.35}$$

for ι=1 or

$$I^{o}[\zeta\,|\,{}^{1}_{0}G^{K0\aleph}_{\dagger,m}] = \frac{1-{}^{1}_{0}\lambda^{2}_{\dagger,m}}{4\zeta^{2}} - \tfrac{1}{4}\,{}^{1}_{0}\nu^{2}_{\dagger,m} + \frac{\breve{O}^{\downarrow}_{m}[\zeta\,|\,{}^{1}_{0}G^{K0\aleph}_{\dagger,m}]}{4\zeta\,\Pi_{m}[\zeta;\bar{\zeta}_{\dagger,m}]} + 2\breve{Q}[\zeta;\bar{\zeta}_{\dagger,m}] \tag{2.35′}$$

for ι=0, where

$$\breve{O}^{\downarrow}_{m}[\xi\,|\,{}^{1}_{\iota}G^{K0\aleph}_{\dagger,m}] = -[4\,{}_{\iota}C^{0}_{0}({}^{1}_{\iota}\lambda_{\dagger,m},{}^{1}_{\iota}\nu_{\dagger,m}) + (-1)^{\iota}\,{}_{\iota}d\,{}_{\iota}\varepsilon_{\dagger,m}]\,\Pi_{m}[\xi;{}_{\iota}\bar{\xi}_{\dagger,m}]$$
$$+ 8\,{}_{\iota}B_{1}[\xi;{}^{1}_{\iota}\lambda_{\dagger,m},{}^{1}_{\iota}\nu_{\dagger,m}]\dot{\Pi}_{m}[\xi;{}_{\iota}\bar{\xi}_{\dagger,m}] \tag{2.36}$$

and

$$\breve{Q}[\xi;\bar{\xi}_{\tau}] \equiv \tfrac{1}{2}\ddot{\Pi}[\xi;\bar{\xi}_{\tau}]/\Pi[\xi;\bar{\xi}_{\tau}] - \dot{\Pi}^{2}[\xi;\bar{\xi}_{\tau}]/\Pi^{2}[\xi;\bar{\xi}_{\tau}] \tag{2.37}$$

is preferable as a starting point for gauge transformations so that we refer to it as 'gauge partial decomposition' (GPD).

Taking into account that

$$\dot{\Pi}^{2}[\xi;\bar{\xi}_{\tau}] = \Pi^{2}_{m-1}[\xi;\bar{\xi}^{(r)}_{\tau}] + 2\sum_{r=1}^{m}\Pi_{m-1}[\xi;\bar{\xi}^{(r)}_{\tau}]\sum_{r'>r}\Pi_{m-1}[\xi;\bar{\xi}^{(r')}_{\tau}], \tag{2.38}$$

where we put

$$\Pi_{m-1}[\xi;\bar{\xi}^{(r)}_{\tau}] = \prod_{r'\neq r}(\xi-\xi_{\tau;r'}), \tag{2.39}$$

the $m^{th}$-order polynomials in the right-hand side of (2.34) and (2.34′) can be related to (2.39) via the following explicit expression:

$$O^{o}_{m}[\xi\,|\,{}^{1}_{\iota}G^{K\Im\aleph}_{\dagger,m}] = \breve{O}^{\downarrow}_{m}[\xi\,|\,{}^{1}_{\iota}G^{K\Im\aleph}_{\dagger,m}] + (-1)^{\iota}4\xi(1-\iota\xi)P_{m-2}[\xi;{}_{\iota}\bar{\xi}_{\dagger,m}], \tag{2.40}$$

with

$$P_{m-2}[\xi;\bar{\xi}_\tau] \equiv \ddot{\Pi}[\xi;\bar{\xi}_\tau] - 4\sum_{r'>r} \Pi_{m-1}[\xi;\bar{\xi}_\tau^{(r)}]\Pi_{m-1}[\xi;\bar{\xi}_\tau^{(r')}]. \qquad (2.41)$$

The second PD

$$I^o[z\,|\,{}_1^1G^{K0\aleph}_{\dagger,m}] = \frac{1-{}_1^1\lambda^2_{\dagger,m}}{4z^2} + \frac{1-{}_1^1\nu^2_{\dagger,m}}{4(1-z)^2} + \frac{O^{\downarrow}_m[z\,|\,{}_1^1G^{K0\aleph}_{\dagger,m}]}{4z(z-1)\Pi_m[z;\bar{z}_{\dagger,m}]} + 2Q[z;\bar{z}_{\dagger,m}] \qquad (2.42)$$

for ι=1 or

$$I^o[\zeta\,|\,{}_0^1G^{K0\aleph}_{\dagger,m}] = \frac{1-{}_0^1\lambda^2_{\dagger,m}}{4\zeta^2} - \tfrac{1}{4}\,{}_0^1\nu^2_{\dagger,m} + \frac{O^{\downarrow}_m[\zeta\,|\,{}_0^1G^{K0\aleph}_{\dagger,m}]}{4\zeta\,\Pi_m[\zeta;\bar{\zeta}_{\dagger,m}]} + 2Q[\zeta;\bar{\zeta}_{\dagger,m}], \qquad (2.42')$$

for ι=0, where

$$Q[\xi;\bar{\xi}_\tau] \equiv \ddot{\Pi}[\xi;\bar{\xi}_\tau]/\Pi[\xi;\bar{\xi}_\tau] - \dot{\Pi}^2[\xi;\bar{\xi}_\tau]/\Pi^2[\xi;\bar{\xi}_\tau], \qquad (2.43)$$

was adopted by us from Quesne's works [37-39, 79] and therefore is referred to below as QPD. One can easily verify that the m-order polynomials $O^{\downarrow}_m[\xi\,|\,{}_\iota^1G^{K0\aleph}_{\dagger,m}]$ in the right-hand side of (2.38) and (2.38′) are related to polynomials (2.36) via the simple formula:

$$O^{\downarrow}_m[\xi\,|\,{}_\iota^1G^{K0\aleph}_{\dagger,m}] = \breve{O}^{\downarrow}_m[\xi\,|\,{}_\iota^1G^{K0\aleph}_{\dagger,m}] - (-1)^\iota 4\xi(1-\iota\xi)\ddot{\Pi}_m[\xi;\bar{\xi}_\tau]. \qquad (2.44)$$

Taking into account that the monomial product $\Pi_m[\xi;{}_\iota\bar{\xi}_{\dagger,m}]$ in the right-hand side of (2.36) satisfies hypergeometric equation (2.21) one can represent (2.44) as

$$O^{\downarrow}_m[\xi\,|\,{}_\iota^1G^{K0\aleph}_{\dagger,m}] = [4\,{}_\iota C^0_0({}_\iota\lambda_{\dagger,m},{}_\iota\nu_{\dagger,m}) - 4\,{}_\iota C^0_0({}_\iota^1\lambda_{\dagger,m},{}_\iota^1\nu_{\dagger,m}) + {}_\iota O^o_0]\,\Pi_m[\xi;{}_\iota\bar{\xi}_{\dagger,m}]$$
$$+8\left\{{}_\iota B_1[\xi;{}_\iota^1\lambda_{\dagger,m},{}_\iota^1\nu_{\dagger,m}] + {}_\iota B_1[\xi;{}_\iota\lambda_{\dagger,m},{}_\iota\nu_{\dagger,m}]\right\}\dot{\Pi}_m[\xi;{}_\iota\bar{\xi}_{\dagger,m}]. \qquad (2.44^*)$$

As demonstrated in Part II, the QPD is especially useful for generating multi-step REs of GRef potentials.

In the linear case (m = 1)

$$Q[\xi;{}_\iota\xi_{\dagger,1}] = \breve{Q}[\xi;{}_\iota\bar{\xi}_{\dagger,1}] = -\frac{1}{(\xi-{}_\iota\xi_{\dagger,1})^2}. \qquad (2.45)$$

and all three decompositions coincide:

$$O_1^{\downarrow}[\xi \,|\, {}^{1}_{\iota}G^{K0\aleph}_{\dagger,1}] = \breve{O}_1^{\downarrow}[\xi \,|\, {}^{1}_{\iota}G^{K0\aleph}_{\dagger,1}] = O_1^{o}[\xi \,|\, {}^{1}_{\iota}G^{K0\aleph}_{\dagger,1}]. \tag{2.46}$$

Substituting the first-order polynomial

$$_{\iota}B_1[\xi; {}_{\iota}\lambda_{\dagger;1}, {}_{\iota}\nu_{\dagger;1}] = \tfrac{1}{2}({}_{\iota}\lambda_{\dagger;1} + {}_{\iota}\nu_{\dagger;1} + \iota + 1)(\xi - {}_{\iota}\xi_{\dagger;1}), \tag{2.47}$$

into differential equation (2.21) for m = 1, one finds the following explicit expression for the outer singular point

$$_{\iota}\xi_{\dagger;1} = \frac{{}_{\iota}\lambda_{\dagger;1} + 1}{\iota({}_{\iota}\lambda_{\dagger;1} + 1) + {}_{\iota}\nu_{\dagger;1} + \iota} \tag{2.48}$$

applicable to any shape-invariant potential. It has been already utilized in Quesne's [6, 13]. linear cases of REs of the DPT potential and isotonic oscillators ($\aleph = 1$). We shall come back to discussion of these particular cases in next Section. An analysis of first-order polynomials (2.46) for a $\aleph \neq 1$ will be presented in separate publication [18].

Coming back to the general case of an arbitrary positive integer m, note that each of the gauge transformations:

$$F[\xi;\varepsilon \,|\, {}^{1}_{\iota}G^{K0\aleph}_{\dagger,m};\sigma] = \Phi[\xi;\varepsilon \,|\, {}^{1}_{\iota}G^{K0\aleph}_{\dagger,m}] / {}_{\iota}\Theta_{\dagger,m}[\xi;\sigma_0\, {}^{1}\lambda(\varepsilon;\dagger), \sigma_1\, {}^{1}_{\iota}\nu(\varepsilon;\dagger)], \tag{2.49}$$

where

$$^{1}\lambda(\varepsilon;\dagger) \equiv \begin{cases} \lambda_0(\varepsilon) & \text{for } c_0 > 0, \\ \lambda_{o;\dagger} & \text{for } c_0 = 0 \end{cases} \tag{2.50}$$

and

$$^{1}_{\iota}\nu(\varepsilon;\dagger) = \begin{cases} \sqrt{f_0 + 1 - {}_{0}a\,\varepsilon} & \text{for } \iota = 0,\ {}_{0}a > 0 \\ {}_{\iota}\nu_{o;\dagger} & \text{for } \iota = 0,\ {}_{0}a = 0 \text{ or } \iota = 1,\ c_1 = 0 \\ \sqrt{-c_1\varepsilon} & \text{for } \iota = 1,\ c_1 > 0 \end{cases} \tag{2.50$'$}$$

converts RSL equation (2.33) into the Fuschian equation

$$\left\{(-1)^{\iota}\, {}_{\iota}\hat{D}(\sigma_0\, {}^{1}\lambda(\varepsilon;\dagger), \sigma_1\, {}^{1}_{\iota}\nu(\varepsilon;\dagger)) + {}_{\iota}C_m[\xi;\varepsilon \,|\, {}^{1}_{\iota}G^{K.0\aleph}_{\dagger,m};\bar{\sigma}]\right\} F[\xi;\varepsilon \,|\, {}^{1}_{\iota}G^{K.0\aleph}_{\dagger,m};\bar{\sigma}] = 0, \tag{2.51}$$

where

$$
{}_{\iota}C_m[\xi;\varepsilon\,|\,{}^{1}_{\iota}\mathcal{G}^{K.0\aleph}_{\dagger,m};\bar{\sigma}]
$$

$$
=\{\tfrac{1}{4}\tilde{O}^{\downarrow}_{m}[\xi\,|\,{}^{1}_{\iota}\mathcal{G}^{K01}_{\dagger,m}]+\tfrac{1}{4}(-1)^{\iota}\,{}_{\iota}d\varepsilon-{}_{\iota}C^{0}_{0}(\sigma_0\,{}^{1}\lambda(\varepsilon;\dagger),\sigma_1\,{}^{1}_{\iota}\nu(\varepsilon;\dagger)\}\Pi_m[\xi;{}_{\iota}\bar{\xi}_{\dagger,m}]
$$

$$
-2\,{}_{\iota}B_1[\xi;\sigma_0\,{}^{1}\lambda(\varepsilon;\dagger),\sigma_1\,{}^{1}_{\iota}\nu(\varepsilon;\dagger)]\,\dot{\Pi}_m[\xi;{}_{\iota}\bar{\xi}_{\dagger,m}]. \qquad (2.52)
$$

Note that the latter formula for the free term in Fuschian equation (2.51) is based on GPDs (2.35) and (2.35').

By setting $\varepsilon$ to ${}_{1}\varepsilon_{\dagger,m}$, fixing the TP parameters $c_0$ and $c_1$ together with the ray identifier $h_{o;0}$ ($h_{o;1}$ = -1 in any case), and choosing the parameter $f_o$ according to (2.8), one comes to the Heine equation [9, 10]

$$
\{-{}_{1}\hat{D}(\lambda_{\dagger,m},\nu_{\dagger,m})+C_m[\xi]\}\,Hi_m[\xi]=0. \qquad (2.53)
$$

with the polynomial solution

$$
Hi_m[\xi]=Hi_m[\xi\,|\,{}^{1}_{1}\mathcal{G}^{K0\aleph}_{\dagger,m};\bar{\sigma}_{\dagger}] \qquad (2.54)
$$

for

$$
C_m[\xi]={}_{1}C_m[\xi;{}_{1}\varepsilon_{\dagger,m}\,|\,{}^{1}_{1}\mathcal{G}^{K0\aleph}_{\dagger,m};\bar{\sigma}]. \qquad (2.54^{*})
$$

To be able to treat the two problems in parallel, we also refer to polynomial solutions of a similar Fuschian equation for $\iota$=0 as GS Heine polynomials while using a more precise term GS *c*-Heine polynomials when we explicitly restrict the discussion to the confluent RSL equation.

It has been proven in [11] for *r*-GRef potentials on the line and then extended to any *r*- or *c*-GRef potential in [19] that RSL equation (2.29) has AEH solutions of the form:

$$
\phi[\xi\,|\,{}_{\iota}\mathcal{G}^{K\Im\aleph}_{\dagger,m};\dagger',m']=\frac{{}_{\iota}\Theta[\xi;{}^{1}_{\iota}\lambda_{\dagger,m|\dagger',m'},{}^{1}_{\iota}\nu_{\dagger,m|\dagger',m'}]}{\sqrt{\prod_{r=1}^{\Im}(\xi-\xi_{T;r})\,\Pi_m[\xi;{}_{\iota}\bar{\xi}_{\dagger,m}]}}\,, \qquad (2.55)
$$

where

$$
{}^{1}_{\iota}\lambda_{c,0|\dagger,m}+{}_{\iota}\lambda_{c,0}-{}_{\iota}\lambda_{\dagger,m}={}_{\iota}\lambda_{c,0}+\sigma_{0;\dagger}\aleph, \qquad (2.56a)
$$

$$ {}^{1}_{\iota}\nu_{\mathsf{t},\mathrm{m}|\mathsf{t}',\mathrm{m}'} = {}_{\iota}\nu_{\mathsf{t},\mathrm{m}} \quad \text{(excluding the DPT potential)}, \tag{2.56b} $$

and $\xi_{\mathrm{T;r}}$ is an outer (possibly complex) root of the TP. The HPG for the generic *r*-GRef or *c*-GRef potential on the line can be then represented as

$$ {}_{\iota}\hat{g}(\Delta\lambda, \Delta\nu; {}_{\iota}\bar{\xi}_{\mathsf{t},\mathrm{m}}) \equiv \xi(1-\iota\xi)\left[\Pi_{\mathrm{m}}[\xi; {}_{\iota}\bar{\xi}_{\mathsf{t},\mathrm{m}}]\frac{d}{d\xi} - \dot{\Pi}_{\mathrm{m}}[\xi; {}_{\iota}\bar{\xi}_{\mathsf{t},\mathrm{m}}]\right] + \tfrac{1}{2}[\Delta\lambda(1-\iota\xi) - \Delta\nu\xi]\,\Pi_{\mathrm{m}}[\xi; {}_{\iota}\bar{\xi}_{\mathsf{t},\mathrm{m}}]. \tag{2.57} $$

For this reason we refer to (2.57) as the ‘generic’ HPG ($\mathcal{g}$-HPG). The appropriate sequences of quantized GS Heine polynomials are thus defined as follows:

$$ {}_{\iota}\upsilon_{\mathsf{t},\mathrm{m};\mathsf{c},\mathrm{v}}\mathrm{Hi}_{\mathrm{m+v+\iota}}[\xi\,|\,{}^{1}_{\iota}\mathcal{G}_{\mathsf{t},\mathrm{m}};\mathsf{c},\mathrm{v}] = {}_{\iota}\hat{g}({}_{\iota}\lambda_{\mathsf{c},\mathrm{v}} - {}_{\iota}\lambda_{\mathsf{t},\mathrm{m}}, {}_{\iota}\nu_{\mathsf{c},\mathrm{v}} - \nu_{\mathsf{t},\mathrm{m}}; {}_{\iota}\bar{\xi}_{\mathsf{t},\mathrm{m}})\,{}_{\iota}\mathrm{P}_{\mathrm{v}}[\xi;\lambda_{\mathsf{c},\mathrm{v}}, \nu_{\mathsf{c},\mathrm{v}}], \tag{2.58} $$

where the scale factor ${}_{\iota}\upsilon_{\mathsf{t},\mathrm{m};\mathsf{c},\mathrm{v}}$ is chosen via the requirement that the leading coefficient of the GS Heine polynomial is equal to 1, i.e.,

$$ \mathrm{Hi}_{\mathrm{n}}[\xi\,|\,{}^{1}_{\iota}\mathcal{G}_{\mathsf{t},\mathrm{m}};\mathsf{t}',\mathrm{m}'] = \Pi_{\mathrm{n}}[\xi; {}_{\iota}\xi_{\mathsf{t},\mathrm{m};\mathsf{t}',\mathrm{m}'}]. \tag{2.59} $$

As discussed in more detail in Part II, a certain advantage of convention (2.59) is that GS Heine polynomials defined in such a way are symmetric with respect to the interchange of the pairs of indexes $\mathsf{t},\mathrm{m}$ and $\mathsf{t}',\mathrm{m}'$. As discussed in subsection 4.3, in the particular case of the isotonic oscillator (K=1, $\Im$=0, ι=0) GS Heine polynomials (2.58) turn into Case-II exceptional Laguerre polynomials.

In addition to the mentioned families of the GRef potentials on the line:

$$ {}_{\iota}\mathrm{T}_{\mathrm{K}}[0] > 0, \tag{2.60} $$

$\mathcal{g}$-HPG (2.57) are also applicable to SUSY partners of *r*- and *c*-GRef potentials on the half-line for $\mathsf{t} = \mathsf{b}$ or $\mathsf{d}$:

$$ {}_{\iota}\mathrm{T}_{\mathrm{K}}[0] = 0, \;\; \sigma_{0;\mathsf{t}} = - \tag{2.60*} $$

(including the isotonic oscillator discussed in more detail in subsection 4.3 below) as well as to the Darboux transform $V[z\,|\,{}^{1}_{1}G^{201}_{\mathbf{d},m}]$ of the DPT potential. On other hand, if $0<\lambda_o<1$ then CDTs with FFs of types $\mathbf{b}$ and $\mathbf{d}$ ($0<\lambda_{o;\mathbf{t}}=1-\lambda_o<1$) convert solutions with a larger characteristic exponent into solutions irregular at the origin. Discrete energy levels in the potential $V[z\,|\,{}^{1}_{\iota}G^{K\Im 1}_{\mathbf{t},m}]$ for $\mathbf{t}=\mathbf{b}$ or $\mathbf{d}$ correspond to the egenfunctions formed by Heine polynomials

$$\begin{aligned}&{}_{\iota}\upsilon_{\mathbf{t},m;\mathbf{b},m'}\mathrm{Hi}_{m+m'+\iota}[\xi\,|\,{}^{1}_{\iota}G_{\mathbf{t},m};\mathbf{b},m']\\ &\quad={}_{\iota}\hat{g}({}_{\iota}\lambda_{\mathbf{b},m'}-{}_{\iota}\lambda_{\mathbf{t},m},{}_{\iota}\nu_{\mathbf{b},m'}-\nu_{\mathbf{t},m};{}_{\iota}\bar{\xi}_{\mathbf{t},m})\,{}_{\iota}P_{m'}[\xi;\lambda_{\mathbf{b},m'},\nu_{\mathbf{b},m'}].\end{aligned} \tag{2. 61}$$

This is a typical example of the breakdown of SUSY quantum mechanics for radial potentials with discrete energy levels embedded into the continuous spectrum (see Appendix A for more details).

Since the characteristic exponents at the origin are energy-independent for any PF potential on the half-line, the first argument of operator (2.60) vanishes for DTs with FFs of type $\mathbf{a}$ or $\mathbf{c}$ which gives

$${}_{\iota}\hat{g}(0,\Delta\nu;{}_{\iota}\bar{\xi}_{\mathbf{a},m})\equiv\xi\,{}_{\iota}\hat{a}(\Delta\nu;{}_{\iota}\bar{\xi}_{\mathbf{a},m}), \tag{2.62}$$

where

$${}_{\iota}\hat{a}(\Delta\nu;{}_{\iota}\bar{\xi}_{\mathbf{t},m})\equiv(1-\iota\xi)\left[\Pi_m[\xi;{}_{\iota}\bar{\xi}_{\mathbf{t},m}]\frac{d}{d\xi}-\dot{\Pi}_m[\xi;{}_{\iota}\bar{\xi}_{\mathbf{t},m}]\right]-\tfrac{1}{2}\Delta\nu\,\Pi_m[\xi;{}_{\iota}\bar{\xi}_{\mathbf{t},m}]$$

$$\text{for } \sigma_{0;\mathbf{t}}=\sigma_{0;\mathbf{t}'} \tag{2.62*}$$

is the 'abridged' HPG ($a$-HPG). On other hand, $o_{0;\mathbf{tt}'}+1=2$ so that

$$\begin{aligned}&\frac{1}{\sqrt{{}_{\iota}T_K[\xi]}}\,{}_{\iota}\Theta[\xi;\sigma_{0;\mathbf{t}'}\lambda_o,{}_{\iota}\nu_{\mathbf{t}',m'}]\,{}_{\iota}\hat{g}(0,\Delta\nu;{}_{\iota}\bar{\xi}_{\mathbf{a},m})\\ &\qquad=\frac{1}{\sqrt{a\,(\xi-\xi_{T;1})}}\,{}_{\iota}\Theta[\xi;\pm\sigma_{0;\mathbf{t}'}\lambda_{o;\mathbf{t}},-{}^{1}_{\iota}\nu_{\mathbf{t}',m'}]\,{}_{\iota}\hat{a}(\Delta\nu;{}_{\iota}\bar{\xi}_{\mathbf{a},m}).\end{aligned} \tag{2.63}$$

We thus conclude that the polynomials

$$\mathrm{Hi}_{m+v+\iota}[\xi\,|\,{}^{1}_{\iota}G^{K\Im 1}_{\dagger,m};c,v] = {}_{\iota}\upsilon^{K\Im 1}_{\dagger,m;c,v}\,{}_{\iota}\hat{a}({}_{\iota}\nu_{\dagger,m} - {}_{\iota}\nu_{\dagger',m'};\bar{\xi}_{\dagger,m})\,{}_{\iota}P_{v}[\xi;\lambda_{v},\nu_{v}] \quad \text{iff} \quad {}_{\iota}T_{K}[0] = 0 \text{ and } \dagger = a \text{ or } c. \tag{2.64}$$

satisfy Heine equation (2.51). In the particular case of the isotonic oscillator (K=1, $\Im$=0, ι=0) GS Heine polynomials (2.64) turn into the Case-I exceptional Laguerre polynomials (see subsection 4.3 below for details).

## 3. Gauss-Reference SL Problems With Energy-Independent Characteristic Exponents

In this paper we are only interested in TPs

$$T_{\iota+1}[\xi;d] = d\,\xi(1-\iota\xi) \tag{3.1}$$

associated with the DPT potential ( $\iota = 1,\ d > 0$) and isotonic oscillator ($\iota = 0,\ d > 0$) so that

$$_{\iota}\wp[\xi] \equiv {}_{\iota}\wp[\xi;\,{}_{\iota}T_{\iota+1}] = \frac{{}_{\iota}d}{4\xi(1-\iota\xi)} \tag{3.1*}$$

and the appropriate Bose invariants take the form:

$$I[z;\varepsilon\,|\,{}_{1}G^{201}] = {}_{1}I^{o}[z\,|\,\lambda_{o},\mu_{o}] + \frac{{}_{1}d\,\varepsilon}{4z(1-z)}\,, \tag{3.2}$$

$$I[\varsigma;\tilde{\varepsilon}\,|\,\tilde{G}^{101}] \equiv I[\varsigma/\nu_{o};\nu_{o}\tilde{\varepsilon}\,|\,{}_{0}G^{101}] = {}_{0}I^{o}[\varsigma;\lambda_{o}] + \frac{{}_{0}d\tilde{\varepsilon}}{4\varsigma}, \tag{3.2'}$$

where we set

$$\lambda_{o} \equiv \sqrt{h_{o;0}+1}, \tag{3.3a}$$

$$_{1}\nu_{o} \equiv \mu_{o} \equiv \sqrt{h_{o;1}+1}, \tag{3.3b}$$

$$_{0}\nu_{o} \equiv \nu_{o} \equiv \sqrt{h_{o;1}}\,. \tag{3.3'}$$

and also got rid of parameter (3.3′) by introducing a new variable:

$$\varsigma = \nu_{o}\zeta \tag{3.4}$$

and re-scaled the energy

$$\tilde{\varepsilon} = \varepsilon / \nu_o \tag{3.5}$$

for the isotonic oscillator. Without loss of generality we can assume that

$$\mu_o \geq \lambda_o \tag{3.6}$$

(changing z for 1−z in the definition of the reference PF otherwise). The reference PFs for Bose invariants (3.2) and (3.2′) are given by the relations

$${}_1I^o[z\,|\,\lambda,\mu] = \frac{1-\lambda^2}{4z^2} + \frac{1-\mu^2}{4(1-z)^2} - \frac{O_0^o(\lambda,\mu)}{4z(1-z)}\,, \tag{3.7}$$

and

$${}_0I^o[\varsigma;\lambda_o] = \frac{1-\lambda_o^2}{4\varsigma^2} - 1/4, \tag{3.7′}$$

respectively, with

$$O_0^o(\lambda,\mu) \equiv \lambda^2 + \mu^2 - 1\,. \tag{3.8}$$

(Note that the coefficient $b_0$ should be changed for $-b_0$ in definition (13a) of the DPT potential in [12].) The remarkable feature of the RSL equation with Bose invariant (3.2) is that characteristic exponents at both finite singular points are energy-independent. This is also true for the singular point $\zeta$=0 of the confluent equation as well as for the constant term in the right-hand-side of (3.2′). To be able to treat both Bose invariants in parallel, we refer to the asymptotic parameters

$$\nu_{\pm}(\varepsilon; h_{o;1}, {}_0a) \equiv \pm\sqrt{h_{o;1} + {}_0a\,\varepsilon}\,, \tag{3.9}$$

as ‘characteristic exponents’, with clear understanding that this is by no means an accurate mathematical term.

To eliminate dependence of the characteristic exponents on energy at the nonzero singular point, we map the h-PT potential onto the negative semi-axis using the variable [25, 34]:

$$\tilde{z} = -sh^2(r/\sqrt{{}_1d})\,, \tag{3.10}$$

instead of the generic variable

$$z = \frac{\tilde{z}}{\tilde{z}-1} = th^2(r/\sqrt{_1d}) \qquad (3.10^*)$$

defined between 0 and 1. The boundary conditions imposed on regular solutions thus require a separate study which is presented in subsection 4.2 below. The change of variable $z_-(r)$ leads to Jacobi polynomials in the variable

$$\eta(r) \equiv 1 - 2z_-(r) = ch(2r/\sqrt{_1d}) \qquad (3.10')$$

first utilized by Dabrowska, Khare, and Sukhatme [107] in their breakthrough study on shape-invariant potentials. More recently Odake and Sasaki [20] (see also [108]) used the representation of the h-PT potential in terms of variable (3.10′) to construct REs quantized via finite sets of orthogonal Heine polynomials (in our terms). Note that Odake and Sasaki's parameter $g$ [20] coincides with the centrifugal-barrier parameter

$$g \equiv s_o = \lambda_o + 1/2 > 1/2 \qquad (3.11a)$$

for both DPT potential and isotonic oscillator (see Appendix A for the restriction imposed on this parameter by the requirement that the potential should be within the validity range of SUSY quantum mechanics). Their parameter $h$ is related to (3.3b) in a similar way:

$$h = \mu_o + 1/2 > 1/2 . \qquad (3.11b)$$

For the h-PT potential, regardless of the choice of variable (3.10) or (3.10*), the energy reading point is determined by the requirement that the indicial equation at the nonzero end has real and complex-conjugated roots at negative and positive energies, respectively, which is equivalent to the condition that the radial potential in question vanishes at infinity. Keeping in mind that two variables are related via the linear fractional transformation Bose invariant (2.2) converted to new variable (3.7) takes the form

$$\begin{aligned} {}_1I^o[z_-;\varepsilon\,|\,\lambda_o,\mu_o] &= \left(\frac{dz}{dz_-}\right)^2 {}_1I^o[z;\varepsilon\,|\,\lambda_o,\mu_o;0,{}_1d,{}_1d] \\ &= \frac{1-\lambda_o^2}{4z_-^2} + \frac{1-\mu_o^2}{4(z_- -1)^2} + \frac{O_0^o(\lambda_o,\mu_o) - {}_1d\,\varepsilon}{4z_-(z_- -1)} \end{aligned} \qquad (3.12)$$

where

$$\mu_o \equiv \sqrt{f_o + 1}\,. \qquad (3.13)$$

We thus come to the same expressions for exponent differences at 0 and 1 as those for the SL equation with Bose invariant (3.2) for the DPT potential.  Odake and Sasaki's parameter g and h are again given by (3. 8a) and (3.8b).  We thus conclude that the Schrödinger equation for both *t*- and *h*-versions of the PT potential can be converted into the same SL equation, namely, we put

$$_1d\,V[z;\lambda_o,\mu_o] \equiv \frac{\lambda_o^2 - 1/4}{z} + \frac{\mu_o^2 - 1/4}{1-z}\,, \qquad (3.14)$$

The only difference lies in the choice of the quantization interval.  In terms of Quesne et al. [13, 108]

$$t\text{-}V_{A'_+,B'}(\pi/2 - x) \equiv {_1d}\,V[\sin^2 x/2; A'_+ - B' - 1/2, A'_+ + B' - 1/2] \qquad (3.15a)$$

and

$$h\text{-}V_{A'_-,B'}(x) \equiv {_1d}V[-sh^2 x/2; A'_- - B' + 1/2, A'_- + B' + 1/2], \qquad (3.15b)$$

respectively, where we put

$$2A'_\pm = \mu_o \pm \lambda_o \pm 1, \quad 2B'_\pm = \mu_o \mp \lambda_o\,. \qquad (3.16)$$

Note that we have to introduce additional subscripts + and – for the parameter A used in [13] and [108] for both versions but dependent differently on the ray identifiers $\lambda_o$ and $\mu_o$.

An extension of Bose invariant (3.2) to the negative semi-axis allows one to analyze behavior of AEH solutions at large |z|.  In fact, taking into account that

$$_1I^o[z;-\varepsilon\,|\,{_1Q^o};-{_1T_K}] = {_1I^o}[z;\varepsilon\,|\,{_1Q^o};{_1T_K}] \text{ for } z < 0 \text{ or } z > 1, \qquad (3.17)$$

we conclude that  RSL equation with Bose invariant (3.2) may have an AEH solution of type † only at the energy

$$_1\varepsilon_{\bar{\sigma}_\dagger,m} = {_1\varepsilon_m}(\sigma_{0;\dagger}\lambda_o, \sigma_{1;\dagger}\mu_o) = (\sigma_{0;\dagger}\lambda_o + 2m + 1)^2 / {_1d} \qquad (3.18)$$

in agreement with (2.15), where we set $c_0 = c_1 = 0$, $f_o = -1$. The common vital feature of the RSL equations with Bose invariants (3.2) and (3.2′).is that both AEH solutions and their energies can be parameterized in terms of the signed exponent differences: $\lambda_o = |\lambda|, \sigma_{0;\dagger} = \operatorname{sign}\lambda$ and $\mu_o = |\mu|, \sigma_{0;\dagger} = \operatorname{sign}\mu$ or $\nu_o = |\nu|, \sigma_{0;\dagger} = \operatorname{sign}\nu$ for $\iota=1$ or 0, respectively. Below we shall use these parameters in formulas which are independent of the solution type. . If the type is important, we shall use a more explicit notation $\lambda_\dagger = \sigma_{0;\dagger}\lambda_o$ and ${}_\iota\nu_\dagger = \sigma_{1;\dagger}\ {}_\iota\nu_o$ (or, if need to be more specific, $\mu_\dagger = \sigma_{1;\dagger}\mu_o$ or $\nu_\dagger = \sigma_{1;\dagger}\nu_o$ for $\iota=1$ or 0, respectively). In particular, representing quadratic formula (3.18) as

$$ {}_1d\ \varepsilon_m(\lambda, \mu) = O_0^o(\lambda,\mu) + 4\,{}_1C_0^0(\lambda, \mu) + 4m(\lambda + \mu + m + 1) \tag{3.18*} $$

one can directly verify that this is indeed the necessary condition for the hypergeometric equation (2.21) with ${}_1\lambda_{\dagger,m} = \lambda_\dagger$, ${}_1\nu_{\dagger,m} = \mu_\dagger$ to have a polynomial solution of order m.

Keeping in mind that

$$ {}_\iota\lambda_{\dagger,m} = \lambda_\dagger,\ \ {}_\iota\nu_{\dagger,m} \equiv {}_\iota\nu_\dagger, \tag{3.19} $$

and K= $\aleph + 1$ for any of three PF potentials generated by means of TP (3.1), one can represent (2.27a) and (2.27b) as

$$ {}_\iota^1\lambda_{\dagger,m} = -\sigma_{0;\dagger}\ \lambda_{o;\dagger} \equiv -{}^{*}\lambda_\dagger, \tag{3.20a} $$

$$ {}_\iota^1\nu_{\dagger,m} = -\sigma_{1;\dagger}\ {}_\iota\nu_{o;\dagger} = -\ {}_\iota^{*}\nu_\dagger \tag{3.20b} $$

where

$$ \lambda_{o;\dagger} \equiv |\lambda_o + \sigma_{0;\dagger}|, \tag{3.21a} $$

$$ {}_\iota\nu_{o;\dagger} \equiv \begin{cases} \nu_o & \text{for } \iota = 0, \\ \mu_{o;\dagger} \equiv |\mu_o + \sigma_{1;\dagger}| & \text{for } \iota = 1 \end{cases} \tag{3.21b} $$

and

$$
{}^{*}\lambda_{\dagger} \equiv \begin{cases} -1-\lambda_{\dagger} & \text{if } -1<\lambda_{\dagger}<0, \\ \lambda_{\dagger}+\sigma_{0;\dagger} & \text{otherwise,} \end{cases} \tag{3.21a*}
$$

$$
{}_{\iota}^{*}\nu_{\dagger} \equiv \begin{cases} -1-{}_{\iota}\nu_{\dagger} & \text{if } -1<{}_{\iota}\nu_{\dagger}<0, \\ {}_{\iota}\nu_{\dagger}+\iota\sigma_{1;\dagger} & \text{otherwise.} \end{cases} \tag{3.21b*}
$$

(Here and below we interpret the symbol $\sigma_r$ in a double way: either as + versus – or as +1 versus –1 depending on the context.) As far as we are only interested in DTs which obey the rules of SUSY quantum mechanics, i.e., we require that

$$
\lambda_{o;\dagger} = \lambda_o + \sigma_{0;\dagger} > 0 \tag{3.22a}
$$

and

$$
\mu_{o;\dagger} = \mu_o + \sigma_{1;\dagger} > 0 \,. \tag{3.22b}
$$

We thus came to the juncture where the theory of RSL equations with GRef Bose invariants merges with Gomez-Ullate, Kamran, and Milson's theory [24, 35] of the second-order differential operators 'exactly solvable' by polynomials (PES) [x)]. Namely, both the parameters of the operator ${}_{\iota}\hat{D}({}_{\iota}\lambda_{\dagger,m}, {}_{\iota}\nu_{\dagger,m})$ and constant (2.23) appearing in differential equation (2.21) become energy-independent so that the latter equation turns into the eigenpolynomial problem for the operator

$$
{}_{\iota}\hat{T}(\lambda,\nu) \equiv {}_{\iota}p[\xi]\frac{d^2}{d\xi^2} + {}_{\iota}q[\xi;\lambda,\nu]\frac{d}{d\xi} + {}_{\iota}r(\lambda,\nu)\,, \tag{3.23}
$$

where

---

[x)]When speaking of rational potentials quantized via Heine polynomials we would prefer to use the term 'conditionally exactly quantized by polynomials' (P-CEQ) to distinguish from the situation when only some unbounded solutions of the Schrödinger equation are expressible in terms of polynomials. In particular, none of the potentials introduced in [53, 109, 110] is P-CEQ despite the fact that some solutions of the Schrödinger equation with each of these potentials can be expressed in terms of Jacobi polynomials.

$$ {}_\iota p[\xi] \equiv -\xi(1-\iota\xi) = -\tfrac{1}{4}\, {}_\iota d\, {}_\iota \wp^{-1}[\xi], \tag{3.23'} $$

$$ {}_\iota q[\xi;\lambda,\nu] \equiv -2\, {}_\iota B_1[\xi;\lambda,\nu], \tag{3.23''} $$

$$ {}_\iota q[\xi;\lambda,\nu] \,/\, {}_\iota p[\xi] = \frac{\lambda+1}{z} + \frac{\nu+\iota}{\iota z - 1}, \tag{3.23*} $$

$$ {}_\iota r(\lambda,\nu) = {}_\iota C_0^0(\lambda,\nu) + \tfrac{1}{4}\, {}_\iota O_0^o(\lambda,\nu). \tag{3.23'''} $$

By applying the generic factorization technique developed in [34], one can show that PES operator (3.23) allows infinitelymany RFs:

$$ {}_\iota \hat{T}(\lambda_\dagger, {}_\iota\nu_\dagger) = \left[ {}^{*}_{\iota}\hat{d}_{\dagger,m} + {}_\iota q[\xi;\lambda_\dagger, {}_\iota\nu_\dagger] \right] {}_\iota\hat{d}_{\dagger,m} + \tfrac{1}{4}\, {}_\iota d\, {}_\iota \varepsilon_{\dagger,m} \tag{3.24} $$

in terms of shift operators ${}_\iota\hat{d}_{\dagger,m} \equiv \hat{d}\{\Pi_m[\xi; {}_\iota\xi_{\dagger,m}]\}$ and their SUSY partners ${}^{*}_{\iota}\hat{d}_{\dagger,m} \equiv {}^{*}\hat{d}\{\Pi_m[\xi; {}_\iota\bar{\xi}_{\dagger,m}]\}$ defined via (B4) in Appendix B and the partner relation

$$ {}^{*}\hat{d}\{\varphi_\tau[\xi \,|\, Q^o]\} \equiv \hat{d}\{\varphi_\tau^{-1}[\xi \,|\, Q^o]\} = \frac{d}{d\xi} + ld\{\varphi_\tau[\xi \,|\, Q^o]\}\ , \tag{3.25} $$

respectively. Note the shift operators in question are rational contrary to the *algebraic* shift operators

$$ {}_\iota\hat{A}_{\dagger,m} \equiv \sqrt{-\,{}_\iota p[\xi]}\ \, {}_\iota\hat{d}_{\dagger,m} \tag{3.26a} $$

and

$$ {}_\iota\hat{B}_{\dagger,m} \equiv \left[ {}_\iota p(\xi;\lambda_\dagger, {}_\iota\nu_\dagger)\, {}^{*}_{\iota}\hat{d}_{\dagger,m} + {}_\iota q[\xi;\lambda_\dagger, {}_\iota\nu_\dagger] \right] / \sqrt{-\,{}_\iota p[\xi]} \tag{3.26b} $$

appearing in the generic factorization scheme [34]. It should be stressed that the RF defined via (3.24) is not the same as the one used in Gomez-Ullate, Kamran, and Milson's theory [24, 35]. They choose the shift operator A(y) to have a polynomial coefficient of the first derivative as explicitly seen from the discussion of $X_m$-Jacobi polynomials at the beginning of subsection 4.2 in [35]. On other hand, in (3.26a) and (3.26b) this coefficient is set to $\sqrt{-\,{}_\iota p[\xi]}$, according to the prescription of the generic factorization scheme [34].

In case of basic AEH solutions ('algebraic ground state condition' in terms of [34])

$$ {}_{\iota}\phi_{\dagger}[\xi] = {}_{\iota}\Theta[\xi; \lambda_{\dagger}, {}_{\iota}\nu_{\dagger}] \tag{3.27} $$

polynomial $\Pi_m[\xi; {}_{\iota}\bar{\xi}_{\dagger,m}]$ turns into a constant and its logarithmic derivative vanishes which gives

$$ \hat{d}\{\Pi_0[\xi; {}_{\iota}\bar{\xi}_{\dagger,0}]\} = {}^{*}\hat{d}\{\Pi_0[\xi; {}_{\iota}\bar{\xi}_{\dagger,0}]\} = \frac{d}{d\xi}. \tag{3.28} $$

Making use of (3.18*), one finds

$$ {}_1 r(\lambda_{\dagger}, \mu_{\dagger}) = \tfrac{1}{4}\, {}_1 d\, {}_1\varepsilon_{\bar{\sigma}_{\dagger},0}. \tag{3.29} $$

and

$$ \tfrac{1}{4}\, {}_1 d( {}_1\varepsilon_{\bar{\sigma},n} - {}_1\varepsilon_{\bar{\sigma},0}) = n(\sigma_0\lambda_o + \sigma_1\mu_o + n + 1) \tag{3.30} $$

so that the eigenvalue problem for PES operator (3.24) turns into the equation for hypergeometric polynomials:

$$ \left[ z(1-z)\frac{d^2}{dz^2} + [\lambda_{\dagger} + 1 - \beta_{\dagger;1} z]\frac{d}{dz} + n\beta_{\dagger} \right] F(-n, \beta_{\dagger;n}; \lambda_{\dagger} + 1; z) = 0, \tag{3.31} $$

where $\beta_{\dagger;n} \equiv \lambda_{\dagger} + \mu_{\dagger} + n + 1$. The change of variable $\eta = 1 - 2z$ then leads to the RF of the equation for Jacobi polynomials for $\ell$ =0 in [87], with the parameters g and h defined via (3.11a) and (3.11b) above.

Coming back to Quesne's rational extensions of the DPT potential, we find that her parameters A and B to exponent differences (3.22a) and (3.22b) as follows

$$ 2A \equiv \lambda_{o;\dagger} + \mu_{o;\dagger} + 1 = 2A' \equiv \lambda_o + \mu_o + 1, \quad |B' - B| = 1 \qquad \text{for } \dagger = a,\ b \tag{3. 32} $$

($A' \equiv A'_+$, $B' \equiv B'_+$) and

$$ 2B \equiv \mu_{o;\dagger} - \lambda_{o;\dagger} = 2B' \equiv \mu_o - \lambda_o\ , |A' - A| = 1 \qquad \text{for } \dagger = c,\ d, \tag{3.32*} $$

in agreement with Quesne's results for $\dagger$ = a or b, m =1 or 2 and $\dagger$ = d, m =2 [13]. It should be stressed that rules of SUSY quantum mechanics are applicable for $\dagger$ = a or b only if either $\mu_o > 1$ or $\lambda_o > 1$, respectively.

One can also verify that factorization energies (3.18) expressed in terms of the parameters A and B:

$$\varepsilon_{\mathbf{t},m} = \begin{cases} (\sigma_{0;\mathbf{t}}B + m - \tfrac{1}{2})^2 & \text{for } \mathbf{t} = \mathbf{a} \text{ or } \mathbf{b} \\ [\sigma_{0;\mathbf{t}}(A - \tfrac{1}{2}) + m - \tfrac{1}{2}]^2 & \text{for } \mathbf{t} = \mathbf{c} \text{ or } \mathbf{d} \end{cases} \tag{3.33}$$

match Quesne's formula's for both linear and quadratic cases.

Substituting density function (3.1*) and the logarithmic derivative

$$ld\,\psi_{\mathbf{t},m}[\xi\,|\,{}_{\iota}G^{K01}_{\downarrow\mathbf{t},m}] = \frac{1+2\lambda_{\mathbf{t}}}{4\xi} - \frac{2\,{}_{\iota}\nu_{\mathbf{t}} + \iota}{4(1-\iota\xi)} + ld\,\Pi_m[\xi;\,{}_{\iota}\bar{\xi}_{\mathbf{t},m}] \tag{3.34}$$

into general expression (B16) for the difference between free terms in the SUSY pairs of the SL equations, one finds

$$\begin{aligned} I^{o}[\xi\,|\,{}^{1}_{\iota}G^{K01}_{\mathbf{t},m}] = I^{o}[\xi\,|\,{}_{\iota}G^{K01}_{\downarrow\mathbf{t},m}] - \frac{1+2\lambda_{\mathbf{t}}}{4\xi^2} - \iota\frac{1+2\,{}_{\iota}\nu_{\mathbf{t}}}{4(1-\xi)^2} \\ - \frac{\iota(\lambda_{\mathbf{t}}+1) + {}_{\iota}\nu_{\mathbf{t}}}{2\xi(1-\iota\xi)} + \frac{(2-\iota-2\iota\xi)\dot{\Pi}[\xi;\,{}_{\iota}\bar{\xi}_{\mathbf{t},m}]}{2\xi(1-\iota\xi)\Pi_m[\xi;\,{}_{\iota}\bar{\xi}_{\mathbf{t},m}]} + 2Q[\xi;\,{}_{\iota}\bar{\xi}_{\mathbf{t},m}]. \end{aligned} \tag{3.35}$$

For $\iota=1$ reference PF (3.28) for can be then represented as

$$I^{o}[z\,|\,{}^{1}_{1}G^{201}_{\mathbf{t},m}] = {}_{1}I^{o}[z\,|\,\lambda_{o;\mathbf{t}},\mu_{o;\mathbf{t}}] + \frac{(1-2z)\dot{\Pi}_m[z;\bar{z}_{\mathbf{t},m}]}{z(1-z)\Pi_m[z;\bar{z}_{\mathbf{t},m}]} + 2Q[z;\bar{z}_{\mathbf{t},m}]; \tag{3.36}$$

whereas conversion of (3.36) to variable (3.4) for K=1, $\iota = 0$ gives

$$\begin{aligned} I^{o}[\varsigma\,|\,{}^{1}_{0}\tilde{G}^{101}_{\mathbf{t},m}] &\equiv \nu_o^{-2}\, I^{o}[\varsigma/\nu_o\,|\,{}_{0}G^{101}] \\ &= \frac{1-\lambda^2_{o;\mathbf{t}}}{4\varsigma^2} - \tfrac{1}{4} - \frac{\sigma_{1;\mathbf{t}}}{2\varsigma} + \frac{\dot{\Pi}_m[\varsigma;\bar{\varsigma}_{\mathbf{t},m}]}{\varsigma\Pi_m[\varsigma;\bar{\varsigma}_{\mathbf{t},m}]} + 2Q[\varsigma;\bar{\varsigma}_{\mathbf{t},m}]. \end{aligned} \tag{3.36′}$$

Alternatively, use of (2.44*) gives

$$O^{\downarrow}_{m}[\xi\,|\,{}^{1}_{\iota}G^{K01}_{\dagger,m}] = [2(\iota\lambda_{\dagger} + \iota\nu_{\dagger} + \iota) + \iota O^{o}_{0}(\lambda_{o},\mu_{o})]\,\Pi_{m}[\xi;\,{}_{\iota}\bar{\xi}_{\dagger,m}]$$

$$+4(2\iota\xi - 1)\dot{\Pi}_{m}[\xi;\,{}_{\iota}\bar{\xi}_{\dagger,m}], \qquad (3.37)$$

so that for K=2, ι = 1

$$O^{\downarrow}_{m}[\xi\,|\,{}^{1}_{1}G^{K01}_{\dagger,m}] = O^{o}_{0}(\lambda_{o;\dagger},\mu_{o;\dagger})\Pi_{m}[\xi;\,{}_{\iota}\bar{\xi}_{\dagger,m}] + 4(2\iota\xi - 1)\dot{\Pi}_{m}[\xi;\,{}_{\iota}\bar{\xi}_{\dagger,m}],$$

$$\text{provided that } |\lambda_{o;\dagger} - \sigma_{0;\dagger}| \geq 1, \ \ |\mu_{o;\dagger} - \sigma_{1;\dagger}| \geq 1. \qquad (3.37^{*})$$

The REs of the DPT potential and isotonic oscillator associated with these reference PFs can be thus represented as

$${}_{1}dV[z\,|\,{}^{1}_{1}G^{201}_{\dagger,m}] = {}_{1}dV[z\,|\,\lambda_{o;\dagger},\mu_{o;\dagger}] - \frac{4(1-2z)\dot{\Pi}_{m}[z;\bar{z}_{\dagger,m}]}{\Pi_{m}[z;\bar{z}_{\dagger,m}]} - 8z(1-z)Q[z;\bar{z}_{\dagger,m}]$$

(3.38)

and

$$\nu_{o}^{-2}\,{}_{0}dV[\varsigma/\nu_{o}\,|\,{}^{1}_{0}G^{101}_{\dagger,m}] = V_{0}[\varsigma;\lambda_{o}] - \frac{4\dot{\Pi}_{m}[\varsigma;\bar{\varsigma}_{\dagger,m}]}{\Pi_{m}[\varsigma;\bar{\varsigma}_{\dagger,m}]} - 8\varsigma Q[\varsigma;\bar{\varsigma}_{\dagger,m}] + 2\sigma_{1;\dagger}$$

(3.38′)

where

$$V_{0}[\varsigma;\lambda_{o}] \equiv \frac{\lambda_{o}^{2} - 1}{\varsigma} + \varsigma\,. \qquad (3.39)$$

In Quesne's linear cases m=1 [13] substituting (2.44) for the QPD in the right-hand side of (3.39) and (3.39′) gives

$${}_{1}dV[z\,|\,{}^{1}_{1}G^{201}_{\dagger,1}] \equiv {}_{1}d\,{}^{1}_{1}V[z\,|\,\lambda_{o;\dagger},\mu_{o;\dagger}]$$

$$= {}_{1}d\,{}_{1}V[z\,|\,\lambda_{o;\dagger},\mu_{o;\dagger}] + \frac{8z_{\dagger,1}(1-z_{\dagger,1})}{(z-z_{\dagger,1})^{2}} + \frac{4(1-2z_{\dagger,1})}{z-z_{\dagger,1}} \qquad (3.40)$$

and

$${}_{0}d\nu_{o}^{2}V[\varsigma/\nu_{o}\,|\,{}^{1}_{0}G^{101}_{\dagger,1}] \equiv {}^{1}_{0}V[\varsigma\,|\,\lambda_{o;\dagger};\dagger,1]$$

$$= {}_{0}V[\varsigma\,|\,\lambda_{o;\dagger}] + \frac{4}{\varsigma-\varsigma_{\dagger,1}} + \frac{8\varsigma_{\dagger,1}}{(\varsigma-\varsigma_{\dagger,1})^{2}} + 2\sigma_{1;\dagger} \qquad (3.40')$$

The second expression can be directly matched to (2.9) in [13] by choosing

$$2\zeta = x^2,\ \lambda_{o;\dagger} = l + \tfrac{1}{2} = -\varsigma_{\dagger,1},\ \nu_o = \omega,\ \zeta_{\dagger,1} = -\frac{l + \tfrac{1}{2}}{\omega},\ {}_0d = 2. \qquad (3.41)$$

We refer to (3.40) and (3.40′) as the regular exceptional Quesne (*r*-XQ) and confluent exceptional Quesne (*c*-XQ) potentials to distinguish them from the REs constructed by means of higher-order Jacobi and Laguerre polynomials [13, 20, 22]. The Schrödinger equation with these two potentials is solved via Heun [3, 4] and *c*-Heun [4, 5] functions. The Heun equation associated with the *r*-XQ potential have been studied in some details by Takemura [40] (see also Section 5 below for some additional comments). Solutions of the Schrödinger equation with the *c*-XQ potential via *c*-Heun functions in general and via *c*-Heun polynomials for bound energy states apparently also deserve a special attention. Another remarkable feature of $X_1$- Jacobi and $X_1$-Laguerre polynomials is that they were introduced without any reference to DTs [1, 2]. As a result, *r*-XQ potential surface (3.40) has a fragment $0 < \lambda_{o;\dagger}, \mu_{o;\dagger} < 1$ which is quantized via $X_1$- Jacobi polynomials despite the fact that it cannot be reached via DTs of the DPT potential.

In fact, substituting (3.22a) and (3.22b) into (2.51) gives

$$_\iota\xi_{\dagger;1} = \frac{\lambda_\dagger + 1}{\iota(\lambda_\dagger + 1) + {}_\iota\nu_\dagger + \iota} \qquad (3.42)$$

$$= \frac{\sigma_{0;\dagger}\lambda_{o;\dagger}}{\iota\sigma_{0;\dagger}\lambda_{o;\dagger} + \sigma_{1;\dagger}\nu_{o\iota;\dagger}}. \qquad (3.42^*)$$

In particular, if $\dagger = \mathsf{a}$ or $\mathsf{b}$ then

$$_1\xi_{\dagger;1} \equiv z_{\dagger;1} = z_-(\lambda_{o;\dagger}, \mu_{o;\dagger}) \qquad (3.43)$$

with

$$z_\pm(\lambda, \mu) = \frac{\lambda}{\lambda \pm \mu} \qquad (3.43^*)$$

for $\iota = 1$ and

$$_0\xi_{\dagger;1}/\nu_o \equiv \varsigma_{\dagger;1} = -\lambda_{o;\dagger} \ \text{ for } \ \dagger = \mathsf{a} \text{ or } \mathsf{b} \qquad (3.43')$$

for $\iota = 0$. It then directly follows from (3.41) and (3.41′) that the potentials $V[z\,|\,{}_{1}^{1}G^{201}_{a,1}]$ and $V[z\,|\,{}_{1}^{1}G^{201}_{b,1}]$ coincide so that both DPT potential and isotonic oscillator have only one isospectral SUSY partner in Quesne's linear case [13]. Note also that ${}_{\iota}\xi_{b;1} = 0$ for $\lambda_o = 1$ which implies that the appropriate AEH solution of type b disappears in the limiting case $\lambda_o \to 1$ for both DPT potential and isotonic oscillator. Since ${}_{1}\xi_{a;1} = 1$ in the limiting case $\mu_o = 1$, the symmetric DPT potential $V_1[z;1,1]$ does not have any isospectral SUSY partners for m=1.

As mentioned above, the fragment $0 < \lambda_{o;\dagger},\, \mu_{o;\dagger} < 1$ of the *r*-XQ potential cannot be reached by any DT. On other hand, as proven by Gomez-Ullate, Kamran and Milson [1] the appropriate Fuschian equations has two infinite sets of X-polynomial solutions:

$$ {}_{1}\phi[z;\lambda,\mu\,|\,+,n] = \frac{{}_{1}\Theta[\xi;\lambda,\mu]}{z - z_{-}(\lambda,\mu)}\; \hat{P}^{(\lambda,\mu)}_{1,1+n}(2z-1)\,, \tag{3.44} $$

$$ {}_{\iota}\phi[z;-\lambda,-\mu\,|\,+,n] = \frac{{}_{1}\Theta[\xi;-\lambda,-\mu]}{z - z_{-}(\lambda,\mu)}\; \hat{P}^{(-\lambda,-\mu)}_{1,1+n}(2z-1) \tag{3.44*} $$

for $0 < \lambda = \lambda_{o;\dagger}, \mu = \mu_{o;\dagger} < 1$, with infinite set (3.44) describing discrete energy levels associated with a higher characteristic exponent.

To match our expression for isospectral SUSY partners of the DPT potential to (11) in [6], one needs first express (3.33) in terms of the new varianle

$$ g(\tfrac{1}{2}\pi - 2x) \equiv \eta(x) = 1 - 2z(x) = \cos 2x \qquad ({}_{1}d = 4), \tag{3.45} $$

taking into account that

$$ \eta - \eta_{\dagger,1} = 2(z_{\dagger,1} - z) \tag{3.45*} $$

and

$$ \lambda_{o;\dagger} = \alpha \equiv A - B - \tfrac{1}{2} > 0\,, \tag{3.46a} $$

$$ \mu_{o;\dagger} = \beta \equiv A + B - \tfrac{1}{2} > 1 \tag{3.46b} $$

$(A > \tfrac{1}{2})$ so that

$$\eta_{\dagger,1} \equiv 1 - 2z_{\dagger,1} = \frac{\lambda_{o;\dagger} + \nu_{o;\dagger}}{\nu_{o;\dagger} - \lambda_{o;\dagger}} = \frac{2A-1}{2B}. \qquad (3.46^*)$$

Substituting (3.45*) and (3.46*), together with

$$z_{\dagger,1}(1 - z_{\dagger,1}) = -\frac{\lambda_{o;\dagger}\,\mu_{o;\dagger}}{(\lambda_{o;\dagger} - \mu_{o;\dagger})^2} = \frac{4B^2 - (2A-1)^2}{16B^2}, \qquad (3.47)$$

into (3.40) then directly leads to Quesne's formulas (11) in [6] or (3.5) in [13].

It should be mentioned in this connection that the Bose invariants of our current interest for m=1 has already appeared in Quesne's pioneering work [6] on the RPs quantized via exceptional Jacobi and Laguerre polynomials. Namely, she started from the Heun and *c*-Heun equations conditionally exactly solvable via the X-OPSs [1, 2] and then converted them to the canonical form via proper gauge transformations. The resultant equations can be then represented in the Schrödinger form using the appropriate Liouville transformations. An analysis of Quesne's formula [6]

$$\tfrac{1}{4} I^o[z;\, {}_1\varepsilon_{c;n-1} \,|\, {}^{1}_{1}\mathcal{G}^{201}_{\dagger,1}] = \frac{G\eta + J}{(1-\eta^2)^2} + \frac{C\eta + \Delta D + {}_1\varepsilon_{c;n-1}}{1-\eta^2} + \frac{C}{\eta - \eta_{\dagger,1}} - \frac{2}{(\eta - \eta_{\dagger,1})^2} \qquad (3.48)$$

for the Bose invariant expressed in terms of variable (2.37) at the energy

$${}_1\varepsilon_{c;n-1} = \tfrac{1}{4}(\lambda_{o;\dagger} + \mu_{o;\dagger} + 2n - 1)^2 \quad \text{for} \quad \dagger = a \text{ or } b. \qquad (3.49)$$

In our terms

$$\frac{G\eta + J}{(1-\eta^2)^2} = \frac{1 - \lambda^2_{o;\dagger}}{16z^2} + \frac{1 - \mu^2_{o;\dagger}}{16(1-z)^2} + \frac{1 - O^o_0(\lambda_{o;\dagger}, \mu_{o;\dagger})}{16z(1-z)}, \qquad (3.50a)$$

$$C = \frac{\mu^2_{o;\dagger} - \lambda^2_{o;\dagger}}{2\lambda_{o;\dagger}\mu_{o;\dagger}} = \frac{2z_{\dagger,1} - 1}{2z_{\dagger,1}(1 - z_{\dagger,1})} = \frac{4B(2A-1)}{(2A-1)^2 - 4B^2}, \qquad (3.50b)$$

$$\Delta D = \frac{\lambda^2_{o;\dagger} + \mu^2_{o;\dagger}}{2\lambda_{o;\dagger}\mu_{o;\dagger}} - \tfrac{5}{4} = \frac{(\lambda_{o;\dagger} - \mu_{o;\dagger})^2}{2\lambda_{o;\dagger}\mu_{o;\dagger}} - \tfrac{1}{4}. \qquad (3.50c)$$

so that our expression for the Bose invariant is indeed consistent with (3.48).

Coming back to the general case, ι = 0 or 1 (m ≥1), note that Fuschian equation (2.51) turns into the RSL equation [33, 87]

$$\{{}_\iota\hat{D}_{\dagger,m_\dagger;\bar{\sigma}'}(\lambda_\dagger, {}_\iota\nu_\dagger) + \tfrac{1}{4}\,{}_\iota d\,\varepsilon\,\Pi_{m_\dagger}[\xi;{}_\iota\bar{\xi}_{\dagger,m_\dagger}]\}\,{}_\iota F_{\dagger,m_\dagger;\bar{\sigma}'}[\xi;\varepsilon;\lambda_\dagger, {}_\iota\nu_\dagger] = 0, \quad (3.51)$$

where both the polynomial coefficients

$$(-1)^\iota 2\,{}_\iota B_{m_\dagger+1}[\xi;\lambda,\nu;{}_\iota\bar{\xi}_{\dagger,m_\dagger}] = [(1+\lambda)(1-\iota\xi)-(\iota+\nu)\xi]\Pi_{m_\dagger}[\xi;{}_\iota\bar{\xi}_{\dagger,m_\dagger}]$$
$$-2\xi(1-\iota\xi)\dot{\Pi}_{m_\dagger}[\xi;{}_\iota\bar{\xi}_{\dagger,m_\dagger}] \quad (3.52)$$

and

$$C^0_{m_\dagger}[\xi;\lambda_\dagger,{}_\iota\nu_\dagger;\bar{\sigma}'] = \tfrac{1}{4}\breve{O}^{\downarrow}_{m_\dagger}[\xi\,|\,{}^1_\iota\mathcal{G}^{K01}_{\dagger,m_\dagger}] + {}_\iota C^0_0(\sigma'_0\lambda_\dagger,\sigma'_1\,{}_\iota\nu_\dagger)\Pi_{m_\dagger}[\xi;{}_\iota\bar{\xi}_{\dagger,m_\dagger}]$$
$$-2\,{}_\iota B_1[\xi;\sigma'_0\lambda_\dagger,\sigma'_1\,{}_\iota\nu_\dagger]\dot{\Pi}_{m_\dagger}[\xi;{}_\iota\bar{\xi}_{\dagger,m_\dagger}] \quad (3.53)$$

(K = ι+1)

of the second-order differential operator

$$ {}_\iota\hat{D}_{\dagger,m_\dagger;\bar{\sigma}'}(\lambda_\dagger,{}_\iota\nu_\dagger) \equiv \xi(1-\iota\xi)\Pi_{m_\dagger}[\xi;{}_\iota\bar{\xi}_{\dagger,m_\dagger}]\frac{d^2}{d\xi^2} \quad (3.54)$$
$$+(-1)^\iota 2\,{}_\iota B_{m_\dagger+\iota}[\xi;\sigma'_0\lambda_\dagger,\sigma'_1\,{}_\iota\nu_\dagger;{}_\iota\bar{\xi}_{\dagger,m_\dagger}]\frac{d}{d\xi} + (-1)^\iota\,{}_\iota C^0_{m_\dagger}[\xi;\lambda_\dagger,{}_\iota\nu_\dagger;\bar{\sigma}']$$

are *energy-independent*, contrary to other shape-invariant potentials. An explicit expression for the m-order GPD polynomial $\breve{O}^{\downarrow}_{m_\dagger}[\xi\,|\,{}^1_\iota\mathcal{G}^{K01}_{\dagger,m_\dagger}]$ is obtained by substituting (3.20a) and (3.20b) into (2.36):

$$ {}_\iota\breve{O}^{\downarrow}_{m_\dagger}[\xi\,|\,{}^1_\iota\mathcal{G}^{K01}_{\dagger,m_\dagger}] = -[4\,{}_\iota C^0_0(-\lambda_\dagger,-{}_\iota\nu_\dagger) - {}_\iota d\,{}_\iota\varepsilon_{\dagger,m_\dagger}]\,\Pi_{m_\dagger}[\xi;{}_\iota\bar{\xi}_{\dagger,m_\dagger}]$$
$$+8\,{}_\iota B_1[\xi;-\lambda_\dagger,-{}_\iota\nu_\dagger]\dot{\Pi}_{m_\dagger}[\xi;{}_\iota\bar{\xi}_{\dagger,m_\dagger}]. \quad (3.55)$$

As expected, the free term in RSL equation (3.53) vanishes at the energy $\varepsilon = {}_\iota\varepsilon_{\dagger,m_\dagger}$ so that the equation has a constant solution associated with basic AEH function (2.25). From our point of view, we deal with an excellent illustration of the problem addressed by Heine [9, 10], with ε chosen in such a way that Fuschian equation (3.53) has a polynomial solution. As discussed in next Section, there is an infinite number of such solutions for both PF beams ${}^{1}_{1}G^{201}_{\dagger,m_\dagger}$ and ${}^{1}_{0}G^{101}_{\dagger,m_\dagger}$. These are certainly not Heine-Stieltjes polynomials since in the latter case all the exponents are required to be positive [10, 111, 113] but otherwise the comment added by Ho, Odake, and Sasaki [87] in the end of their paper does not seem to be appropriate.

RSL equation (3.51) can be also interpreted as an eigenpolynomial problem for the P-CES differential operator

$$\begin{aligned}{}^{*}_{\iota}\hat{T}_{\dagger,m_\dagger;\bar{\sigma}'}(\lambda_{o;\dagger},\mu_{o;\dagger};\bar{\xi}_{\dagger,m_\dagger}) \equiv p[\xi]\frac{d^2}{d\xi^2} + {}_{\iota}q[\xi;\sigma'_0\lambda_\dagger,\sigma'_1\,{}_{\iota}\nu_\dagger;\bar{\xi}_{\dagger,m_\dagger}]\frac{d}{d\xi} \\ + {}_{\iota}C^{0}_{m_\dagger}[\xi;\lambda_\dagger,{}_{\iota}\nu_\dagger;\bar{\sigma}'],\end{aligned} \tag{3.56}$$

where the coefficient of the second derivative and the free term are defined via (3.23′) and (3.53), respectively, and

$${}_{\iota}q[\xi;\sigma'_0\lambda_\dagger,\sigma'_1\,{}_{\iota}\nu_\dagger;\bar{\xi}_{\dagger,m_\dagger}] \equiv {}_{\iota}p[\xi]\left[\frac{\sigma'_0\lambda_\dagger+1}{\xi} - \frac{\sigma'_1\,{}_{\iota}\nu_\dagger+1}{1-\iota\xi} - \sum_r \frac{2}{\xi-\xi_{\dagger,m_\dagger;r}}\right]. \tag{3.57}$$

The derived expression for operator (3.56) quantized via orthogonal polynomials is expected to match (3.5) in [87] though we were unable to explicitly confirm this so far. We could neither directly relate the RF scheme given by (3.2)-(3.4) to that utilized in Gomez-Ullate, Kamran, and Milson’s theory of exceptional orthogonal polynomials.

In Quesne’s linear case (m=1) for ι=1

$$\begin{aligned}{}_{\iota}C^{0}_{m_\dagger}[\xi;\lambda_\dagger,{}_{\iota}\nu_\dagger;\bar{\sigma}'] = [\tfrac{1}{4}O^{o}_{0}(\lambda_{o;\dagger},\mu_{o;\dagger}) + {}_{1}C^{0}_{0}(\sigma'_0\lambda_\dagger,\sigma'_1\mu_\dagger)](z-z_{\dagger,1}) \\ -\sigma'_0\lambda_\dagger(z-1)-\sigma'_1\mu_\dagger z\end{aligned} \tag{3.58}$$

and the appropriate Heine equations merge into a single Heun equation discussed in more detail in Section 5 below. By expressing this equation in terms of $\eta$ and setting $\bar{\sigma}'$ to $\bar{\sigma}_\dagger$, we come to the starting point of Quesne’s breakthrough study [6] on rational extensions of the DPT potentials quantized via $X_1$-Jacobi polynomials, with

$$Q(\eta) = \frac{\lambda_{o;\dagger}+1}{\eta-1} + \frac{\mu_{o;\dagger}+1}{\eta+1} - \frac{2}{\eta-\eta_{\dagger;1}} \tag{3.59a}$$

and

$$R(\eta) = \frac{\lambda_{o;\dagger}-\mu_{o;\dagger}}{\eta_{\dagger;1}-\eta} + \frac{\lambda_{o;\dagger}-\mu_{o;\dagger}+{}_1d(\varepsilon_{c;n-1}-\varepsilon_{c;0})}{1-\eta^2} \tag{3.59b}$$

in our terms, taking into account that

$$\varepsilon_{c;0} = (\lambda_{o;\dagger}+\mu_{o;\dagger}+1)^2 = O_0^0(\lambda_{o;\dagger},\mu_{o;\dagger}) + 4\,{}_1C_0^0(\lambda_{o;\dagger},\mu_{o;\dagger}) \tag{3.60}$$

for $\dagger$ = $\mathbf{a}$ or $\mathbf{b}$.

So far we simply assumed that the RSL equations with Bose invariants (3.2) and (3.2′) do have AEH solutions of all four types $\mathbf{a}$, $\mathbf{b}$, $\mathbf{c}$, and $\mathbf{d}$. In next Section we discuss this question in a more systematic way by analyzing explicit formulas for energies of all possible AEH solutions obtained by truncating the appropriate hypergeometric and confluent hypergeometric series.

## 4. Almost-Everywhere-Holomorphic Solutions of the Gauss-Reference SL Equations With Energy-Independent Characteristic Exponents

### 4.1 The DPT potential

By making one of four gauge transformations

$$\Phi[z;\varepsilon\,|\,{}_1\mathcal{G}^{201}] = {}_1\Theta[z;\sigma_0\lambda_o,\sigma_1\mu_o]\;\; F[z;{}_1d\varepsilon\,|\,{}_1\mathcal{G}^{201};\bar{\sigma}] \tag{4.1.1}$$

one comes to the hypergeometric equation

$$[{}_1\hat{D}(\sigma_0\lambda_o,\sigma_1\mu_o) + \tfrac{1}{4}(\sigma_0\lambda_o+\sigma_1\mu_o+1)^2 + \tfrac{1}{4}\,{}_1d\varepsilon]\; F[z;{}_1d\varepsilon\,|\,{}_1\mathcal{G}^{201};\bar{\sigma}] = 0\,, \tag{4.1.2}$$

where the second-order differential operator is defined via (2.22) with the explicit expression for free term found by combining (2.23), (2.23*), and (3.5). The solution

$$F_0[z;{}_1d\varepsilon\,|\,{}_1\mathcal{G}^{201};\bar{\sigma}] = F[\alpha({}_1d\varepsilon\,|\,\sigma_0\lambda_o,\sigma_1\mu_o),\ \beta({}_1d\varepsilon\,|\,\sigma_0\lambda_o,\sigma_1\mu_o);\sigma_0\lambda_o+1;z]\,, \tag{4.1.3}$$

with

$$\alpha(\varepsilon\,|\,\lambda,\mu) \equiv \tfrac{1}{2}(\lambda+\mu+1-\varepsilon)\,, \tag{4.1.4a}$$

and

$$\beta(\varepsilon\,|\,\lambda,\mu) \equiv \tfrac{1}{2}(\lambda+\mu+1+\varepsilon)\,, \tag{4.1.4b}$$

is chosen to satisfy the boundary condition

$$F_0[0;{}_1d\varepsilon\,|\,{}_1\mathcal{G}^{201};\bar{\sigma}_{\dagger}] = 1 \tag{4.1.5}$$

so that $\Pi_m[z;\bar{z}_{\dagger,m}]$ is a scaled hypergeometric polynomial

$$\Pi_m[z;\bar{z}_{\dagger,m}] = F[-m,\ \tfrac{1}{2}(\lambda_{\dagger}+\mu_{\dagger}+1+{}_1d\,{}_1\varepsilon_{\dagger,m});\sigma_0\lambda_o+1;z]/f_{\dagger,m}\,, \tag{4.1.6}$$

where

$$f_{\dagger,m} = \frac{\alpha({}_1d\,{}_1\varepsilon_{\dagger,m}\,|\,\lambda_{\dagger},\mu_{\dagger})\ \beta({}_1d\,{}_1\varepsilon_{\dagger,m}\,|\,\lambda_{\dagger},\mu_{\dagger})}{(\lambda_{\dagger}+m)_m}\,, \tag{4.1.6*}$$

except either $\mu_o$ or $\lambda_o$ is a positive integer not larger than m. The latter constraints come from the requirement that the polynomial $\Pi_m[z;\bar{z}_{\dagger,m}]$ may not vanish at the ends of the quantization interval [0, 1]:

$$\Pi_m[0;\bar{z}_{\dagger,m}] \neq 0, \tag{4.1.7a}$$

$$\Pi_m[1;\bar{z}_{\dagger,m}] \neq 0\,. \tag{4.1.7b}$$

For $\dagger = \mathsf{b}$ or $\mathsf{d}$ condition (4.1.7a) holds unless the third argument of hypergeometric function (4.1.3) is either zero or a negative integer –k with absolute value not larger than m. Similarly, the second condition holds for $\dagger = \mathsf{a}$ or $\mathsf{d}$ as far as the third argument of hypergeometric function

$$F_1[z;{}_1d\varepsilon\,|\,{}_1\mathcal{G}^{201};\bar{\sigma}_{\dagger}] = F[\alpha({}_1d\varepsilon\,|\,\lambda_{\dagger},\mu_{\dagger}),\ \beta({}_1d\varepsilon\,|\,\lambda_{\dagger},\mu_{\dagger});\mu_{\dagger}+1;1-z] \tag{4.1.8}$$

differs from a nonpositive integer –k, again with absolute value not larger than m. The corresponding AEH solution becomes a normalizable function in both cases $\dagger = \mathsf{a}$ and $\dagger = \mathsf{b}$ while turning into a regular solution at the appropriate end for $\dagger = \mathsf{d}$.

We thus conclude that condition (3.18) is both necessary and sufficient for an AEH solution to exist at the energy ${}_1\varepsilon_{\mathbf{t},m}$ except either $\lambda_o$ for $\mathbf{t} = \mathbf{b}$ or $\mathbf{d}$ or $\mu_o$ for $\mathbf{t} = \mathbf{a}$ or $\mathbf{d}$ is a positive integer not larger than m. As far as the latter condition holds $\Pi_m[z;\bar{z}_{\mathbf{t},m}(\lambda_o,\mu_o)]$ is proportional to the Jacobi polynomial:

$$P_m^{(\mu_{\mathbf{t}},\lambda_{\mathbf{t}})}(1-2z) = \tilde{k}_m(\lambda_{\mathbf{t}},\mu_{\mathbf{t}})\Pi_m[z;\bar{z}_{\mathbf{t},m}(\lambda_o,\mu_o)], \tag{4.1.9}$$

keeping in mind that

$$P_m^{(\mu_{\mathbf{t}},\lambda_{\mathbf{t}})}(\eta) = k_m(\lambda_{\mathbf{t}},\mu_{\mathbf{t}})\Pi_m[\eta;\bar{\eta}_{\mathbf{t},m}(\lambda_o,\mu_o)] \tag{4.1.9*}$$

where [113]

$$k_m(\lambda,\mu) = 2^{-m}\tilde{k}_m(\lambda,\mu) = \frac{(\lambda+\mu+m)_m}{m!\ 2^m}, \tag{4.1.10}$$

with $(\alpha)_m$ standing for the rising factorial.

Since solutions regular near the point z=0 do not have nodes between 0 and 1 for $\varepsilon < {}_1\varepsilon_{\mathbf{c},0}$, they can be used as FFs to construct a one-parameter isospectral family of CEQ potentials as done by Conteras-Astorga and Fernandez [114]. More recently Odake and Sasaki [115] came up with a similar suggestion but, in addition, made a very important observation -- the isospectral family of CEQ potentials constructed in such a way retains its form under DTs with nodeless normalizable FFs. It should be stressed in this connection that the shape-invariance of isospectral SUSY partners of the DPT potential has nothing to do with the fact that its REs are quantized via $X_m$-Jacobi polynomials. One can construct a similar one-parameter isospectral family of shape-invariant P-CEQ potentials by applying DTs with regular FFs to any shape-invariant potential. The first example of a shape-invariant potential of this kind can be traced back to the breakthrough work of Sukumar [73] who constructed an isospectral SUSY partner of the harmonic oscillator using the parabolic cylinder function -- the solution of type $\mathbf{b}$ according to (19.8.1) in [73]. All unbroken SUSY partners of shape-invariant potentials constructed by Junker and Roy [8, 118] are shape-invariant.

We can further extend this procedure by multi-step DTs with regular FFs to construct *infinite multi-index* sequences of continuous one-parameter isospectral families of shape-invariant CEQ potentials. From our point of view, this fact largely deflates significance of Gendenshtein's arguments [105]. It is not obvious that shape-invariance of the potentials constructed in [114, 115] makes them in some way more special compared with similar continuous one-parameter isospectral families of SUSY partners of the generic *r*-and *c*-GRef potentials.

The most important common feature of all shape-invariant GRef potentials is that the density function in the appropriate RSL as well as the ground eigenfunction have no zeros either on the real axis or away from it, when the latter are analytically continued into the complex plane. This common feature of these potential makes them form-invariant under DTs with nodeless normalizable FFs since the mentioned transformations do not create new singular points. However, the latter feature is possibly an interesting observation but by no means a fundamentally significant criterion. For this reason we are not planning to emphasize furthermore that all the isospectral REs of the DPT potential and isotonic oscillator discussed both here and in Part II turn out to be shape-invariant. Since energies of non-normalizable AEH solutions grows as $m_{\mathsf{t}}^2$ at large $m_{\mathsf{t}}$, we immediately conclude that the DPT potential may have only a finite number of REs for the given values of exponent differences. An analysis of the energy frequencies:

$$ {}_1d({}_1\varepsilon_{\mathsf{a},m_{\mathsf{a}}} - {}_1\varepsilon_{\mathsf{c},0}) = 4(m_{\mathsf{a}} - \mu_o)(\lambda_o + m_{\mathsf{a}} + 1)\,, \tag{4.1.11a} $$

$$ {}_1d({}_1\varepsilon_{\mathsf{b},m_{\mathsf{b}}} - {}_1\varepsilon_{\mathsf{c},0}) = 4(m_{\mathsf{b}} - \lambda_o)(\mu_o + m_{\mathsf{a}} + 1)\,, \tag{4.1.11b} $$

shows that the solutions of type $\mathsf{a}$ and $\mathsf{b}$ lie below the ground energy level for

$$ 0 \le m_{\mathsf{a}} < \mu_o \tag{4.1.12a} $$

and

$$ 0 \le m_{\mathsf{b}} < \lambda_o\,, \tag{4.1.12b} $$

accordingly.[x)] It is worth mentioning that that all regular-at-one-end AEH solutions formed by orthogonal Jacobi polynomials of nonzero order ( $\mathsf{t} = \mathsf{a},\ \lambda_o,\ \mu_o < 1$ or $\mathfrak{t} = \mathfrak{b}$, $\lambda_o < 1$) lie above the ground energy level in support of the assumption that the conventional results [116, 117] for positions of zeros of regular solutions are applicable to discrete energy solutions embedded in the continuous energy spectrum.

In the limiting cases of integer $\mu_o$ or $\lambda_o$ the appropriate AEH solutions of types $\mathfrak{a}$ and $\mathfrak{b}$ for $m_{\mathfrak{a}} = \mu_o$ or, subsequently, $m_{\mathfrak{b}} = \lambda_o$ turn into the ground eigenfunction. A simple illustration of this rule has been already presented in previous Section, namely, as either $\lambda_o$ or $\mu_o$ tends to 1 the node of the first order polynomial $\Pi_1[z; z_{\mathfrak{t},1}]$ approaches either 0 or 1 for $\mathfrak{t} = \mathfrak{b}$ or $\mathfrak{a}$, respectively,

So far, it was taken for granted that the order of the polynomial in question is equal to $m_{\mathfrak{t}}$. However, if the second parameter of the hypergeometric function is also a negative integer larger than $-m_{\mathfrak{a}}$, i. e., if

$$\beta({}_1d_1\varepsilon_{\mathfrak{a},m_{\mathfrak{a}}} \mid -\mu_o, \lambda_o) = \lambda_o - \mu_o + m_{\mathfrak{a}} + 1 = k_{\mathfrak{a}} - m_{\mathfrak{a}} > -m_{\mathfrak{a}} \quad \text{for } 0 < k_{\mathfrak{a}} \le m_{\mathfrak{a}}, \tag{4.1.13}$$

then the polynomial order is equal to $n_{\mathfrak{a}} = m_{\mathfrak{a}} - k_{\mathfrak{a}}$, not $m_{\mathfrak{a}}$, which implies that the appropriate RE of the DPT potential disappears for any integer $\mu_o - \lambda_o$ larger than 1. We thus come to the degenerate case described by (89) in [42], with $\alpha = -\mu_o$, $\beta = \lambda_o$, $n = m_{\mathfrak{a}}$, $m = m_{\mathfrak{a}} - k_{\mathfrak{a}}$ in our terms. If $\mu_o - \lambda_o$ is an integer larger than 2 the sequence of REs of type $\mathfrak{a}$ has a gap for $m_{\mathfrak{a}}$ within the range

$$[(\mu_o - \lambda_o)/2] \le m_{\mathfrak{a}} \le \mu_o - \lambda_o - 1. \tag{4.1.14}$$

Since

---

x) To prove that FFs used in [20] to construct REs of the DPT potential are indeed nodeless, Odake and Sasaki [23] pointed to the fact that coefficients of hypergeometric polynomials $F(-m_{\mathfrak{a}}, \lambda_o - \mu_o + m_{\mathfrak{a}} + 1; \lambda_o + 1; z)$ are all positive for $0 < \lambda_o + 1 < \mu_o - m_{\mathfrak{a}}$. However, the range $\mu_o - \lambda_o - 1 < m_{\mathfrak{a}} < \mu_o$ is not covered by the proof.

$$(k_{\mathfrak{b}}-\lambda_o)F[-m_{\mathfrak{b}},\nu_o-\lambda_o+m_{\mathfrak{b}}+1;1-\lambda_o;z]=\frac{1}{\prod_{r=1}^{k_{\mathfrak{b}}-1}(r-\lambda_o)}z^{k_{\mathfrak{b}}} \tag{4.1.15}$$

$$\times\sum_{j=1}^{m_{\mathfrak{b}}-k_{\mathfrak{b}}}\frac{(j+k_{\mathfrak{b}}-m_{\mathfrak{b}})_{j+k_{\mathfrak{b}}}\,(\nu_o-\lambda_o+m_{\mathfrak{b}}+j+k_{\mathfrak{b}}+1)_{j+k_{\mathfrak{b}}}}{\prod_{r'=k_{\mathfrak{b}}+1}^{j+k_{\mathfrak{b}}}(r'-\lambda_o)}z^j$$

for integer $\lambda_o$, $k_{\mathfrak{b}}$ outer singularities merge with zero, the limiting eigenfunction becomes squarely integrable, and the solution of type $\mathfrak{b}$ disappears.

As proven in Appendix C, there is also a finite regular sequence of nodeless AEH solutions of type $\mathfrak{d}$ formed by Jacobi polynomials of even order

$$2\le m_{\mathfrak{d}}=2j_{\mathfrak{d}}<\lambda_o\le\mu_o. \tag{4.1.16}$$

As expected, they all lie below the ground energy level:

$${}_1d({}_1\varepsilon_{\mathfrak{d},m_{\mathfrak{d}}}-{}_1\varepsilon_{\mathfrak{c},0})=-4(\lambda_o+\mu_o-m_{\mathfrak{d}})(m_{\mathfrak{d}}+1)<0. \tag{4.1.17}$$

To be able to apply the same generators of J-Heine polynomials to both versions of the PT potential, we re-define the HPGs for ι = 1 in a uniform way:

$$\hat{a}(\mu_{\mathfrak{t}};\bar{z}_{\mathfrak{t},m_{\mathfrak{t}}})=(1-z)\left[\Pi_{m_{\mathfrak{t}}}[z;\bar{z}_{\mathfrak{t},m_{\mathfrak{t}}}]\frac{d}{dz}-\dot{\Pi}_{m_{\mathfrak{t}}}[z;\bar{z}_{\mathfrak{t},m_{\mathfrak{t}}}]\right]+\mu_{\mathfrak{t}}\,\Pi_{m_{\mathfrak{t}}}[z;\bar{z}_{\mathfrak{t},m_{\mathfrak{t}}}], \tag{4.1.18a}$$

$$\hat{b}(\lambda_{\mathfrak{t}};\bar{z}_{\mathfrak{t},m_{\mathfrak{t}}})\equiv z\left[\Pi_{m_{\mathfrak{t}}}[z;\bar{z}_{\mathfrak{t},m_{\mathfrak{t}}}]\frac{d}{dz}-\dot{\Pi}_{m_{\mathfrak{t}}}[z;\bar{z}_{\mathfrak{t},m_{\mathfrak{t}}}]\right]-\lambda_{\mathfrak{t}}\,\Pi_{m_{\mathfrak{t}}}[z;\bar{z}_{\mathfrak{t},m_{\mathfrak{t}}}], \tag{4.1.18b}$$

$$\hat{c}(\bar{z}_{\mathfrak{t},m_{\mathfrak{t}}})\equiv\Pi_{m_{\mathfrak{t}}}[z;\bar{z}_{\mathfrak{t},m_{\mathfrak{t}}}]\frac{d}{dz}-\dot{\Pi}_{m_{\mathfrak{t}}}[z;\bar{z}_{\mathfrak{t},m_{\mathfrak{t}}}], \tag{4.1.18c}$$

$$\hat{d}(\lambda_{\mathfrak{t}},\mu_{\mathfrak{t}};\bar{z}_{\mathfrak{t},m_{\mathfrak{t}}})=\hat{g}(-\lambda_{\mathfrak{t}},-\mu_{\mathfrak{t}};\bar{z}_{\mathfrak{t},m_{\mathfrak{t}}}), \tag{4.1.18d}$$

where generic HPG (4.1.18d) is defined via (2.57). If negative, the signed exponent differences have to be restricted by the constraints $\lambda_{\mathfrak{b}}$, $\lambda_{\mathfrak{d}}$, $\mu_{\mathfrak{a}}$, $\mu_{\mathfrak{d}}<-1$. The monomial products in the right-hand side of (4.1.18a)-(4.1.18c) and (2.57) are related to Jacobi polynomials via (4.1.9), with

$$\bar{z}_{\dagger,m_\dagger} = \bar{z}_{m_\dagger}(\lambda_\dagger,\mu_\dagger) \tag{4.1.19}$$

The appropriate 16 infinite sequences of polynomial solutions can be then represented as

$$\mathrm{Hi}^{(m_\dagger+2)}_{m_\dagger+m}[z;\lambda_\dagger,\mu_\dagger;+-] \tag{4.1.20a}$$

$$\equiv \frac{1}{(\mu_\dagger - m + m_\dagger)\tilde{k}_m(\lambda_\dagger,-\mu_\dagger)}\hat{a}(\mu_\dagger;\bar{z}_{\dagger,m_\dagger})P_m^{(-\mu_\dagger,\lambda_\dagger)}(2z-1),$$

$$\mathrm{Hi}^{(m_\dagger+2)}_{m_\dagger+m}[z;\lambda_\dagger,\mu_\dagger;-+] \tag{4.1.20b}$$

$$\equiv \frac{1}{(m - m_\dagger - \lambda_\dagger)\tilde{k}_m(-\lambda_\dagger,\mu_\dagger)}\hat{b}(\lambda_\dagger;\bar{z}_{\dagger,m_\dagger})P_m^{(\mu_\dagger,-\lambda_\dagger)}(2z-1),$$

$$\mathrm{Hi}^{(m_\dagger+2)}_{m+m_\dagger-1}[z;\lambda_\dagger,\mu_\dagger;++] \tag{4.1.20c}$$

$$\equiv \frac{1}{(m - m_\dagger)\tilde{k}_m(\lambda_\dagger,\mu_\dagger)}\hat{c}(\bar{z}_{\dagger,m_\dagger})P_m^{(\mu_\dagger,\lambda_\dagger)}(2z-1) \quad \text{for } m \neq m_\dagger,$$

and

$$\mathrm{Hi}^{(m_\dagger+2)}_{m+m_\dagger+1}[z;\lambda_\dagger,\mu_\dagger;--] \tag{4.1.20d}$$

$$\equiv \frac{1}{(m - m_\dagger - \lambda_\dagger - \mu_\dagger)\tilde{k}_m(-\lambda_\dagger,-\mu_\dagger)}\hat{d}(-\lambda_\dagger,-\mu_\dagger;\bar{z}_{\dagger,m_\dagger})P_m^{(-\mu_\dagger,-\lambda_\dagger)}(2z-1),$$

again assuming that $\lambda_b$, $\lambda_d$, $\mu_a$, $\mu_d < -1$. We refer to 4 infinite sequences of GS Heine polynomials covered by the same formula as belonging to the same series a, b, c, or d, respectively. Altogether we thus have 16 sequences keeping in mind that zeros of Jacobi polynomials may behave differently dependent on the particular quadrant in which polynomial indexes lie. The common feature of GS Heine polynomials from the same series is that they are all formally generated using the HPG of the same type.

As expected, Heine polynomials (4.1.20c) form a parse sequence missing a polynomial of order $2m_{\dagger}-1$. Another interesting feature of these sequence starts from the Jacobi polynomial $P^{(\mu_{\dagger},\lambda_{\dagger})}_{m_{\dagger}-1}(y)$, with y = 2z –1, as a direct consequence of the form-invariance of the DPT potential under CDTs with basic FFs.

One can directly relate the orders of GS Heine polynomials (4.1.20a)-(4.1.20d) to energies of the appropriate AEH solutions

$$\phi[z;\lambda_{o;\dagger},\mu_{o;\dagger}\,|\,\bar{\sigma}_{\dagger},m_{\dagger};\bar{\sigma},n]=\frac{{}_1\Theta[\xi;\sigma_0\lambda_{o;\dagger},\sigma_1\mu_{o;\dagger}]}{\Pi_{m_{\dagger}}[z;\bar{z}_{m_{\dagger}}(\lambda_{\dagger},\mu_{\dagger})]}\mathrm{Hi}_n^{(m_{\dagger}+2)}[z;\lambda_{\dagger},\mu_{\dagger};\bar{\sigma}] \tag{4.1.21}$$

In fact, an analysis of asymptotic behavior of this solution at large |z| shows that its energy $\varepsilon_{\dagger',m'}$ is related to the order n of the GS Heine polynomial via the following necessary condition:

$$\begin{aligned}{}_1d_1\varepsilon_{\dagger',m'} = O^{o}_{m_{\dagger};m_{\dagger}}\,|\,{}^{1}_{1}G^{201}_{\dagger,m_{\dagger}}) - O^{o}_{0}(\lambda_{o;\dagger},\mu_{o;\dagger}) - 8m_{\dagger}\\ +(\sigma_0\,\lambda_{o;\dagger}+\sigma_1\,\mu_{o;\dagger}+2n-2m_{\dagger}+1)^2,\end{aligned} \tag{4.1.22}$$

where $O^{o}_{m_{\dagger};m_{\dagger}}\,|\,{}^{1}_{1}G^{201}_{\dagger,m_{\dagger}})$ is the leading coefficient of the polynomial $O^{o}_{m_{\dagger}}[z\,|\,{}^{1}_{1}G^{201}_{\dagger,m_{\dagger}}]$. To find an explicit expression for this coefficient, first note that the leading coefficient of polynomial (2.41) is equal to

$$P_{m_{\dagger}-2;m_{\dagger}-2}(\bar{z}_{\dagger,m_{\dagger}}) = -m_{\dagger}(m_{\dagger}-1) \tag{4.1.23}$$

and hence, according to (2.40),

$$O^{o}_{0}(\lambda_{o;\dagger},\mu_{o;\dagger}) = \breve{O}^{\downarrow}_{m_{\dagger};m_{\dagger}}|\,{}^{1}_{1}G^{201}_{\dagger,m_{\dagger}}) - 4m_{\dagger}(m_{\dagger}-1)\,. \tag{4.1.24}$$

On other hand, as a direct consequence of (2.44) one finds

$$\breve{O}^{\downarrow}_{m_{\dagger};m_{\dagger}}|\,{}^{1}_{1}G^{201}_{\dagger,m_{\dagger}}) = O^{\downarrow}_{m_{\dagger};m_{\dagger}}\,|\,{}^{1}_{1}G^{201}_{\dagger,m_{\dagger}}) + 4m_{\dagger}(m_{\dagger}-1)\;, \tag{4.1.24*}$$

where

$$O^{\downarrow}_{m_{\dagger};m_{\dagger}}|\,{}^{1}_{1}G^{K01}_{\dagger,m_{\dagger}}) = O^{o}_{0}(\lambda_{o;\dagger},\mu_{o;\dagger}) + 8m_{\dagger} \tag{4.1.25}$$

is the leading coefficient of polynomial (3.37*). Combining (4.1.24) and (4.1.24*) with (4.1.25) gives

$$O^{o}_{0}(\lambda_{o;\dagger},\mu_{o;\dagger}) = O^{\downarrow}_{m_\dagger;m_\dagger} \mid {}^{1}_{1}G^{K01}_{\dagger,m_\dagger}) = O^{o}_{0}(\lambda_{o;\dagger},\mu_{o;\dagger}) + 8m_\dagger \qquad (4.1.25')$$

so that

$$(\sigma_{0;\dagger'}\ \lambda_o + \sigma_{1;\dagger'}\ \mu_o + 2m' + 1)^2 = (\sigma_0\ \lambda_{o;\dagger} + \sigma_1\ \mu_{o;\dagger} + 2n - 2m_\dagger + 1)^2. \qquad (4.1.26)$$

If $\lambda_o > 1$ then $\bar{\sigma}_\dagger{}' = \bar{\sigma}$ and either

$$n = m' + m_\dagger - \tfrac{1}{2}(\sigma_{0;\dagger'}\sigma_{0;\dagger} - \sigma_{1;\dagger'}\sigma_{1;\dagger}) \qquad (4.1.27a)$$

or

$$n = m_\dagger + 1 - \lambda_{\dagger'} - \mu_{\dagger'} - \tfrac{1}{2}(\sigma_{0;\dagger'}\sigma_{0;\dagger} + \sigma_{1;\dagger'}\sigma_{1;\dagger}). \qquad (4.1.27b)$$

The first formula for the polynomial order matches the general case represented by GS Heine polynomials (4.1.20a)-(4.1.20d) whereas the second is applicable for some integer values of the parameter $\lambda_{\dagger'} + \mu_{\dagger'}$.

For $m_\dagger = 0$ HPGs (4.1.20a)–(4.1.20d) turn into the ladder operators

$$\hat{a}(\mu) \equiv (1-z)\frac{d}{dz} + \mu = (1-y)\frac{d}{dy} + \mu\,, \qquad (4.1.28a)$$

$$\hat{b}(\lambda) \equiv z\frac{d}{dz} - \lambda = (\eta - 1)\frac{d}{d\eta} - \lambda\,, \qquad (4.1.28b)$$

$$\hat{c} \equiv \frac{d}{dz} = 2\frac{d}{dy} = -2\frac{d}{d\eta}\,, \qquad (4.1.28c)$$

$$\hat{d}(\lambda,\mu) = z(1-z)\frac{d}{dz} + \lambda(z-1) + \mu z = \tfrac{1}{2}\left[(1-y^2)\frac{d}{dy} + \lambda(y-1) + \mu(y+1)\right] \qquad (4.1.28d)$$

which change each index of Jacobi polynomials exactly by 1:

$$\hat{a}(\mu)P_n^{(-\mu,\lambda)}(y) = (\mu - n)P_n^{(-\mu-1,\lambda+1)}(y)\,, \qquad (4.1.29a)$$

$$\hat{b}(\lambda)P_n^{(\mu,-\lambda)}(y) = (n - \lambda)P_n^{(\mu+1,-\lambda-1)}(y)\,, \qquad (4.1.29b)$$

---

x) Note that z-1 should be changed for 1-z in the left-hand side of similar expression (74) in [42].

$$\hat{c}\,P_n^{(\mu,\lambda)}(y) = \frac{1}{2}(\lambda+\mu+n+1)P_{n-1}^{(\mu+1,\lambda+1)}(y)\,, \tag{4.1.29c}$$

$$\hat{d}(\lambda,\mu)P_n^{(-\mu,-\lambda)}(y) = -2n\,P_{n+1}^{(-\mu-1,-\lambda-1)}(y)\,. \tag{4.1.29d}$$

The first two of the latter ladder relations directly follow from (83) in [35]:[x)]

$$(x-1)\dot{P}_n^{(\alpha,\beta)}(x) + \alpha P_n^{(\alpha,\beta)}(x) = (\alpha+n)P_n^{(\alpha-1,\beta+1)}(x)\,, \tag{4.1.30}$$

with $x = y$, $\alpha = -\mu$, $\beta = \lambda$ and $x = \eta$, $\alpha = \mu$, $\beta = -\lambda$, respectively; whereas the third and fourth ladder relation lead to forward and backward shift relations (E.15) and (E.16), respectively, in Appendix E.2 in [87].

To explicitly relate Heine polynomials (4.1.20a) and (4.1.20b) to $X_m$-Jacobi polynomials, let us express operators (4.1.18a) for $\bar{\sigma}_{\dagger} = +-$ and (4.1.18b) for $\bar{\sigma}_{\dagger} = -+$ in terms of variables $y = 2z - 1$ and $\eta = 1-2z$. respectively. Taking into account (3.9), we can represent the exceptional polynomial generators (XPGs) as

$$\hat{X}_{J1}(\lambda_o,\mu_o;m_a) \equiv -\tilde{k}_{m_a}(\lambda_o,-\mu_o)\,\hat{a}(-\mu_o;\bar{z}_{a,m_a}) \tag{4.1.31a}$$

$$= (y-1)\left[P_{m_a}^{(-\mu_o,\lambda_o)}(y)\frac{d}{dy} - \dot{P}_{m_a}^{(-\mu_o,\lambda_o)}(y)\right] - 2\mu_o\,P_{m_a}^{(-\mu_o,\lambda_o)}(y), \tag{4.1.31a*}$$

and

$$\hat{X}_{J2}(\lambda_o,\mu_o;m_b) \equiv (-1)^m\tilde{k}_{m_b}(\lambda_o,-\mu_o)\hat{b}(-\lambda_o;\bar{z}_{b,m_b}) \tag{4.1.31b}$$

$$= (\eta-1)\left[P_{m_b}^{(-\lambda_o,\mu_o)}(\eta)\frac{d}{d\eta} - \dot{P}_{m_b}^{(-\lambda_o,\mu_o)}(\eta)\right] + 2\lambda_o\,P_{m_b}^{(-\lambda_o,\mu_o)}(\eta), \tag{4.1.31b*}$$

where dot stands for the derivative of the Jacobi polynomial with respect to its argument and $(\lambda)_m$ is the rising factorial. Making use of (4.1.30) thus gives

$$\hat{X}_{J1}(\lambda_o,\mu_o;m) = (\mu_o - m)P_m^{(-\mu_o-1,\lambda_o+1)}(y) + (y-1)\,P_m^{(-\mu_o,\lambda_o)}(y)\frac{d}{dy} \tag{4.1.32a}$$

and

$$\hat{X}_{J2}(\lambda_o,\mu_o;m) = (\lambda_o - m)P_m^{(-\lambda_o-1,\mu_o+1)}(\eta) + (\eta-1)P_m^{(-\lambda_o,\mu_o)}(\eta)\frac{d}{d\eta} \quad (4.1.32b)$$

(cases I and II in Quesne's terms [13]). Comparing (4.1.32a) and (4.1.32b) with (72) in [42] gives

$$\hat{X}_{J1}(\lambda_o,\mu_o;m)P_v^{(\mu_o,\lambda_o)}(y) = (-1)^m(\mu_o+v)\hat{P}_{m,m+v}^{(\mu_o-1,\lambda_o+1)}(y) \quad (4.1.33a)$$

and

$$\hat{X}_{J2}(\lambda_o,\mu_o;m)P_v^{(\lambda_o,\nu_{o1})}(y) = (-1)^m(\lambda_o+v)\hat{P}_{m,m+v}^{(\lambda_o-1,\mu_o+1)}(y) \quad (4.1.33b)$$

so that Heine polynomials (2.53a) and (2.53b) with $\bar{\sigma}=+-$ and $\bar{\sigma}=-+$, respectively, and $m'=v$ are nothing but scaled $X_m$- Jacobi polynomials:

$$\frac{(\lambda_o-\mu_o+2v)_v}{v!}Hi_{m_a+v}^{(m_a+2)}[z;\lambda_o,-\mu_o;+-] \quad (4.1.34a)$$
$$= (-1)^{m_a}(\mu_o+v)\upsilon(\lambda_o,\mu_o;a)\ \hat{P}_{m_a,m_a+v}^{(\mu_o-1,\lambda_o+1)}(2z-1)$$
$$\text{for } \lambda_o \geq 1,\ \mu_o \geq 2$$

and

$$\frac{(\mu_o-\lambda_o+2v)_v}{v!}Hi_{m_b+v}^{(m_b+2)}[z;-\lambda_o,\mu_o;-+]$$
$$= (\lambda_o+v)\upsilon(\lambda_o,\mu_o;b)\hat{P}_{m_b,m_b+v}^{(\lambda_o-1,\mu_o+1)}(1-2z) \quad (4.1.35b)$$
$$\text{for } \lambda_o \geq 2,\ \mu_o \geq 1.$$

If necessary, explicit expressions for the scale factors $\upsilon(\lambda_o,\mu_o\,|\,\bar{\sigma};\bar{\sigma})$ with $\bar{\sigma}=+-$ and $\bar{\sigma}=-+$ can be easily obtained by computing the leading coefficients of the appropriate $X_m$-Jacobi polynomials. With $\lambda_o = g+\ell+1/2$, $\mu_o = h+\ell-3/2$, the derived expression matches (2.3) in [87] for the exceptional Jacobi polynomials of types J1 and J2, respectively.

It should be stressed again that the indexes of the $X_m$-Jacobi polynomials appearing in the right-hand side of (4.1.35a) and (4.1.35b) may not be smaller than 1 for the appropriate potential to retain within the validity range of SUSY quantum mechanics whereas the precise definition of these polynomials $\hat{P}^{(\alpha,\beta)}_{m,m+v}(y)$ given by Gomez-Ullate, Kamran, and Milson [35] covers the region $\alpha, \beta \geq -1$. This implies that the formalism of REs of the DPT potential cannot be used as a conceptual tool for introducing this new class of orthogonal polynomials. In particular, this conclusion is applicable to multi-index exceptional Jacobi polynomials which will be analyzed in detail in Part II.

In addition to REs of the DPT potential quantized via the $X_m$-Jacobi polynomials, there are CEQ RPs with excited bound states described by orthogonal Heine polynomials (4.1.24d), The first term in this sequence of exactly solvable potentials has been already discovered by Quesne [13] in her pioneering analysis of REs of the DPT potential using second-order polynomials. In following [13], we refer to this sequence as Case III. Compared with the Cases I and II, these potentials are quantized via the so-called ‘sparse sequences’ of orthogonal Heine polynomials, namely, each begins with a constant but then has a gap followed by orthogonal Heine polynomials such that

$$\int_0^1 \frac{z^{\lambda_o}(1-z)^{\mu_o}}{\Pi^2_{2j_d}[z;\bar{z}_{d,2j}]} \mathrm{Hi}^{(2j_d+2)}_{2j_d+v}[z;-\lambda_o,-\mu_o;—]\,dz = 0. \tag{4.1.36}$$

Similarly to the $\mathfrak{a}$-series, the number of REs of the DBT potential specified by condition (4.1.16) grows with increase of $\mu_o$ at fixed $\lambda_o$. In the limit $\mu_o \to \infty$ this finite family of PF potentials turns into the infinitely many sequence of REs of the isotonic oscillator discovered by Grandati [31] using the LKH theorem. We shall come back to the discussion of this so-called ‘L3-series’ in sub-section 4.3 below.

### 4.2 The h-PT Potential

If the Schrödinger equation with the h-PT potential is converted into the SL problem with Bose invariant $I[\tilde{z};\varepsilon\,|\,_1\tilde{G}^{201}]$ using change of variable (3.10), instead of (3.10*), then energies of the AEH solutions

$$\phi_{\tilde{t}_\pm, m_{\tilde{t}_\pm}}[\tilde{z} \mid {}_1\tilde{G}^{201}_{\downarrow \tilde{t}_\pm, m_{\tilde{t}_\pm}}] = |\tilde{z}|^{(1\pm\lambda_o)/2}(1-\tilde{z})^{(1+o_{\tilde{t}_\pm}\mu_o)/2} \times P^{(o_{\tilde{t}_\pm}\mu_o, \pm\lambda_o)}_{m_{\tilde{t}_\pm}}(2\tilde{z}-1)/k_m(\pm\lambda_o, o_{\tilde{t}_\pm}\mu_o) \quad (4.2.1)$$

are described by the same formulas

$$ {}_1d\,\varepsilon_{\tilde{t}_\pm, m_{\tilde{t}_\pm}} = (\pm\lambda_o + o_{\tilde{t}_\pm}\mu_o + 2m_{\tilde{t}_\pm} + 1)^2 \quad (4.2.2)$$

as those used for the h-PT potential expressed in terms of variable (3.10*), with the only difference that the labeling is done by means of a new indexes $o_{\tilde{t}} = \pm$, instead of $\sigma_{1;t}$. To be more precise, comparing the asymptotic behavior of the right- and left-hand sides of the AEH solution

$$\sqrt{\left|\frac{dz}{d\tilde{z}}\right|}\,\phi_{\tilde{t}_\pm, m_{\tilde{t}_\pm}}[\tilde{z} \mid {}_1\tilde{G}^{201}_{\downarrow \tilde{t}_\pm, m_{\tilde{t}_\pm}}] = z^{(1\pm\lambda_o)/2}(1-z)^{(1+\lambda_{1;\tilde{t}_\pm, m_{\tilde{t}_\pm}})/2} \times P^{(o_{\tilde{t}_\pm}\mu_o, \pm\lambda_o)}_{m_{\tilde{t}_\pm}}(2\tilde{z}-1)/P^{(o_{\tilde{t}_\pm}\mu_o, \pm\lambda_o)}_{m_{\tilde{t}_\pm}}(1) \quad (4.2.3)$$

in the limit $\tilde{z} \to -\infty$ ($z \to 1$) one finds generic form:

$$\lambda_{1;\tilde{t}_\pm, m_{\tilde{t}_\pm}} = \mp\lambda_o - o_{\tilde{t}_\pm}\mu_o - 2m_{\tilde{t}_\pm} - 1. \quad (4.2.4)$$

While the boundary condition at the origin remains unchanged regardless of the choice of the variable $\tilde{z}$ or z, square integrability of a regular solution at -∞ is determined by the requirement that

$$0 \le m_{\tilde{t}_\pm} < -\tfrac{1}{2}(o_{\tilde{t}_\pm}\mu_o \pm \lambda_o + 1) \quad (4.2.5)$$

which depends both on the ray identifiers $\lambda_o$ and $\mu_o$ and on the polynomial order m. There are AEH solutions of 6 types which we label as

$$\tilde{t}_+ = \tilde{a} \quad \text{if} \quad o_{\tilde{t}_+} = +; \quad (4.2.6a)$$

$$\tilde{t}_+ = \tilde{a}' \quad \text{if} \quad o_{\tilde{t}_+} = - \text{ and } 2m_{\tilde{t}_+} + 1 > \mu_o - \lambda_o = \lambda_{o;\tilde{t}_+} - \mu_{o;\tilde{t}_+} + 2; \quad (4.2.6a')$$

$$\tilde{t}_- = \tilde{b} \quad \text{if } o_{\tilde{t}_-} = - \text{ and } 2m_{\tilde{t}_-} + 1 < \lambda_o + \mu_o = \lambda_{o;\tilde{t}_-} + \mu_{o;\tilde{t}_-}\,; \tag{4.2.6b}$$

$$\tilde{t}_- = \tilde{b}' \quad \text{if } o_{\tilde{t}_-} = + \text{ and } 1 < 2m_{\tilde{t}_-} + 1 < \lambda_o - \mu_o = \lambda_{o;\tilde{t}_-} - \mu_{o;\tilde{t}_-} + 2; \tag{4.2.6b$'$}$$

$$\tilde{t}_+ = \tilde{c} \quad \text{if } o_{\tilde{t}_+} = - \text{ and } 1 < 2m_{\tilde{t}_+} + 1 < \mu_o - \lambda_o = \mu_{o;\tilde{t}_+} - \lambda_{o;\tilde{t}_+} + 2; \tag{4.2.6c}$$

$$\tilde{t}_- = \tilde{d} \quad \text{if } o_{\tilde{t}_-} = - \text{ and } 2m_{\tilde{t}_-} + 1 > \lambda_o + \mu_o = \lambda_{o;\tilde{t}_-} + \mu_{o;\tilde{t}_-} + 2\,; \tag{4.2.6d}$$

$$\tilde{t}_- = \tilde{d}' \quad \text{if } o_{\tilde{t}_-} = + \text{ and } 2m_{\tilde{t}_-} + 1 > \lambda_o - \mu_o = \lambda_{o;\tilde{t}_-} - \mu_{o;\tilde{t}_-} + 2\,. \tag{4.2.6d$'$}$$

It directly follows from (4.2.6b′) that there is no AEH solution of type $\tilde{b}'$ co-existent with the discrete energy spectrum so that we will drop solutions of this type from further consideration.

Let us consider the Bagchi-Quesne-Roychoudhury (BQR) potential [108] as an illustration. Keeping in mind that the indexes $\sigma_{0;\tilde{t}}$ and $o_{\tilde{t}}$ have the same sign for both principal regular solutions $\tilde{a}$ and $\tilde{b}$, we represent the constants A≡A– and B≡B– used to parameterize potential (9) in [108] as

$$A_- \equiv \tfrac{1}{2}(\mu_{o;\tilde{t}} - \lambda_{o;\tilde{t}} - 1)\,, \tag{4.2.7a}$$

$$B_- \equiv \tfrac{1}{2}(\mu_{o;\tilde{t}} + \lambda_{o;\tilde{t}})\,. \tag{4.2.7b}$$

For $\tilde{t} = \tilde{a}$ or $\tilde{b}$

$$A_- = \tfrac{1}{2}(\mu_o - \lambda_o - 1)\,, \tag{4.2.7a*}$$

$$B_- = B'_- + \sigma_{0;\tilde{t}} \tag{4.2.7b*}$$

and the appropriate factorization energies depend only on the parameter $B_-$, which is also true for the AEH solution of type $\tilde{d}$. There are also P-CEQ potentials constructed using FFs types $\tilde{a}'$ and $\tilde{d}'$ with factorization energies dependent only on the parameter $A_-$. By setting ${}_1d$ set to –4 in (3.32) one can represent the CES potential in question as

$$V_{BQR}[\tilde{z};\lambda,\mu] = V[\tilde{z};\lambda,|\mu|] - \frac{2\tilde{z}_1(\lambda,\mu)[1-\tilde{z}_1(\lambda,\mu)]}{[\tilde{z}-\tilde{z}_1(\lambda,\mu)]^2} - \frac{\tilde{\eta}_1(\lambda,\mu)}{\tilde{z}-\tilde{z}_1(\lambda,\mu)} \tag{4.2.8}$$

for $0 < \lambda \neq 1,\ -\lambda < \mu \neq 0$

where

$$\tilde{z}_1(\lambda,\mu) = \frac{\lambda}{\lambda+\mu} > 0 \qquad (4.2.9)$$

and

$$\tilde{\eta}_1(\lambda,\mu) \equiv 1 - 2\tilde{z}_1(\lambda,\mu) < 1. \qquad (4.2.9')$$

Note that the potential has two distinct ('primary' and 'secondary') branches: $\mu > 0$ and $-\lambda < \mu < 0$. The secondary branch may have no more than one bound state, if any. At $\mu \to 0$ the singular points $\tilde{z}=1$ and $\tilde{z}=\tilde{z}_1(\lambda,\mu)$ merge and the BQR potential collapses into the DPT potential. By substituting the Jacobi polynomial in the right-hand side of (4.2.1) for $\tilde{z}$-1 and setting $\mu_o$ to 1 one can verify that

$$\lambda_{1;\tilde{t}_\pm,1} = -(\pm\lambda_o + 3 + o_{\tilde{t}_\pm}) = -(\pm\lambda_o + 1 - o_{\tilde{t}_\pm}) \qquad \text{with } o_{\tilde{t}_\pm} = -1 \qquad (4.2.10)$$

and

$$\sigma_{0;\tilde{t}_\pm} = \pm, \quad \sigma_{1;\tilde{t}_\pm} = -\ \sigma_{1;t_\pm} = -. \qquad (4.2.10^*)$$

By making change of variable

$$1 - 2\tilde{z}(x) = \tilde{\eta}(x) = \operatorname{ch} x \qquad (4.2.11)$$

one can verify that the primary branch does match potential (9) in [108]. The region $\lambda > 1,\ \mu > 0$ can be reached via the DT of the h-PT potential with the FF of type $\tilde{a}$ which implies that the BQR potential within this parameter range has the discrete energy spectrum:

$$\varepsilon_{c,v} = -(A_- - v)^2 \quad \text{for } v = 0, 1, ..., v_{max} \ (A_- - 1 \le v_{max} < A_-)\,. \qquad (4.2.12)$$

The same formula is applicable to the region $\lambda > 0,\ \mu > \lambda + 1$ which can be reached via the DT of the h-PT potential with the FF of type $\tilde{b}$.

The secondary branch has a single bound energy level at the energy

$$\varepsilon_{\tilde{d}',1} = -\tfrac{1}{4}(\mu_o - \lambda_o + 3)^2 \quad \text{for } -\lambda < \mu < 1 - \lambda < 0 \qquad (4.2.13)$$

and no discrete energy spectrum if $\mu$ is between $1 - \lambda < 0$ and 0.

Note also that

$$B_{-} \pm \tfrac{1}{2} - A_{-} - \tfrac{1}{2} = \lambda_{o;\tilde{t}} \pm \tfrac{1}{2} = \lambda_o \mp \tfrac{1}{2} = s_o \mp \tfrac{1}{2} - \tfrac{1}{2}, \qquad (4.2.14)$$

where the upper and lower signs correspond to solutions of type $\tilde{b}$ regular at $\infty$ ($\tilde{t} = \tilde{b}$) and at 0 ($\tilde{t} = \tilde{a}$), accordingly. Substituting (4.2.10) into FF (2.25) with m set to 1 and taking into account that the appropriate wavefunctions are related to (2.25) via (B6) in Appendix B below, one can thus directly verify that our expressions for FFs of DTs agree with the formula derived in [108] based on the corresponding superpotentials.

Coming back to the general case $m_{\tilde{t}} \geq 1$, we can write down the following necessary and sufficient conditions for existence of nodeless regular AEH solutions:

$$\varepsilon_{\tilde{a},m_{\tilde{a}}} - \varepsilon_{\tilde{c};0} = -4(\lambda_o + m_{\tilde{a}} + 1)(\mu_o + m_{\tilde{a}}) < 0 \text{ for any } m_{\tilde{a}} \geq 0, \qquad (4.2.15a)$$

$$\varepsilon_{\tilde{a}',m_{\tilde{a}'}} - \varepsilon_{\tilde{c};0} = -4m_{\tilde{a}'}(\lambda_o - \mu_o + m_{\tilde{a}'} + 1) < 0, \qquad (4.2.15a')$$

$$\text{for } 1 < \mu_o - \lambda_o - 1 < m_{\tilde{t}_+}$$

$$\varepsilon_{\tilde{b},m_{\tilde{b}}} - \varepsilon_{\tilde{c},0} = -4(m_{\tilde{b}} - \mu_o + 1)(m_{\tilde{b}} - \lambda_o) < 0$$

$$\text{for } 0 \leq m_{\tilde{b}} < \lambda_o < \tfrac{1}{2}(\lambda_o + \mu_o - 1) < \mu_o - 1. \qquad (4.2.15b)$$

Similarly, nodeless AEH solutions irregular at both end points may exist only if either

$$\varepsilon_{\tilde{d},m_{\tilde{d}}} - \varepsilon_{\tilde{c},0} = -4(\mu_o - m_{\tilde{d}} - 1)(\lambda_o - m_{\tilde{d}}) < 0 \quad \text{for } m_{\tilde{d}} > \mu_o - 1 > \tfrac{1}{2}(\lambda_o + \mu_o - 1) > \lambda_o \qquad (4.2.15d)$$

or

$$\varepsilon_{\tilde{d}',m_{\tilde{d}'}} - \varepsilon_0 = -4(m_{\tilde{d}'} + 1)(\mu_o + \lambda_o - m_{\tilde{d}'}) < 0 \text{ for } m_{\tilde{d}'} > \lambda_o. \qquad (4.2.15d')$$

As expected, the listed conditions match the constrained imposed above on parameters of the BQR potential.

Eigenfunctions of bound states at the energies

$$\varepsilon_v = -(\mu_o - \lambda_o - 2v - 1)^2 \text{ for } 1 \leq 2v + 1 < \mu_o - \lambda_o. \qquad (4.2.16)$$

(v=0, 1, ..., $n_c$ -1) are described by $n_c$ Romanovsky-Jacobi polynomials

$$P_v^{(-\mu_o,\lambda_o)}(2\xi-1)=\frac{(\lambda_o-\mu_o+2v)_v}{v!}\Pi_v[\xi;\bar{\xi}_{c,v}] \qquad (4.2.17)$$

mutually orthogonal on the negative semi-axis with the weight

$$w[\xi<0;\lambda_o,\mu_o]=|\xi|^{\lambda_o}(1-\xi)^{-\mu_o}. \qquad (4.2.18)$$

Substituting (4.2.4) into (C5) in Appendix C shows that the constraint

$$u_2(\lambda_o,\lambda_{1;\tilde{t}_\pm,m_{\tilde{t}_\pm}};m_{\tilde{t}_\pm})-1=\tfrac{1}{2}(\mu_o-\lambda_o-1-|\lambda_{1;\tilde{t}_\pm,m_{\tilde{t}_\pm}}|)\le 0 \qquad (4.2.19)$$

is equivalent to the requirement that the appropriate AEH solution lies below the ground energy level. If $\alpha$ and $\beta$ in (C2) are both positive ($\alpha=\mu_o$, $\beta=\lambda_o$) then the second constraint automatically holds in agreement with our conclusion that all AEH solutions of type $\tilde{a}$ must be nodeless. Similarly, conditions (4.2.15a′) and (4.2.15b) indicate that both raising factorials in the second constraint in (C2) are composed of either all positive ($\tilde{t}=\tilde{a}'$) or all negative ($\tilde{t}=\tilde{b}$) numbers, respectively, which confirms that AEH solutions regular either at zero or $-\infty$ do not have nodes on the negative semi-axis as far as they lie below the ground energy level. Contrary to the DPT potential, we were unable to formulate any simple rule to select nodeless solutions irregular at both ends.

As an illustration, let us consider the quadratic case similar to that discussed by Quesne [13] for the DPT potential. Two zeros of the appropriate second-order Jacobi polynomial are given by the quadratic equation:

$$(\lambda_{\tilde{t}}+1)_2-2(\lambda_{\tilde{t}}+o_{\tilde{t}}\mu_o+3)(\lambda_{\tilde{t}}+2)z_{\tilde{t};k}+(\lambda_{\tilde{t}}+o_{\tilde{t}}\mu_o+3)_2 z_{\tilde{t};k}^2=0 \qquad (4.2.20)$$

with discriminant

$$\Delta_{\tilde{t}}=4(\lambda_{\tilde{t}}+2)(o_{\tilde{t}}\mu_o+2)(\lambda_{\tilde{t}}+o_{\tilde{t}}\mu_o+3). \qquad (4.2.21)$$

In the simplest case $\sigma_{o;\tilde{t}}=o_{\tilde{t}}=+1$ the equation has two positive roots for any positive values of the parameters $\lambda_o$ and $\mu_o$ so that there is always an AEH solution of type $\tilde{a}$. An AEH solution of type $\tilde{a}'$ is associated with two complex roots appearing if $2 < \mu_o < \lambda_o + 3$. (The condition $\mu_o > 2$ is necessary for the existence of the discrete energy spectrum.)

An analysis of AEH solutions irregular at the origin can be simplified by noticing that the second constraint in (C2) is equivalent to the requirement that the leading coefficient and free term have the same sign. Since discriminant (4.2.21) and the linear coefficient have opposite sign for $o_{\tilde{t}} = +1$, the quadratic equation has either a pair of complex-conjugated or positive roots for $\tilde{t} = \tilde{d}'$ (assuming that both leading coefficient and free term remain positive). The quadratic equation has negative discriminant for $\lambda_o > 2$ and either $\mu_o > 2$, $\lambda_o + \mu_o > 5$ if $\tilde{t} = \tilde{b}$ or $4 < \lambda_o + \mu_o < 5$ $(2 < \mu_o < 3)$ if $\tilde{t} = \tilde{d}$.

To summarize, REs of the h-PT potential can be constructed by means of DTs with FFs of 5 different types $\tilde{t}$. Each RE is quantized via a finite subset of orthogonal polynomials which belong to one of 16 infinite sequences of Heine polynomials defined by (4.1.20a)-(4.1.20d). The potentials constructed by means of DTs with regular FFs ($\tilde{t} = \tilde{a}, \tilde{a}'$, and $\tilde{b}$) form a family of isospectral potentials conditionally exactly quantized via 'abnormal' orthogonal Heine polynomials from series a, b, and c. Odake and Sasaki [20] discovered the regular sequence of type $\tilde{b}$ generated using the same forward and backward shift operators as the J2 case [87]. Similarly, the regular sequence of type $\tilde{a}$ is generated using the same forward and backward shift operators as Case J1. Contrary to two other regular sequences of orthogonal Heine polynomials, the sequence of type $\tilde{a}'$ ($m_t > 1$) starts from the polynomial of order $m_t - 1$ (not $m_t$).

For $m_t > 1$ there are also two sparse finite sequences of orthogonal Heine polynomials: $\tilde{t} = \tilde{d}$ and $\tilde{d}'$ which belong to series b and d, respectively. In particular, this implies that the regular parts of these sequences start from the polynomial of order $m_t$ and $m_t + 1$, accordingly. Again all the CD transforms of Romanovasky-Jacobi polynomials from these sequences form a subset orthogonal to a constant when integrated over the negative semi-axis with the density function ${}_1\wp[\xi]$ defined via (3.1*).

### 4.3 The Isotonic Oscillator

An analysis of PES equation (2.21) is slightly complicated by the fact that one has to reflect the argument of the GL polynomial for AEH solutions of types **a** and **d**. In fact, by making the change of variable

$$\varsigma_{\pm} = \pm\nu_o\xi \equiv \pm\nu_o\zeta \tag{4.3.1}$$

in (2.21) one comes to the confluent hypergeometric equation

$$\left\{\varsigma_{\pm}\frac{d^2}{d\varsigma_{\pm}^2} + [\sigma_0\lambda_o + 1 - \varsigma_{\pm}]\frac{d}{d\varsigma_{\pm}} - \alpha(\varepsilon;\sigma_0,\pm)\right\} F[\alpha(\varepsilon;\sigma_0,\pm);\sigma_0\lambda_o + 1;\varsigma_{\pm}] = 0 \tag{4.3.2}$$

where

$$\alpha(\varepsilon;\bar{\sigma}) = -\tfrac{1}{2}(\sigma_0\lambda_o + 1 - \tfrac{1}{2}\sigma_1\,{}_0d\,\varepsilon/\nu_o)\,. \tag{4.3.3}$$

The specific feature of all *c*-GRef potentials that the appropriate PES equation like (4.3.2) has to be solved on the negative semi-axis for solutions of types **a** and **d**.

To our knowledge, it was Junker and Roy [8, 118] who first suggested to use confluent hypergeometric functions nodeless on the negative semi-axis as FFs for DTs of both Morse and isotonic oscillators.

The sought-for energies of AEH solutions are thus given by the simple formula

$${}_0d\,\varepsilon_{\dagger,m_\dagger}/\nu_o \equiv \tilde{\varepsilon}_{\dagger,m_\dagger} = 2\sigma_{1;\dagger}(\lambda_\dagger + 2m_\dagger + 1) \tag{4.3.4}$$

or, using scaled energies (3.5),

$$\tilde{\varepsilon}_{\dagger,m_\dagger} = 2\sigma_{1;\dagger}(\sigma_{0;\dagger}\lambda_{o;\dagger} + 2m_\dagger) \text{ for } \lambda_o > 1 \text{ or } \sigma_{0;\dagger} = +1. \tag{4.3.4*}$$

In case of FFs regular at one of the end points one can re-write (4.3.4*) as

$$\tilde{\varepsilon}_{\dagger,m_\dagger} = -2(\lambda_{o;\dagger} - 2\sigma_{1;\dagger}m_\dagger) \text{ for } \dagger = \mathbf{b}\ (\lambda_o > 1) \text{ or } \mathbf{a}\,. \tag{4.3.5}$$

In Quesne's linear case the derived formula for scaled factorization energies directly turns into (2.11) in [13], with parameters ${}_0$d and $\lambda_{o;\dagger}$ given by (3.41) above. Similarly, the scaled factorization energy for AEH solutions irregular at both end points can be represented as

$$\tilde{\varepsilon}_{\mathsf{d},m_{\mathsf{d}}} = 2(\lambda_{o;\mathsf{d}} - 2m_{\mathsf{d}}) \quad \text{for } \lambda_o > 1. \tag{4.3.5*}$$

Energy shifts relative to the ground energy can be thus represented as

$${}_0d({}_0\tilde{\varepsilon}_{\mathsf{a},m_{\mathsf{a}}} - {}_0\tilde{\varepsilon}_{\mathsf{c},0}) = -2(m_{\mathsf{a}} + 1 + \lambda_o), \tag{4.3.6a}$$

$${}_0d({}_0\tilde{\varepsilon}_{\mathsf{b},m_{\mathsf{b}}} - {}_0\tilde{\varepsilon}_{\mathsf{c},0}) = 2(m_{\mathsf{b}} - \lambda_o) \quad \text{for } \lambda_o > 1, \tag{4.3.6b}$$

$${}_0d({}_0\tilde{\varepsilon}_{\mathsf{d},m_{\mathsf{d}}} - {}_0\tilde{\varepsilon}_{\mathsf{c},0}) = -2(m_{\mathsf{d}} + 1) \quad \text{for } \lambda_o > 1. \tag{4.3.6*}$$

In agreement with Odake and Sasaki's observation [23] that all the coefficients of the polynomial $L_m^{(\lambda_0)}(-\varsigma)$ in $\varsigma$ are all positive for $\lambda_0 > 0$, there is an infinite number of AEH solutions of type $\mathsf{a}$ as prescribed [31] by the KLH theorem. Note also that, according to (4.3.11b), the potential $V[\zeta \,|\, {}^{1}_{0}\mathcal{G}^{101}_{\mathsf{b},m_{\mathsf{b}}}]$ disappears in the limit $\lambda_o \to m_{\mathsf{b}}$ (similarly to the case $\iota = 1$).

We can now directly relate our results to Levai and Roy's pioneering work [7] using Sukumar's notation [68] for solution types:

$$T_1 = \mathsf{c},\ T_2 = \mathsf{d},\ T_3 = \mathsf{a},\ T_4 = \mathsf{b} \tag{4.3.7}$$

.Taking into account that $\gamma = s_o - 1 = \lambda_o - \frac{1}{2}$ in our terms, one can easily verify that the derived formulas for scaled energy shifts match those in [7] with N standing for $m_{\dagger}$ here.

It should be stressed that Levai and Roy [7] could not properly treat DTs with FFs irregular at both ends because Junker and Roy [8] constructed nodeless solutions of these type as superpositions of two positive regular solutions $\psi_{\mathsf{a}}(x;\varepsilon)$ and $\psi_{\mathsf{b}}(x;\varepsilon)$ [74, 75, 119]:

$$\psi_{\mathsf{d}}(x;\varepsilon;\sigma) \equiv \psi_{\mathsf{a}}(x;\varepsilon) + \sigma\psi_{\mathsf{b}}(x;\varepsilon), \tag{4.3.8}$$

with a positive coefficient $\sigma > 0$ at energies $\varepsilon$ below the ground level. (Approximately at the same time with Deift and Trubowitz' works [74, 75] the author [119] suggested FFs of such a type for Darboux tranforms of CB potentials based on the analysis of zeros of the appropriate Jost functions.) It is not an easy task to find nodeless AEH solutions irregular on both end which can be represented in form of (4.3.8).

A conceptually new approach was more recently suggested by Gomez-Ullate, Kamran and Milson [25] who constructed nodeless FFs irregular at both ends of the positive semi-axis by analyzing zeros of GL polynomials $L_m^{(\lambda)}(-\varsigma)$ in the reflected argument $-\varsigma$ ($\varsigma_+ \equiv \varsigma$), with the Morse oscillator as an example. (In this connection we should again mention Grandati's concise formulation [31, 32] of the KLH theorem [27-30], which helped to rectify Gomez-Ullate, Kamran, and Milson's results [25].) Though the new technique gives equivalent results for regular AEH solutions below the ground energy level (as it was demonstrated in subsection 4.1 for the DPT potential) it allows one to construct REs of shape-invariant potentials with inserted ground states. It is worth mentioning in this connection that the potential constructed in such a way are not shape-invariant like all broken-symmetry SUSY partners constructed by Junker and Roy [8, 118].

One of the most important consequences of the KLH theorem for the isotonic oscillator is that any AEH solution regular at the origin (type **a**) is nodeless and therefore can be used to construct a rational CES potential. Note that Levai and Roy [7] have already come to this conclusion from a different direction by proving that all AEH solutions regular at the origin lie below the ground energy level and therefore do not have nodes on the positive semi-axis.

As an illustration let us discuss Quesne's quadratic case in more detail. First note that zeros of the second-order GL polynomial [121]

$$L_2^{(\sigma_{0;\dagger}\lambda_o)}[\sigma_{1;\dagger}\varsigma] = \tfrac{1}{2}\varsigma^2 - (\sigma_{0;\dagger}\lambda_o + 2)\sigma_{1;\dagger}\varsigma + \tfrac{1}{2}(\sigma_{0;\dagger}\lambda_o + 1)_2 \qquad (4.3.9)$$

are given by roots of the quadratic equation

$$\varsigma_{\dagger;1}^{2} - 2(\sigma_{0;\dagger}\lambda_o + 2)\sigma_{1;\dagger}\varsigma_{\dagger;1} + (\sigma_{0;\dagger}\lambda_o + 1)_2 = 0 \qquad (4.3.10)$$

with discriminant

$$\Delta_\dagger = 4(\sigma_{0;\dagger}\lambda_o + 2)\,. \qquad (4.3.11)$$

which is equal to

$$\Delta_\dagger = 4(\sigma_{0;\dagger}\lambda_{o;\dagger} + 1)\,. \qquad (4.3.11^*)$$

in the region of our interest: $\lambda_{o;\dagger} = \lambda_o + \sigma_{0;\dagger} > 0$. Under the latter condition quadratic equation (4.3.10) has a pair of complex-conjugated roots iff $\dagger$ = b or d and $\lambda_o > 2$. Otherwise both real roots are either negative for $\dagger$ = a or positive for bound energy states ($\dagger$ = c). (For $1 < \lambda_o < 2$ the GL polynomial associated with the AEH solution of type b or d has zeros of opposite sign since the appropriate factorization energy lies between the ground and first-excited energy levels.) One can verify that the energy of the AEH solution of type a given by (4.3.5) agrees with the energy of eigenfunction (2.16) in [13]. Similarly, energies of AEH solutions of type $\dagger$ = b and d described by (2.18) in [13] match (4.3.5) and (4.3.5*), respectively.

As mentioned in Section 2, all 16 infinite sets of GS Heine polynomials associated with Bose invariant (3.2′) can be obtained by using universal *g*- and *a*-HPGs defined via (2.57) and (2.61*).

$$\nu_o^{m_\dagger}\,{}_0\hat{g}(2\lambda_{\dagger'}, \Delta\sigma_{1;\dagger'\dagger}\,\nu_o; \bar{\zeta}_{\dagger,m_\dagger}) \equiv \varsigma\{\Pi_{m_\dagger}[\varsigma;\bar{\varsigma}_{\dagger,m_\dagger}]\frac{d}{d\varsigma} - \dot{\Pi}_{m_\dagger}[\varsigma;\bar{\varsigma}_{\dagger,m_\dagger}]\}$$
$$+ \tfrac{1}{2}(\Delta\sigma_{0;\dagger'\dagger}\lambda_o - \Delta\sigma_{1;\dagger'\dagger}\,\varsigma)\Pi_{m_\dagger}[\varsigma;\bar{\varsigma}_{\dagger,m_\dagger}] \quad \text{for } \sigma_{0;\dagger} = -\sigma_{0;\dagger'} \qquad (4.3.12)$$

and

$$\nu_o^{m_\dagger - 1}\,{}_0\hat{a}(\Delta\nu_{\dagger',m';\dagger,m_\dagger}; \bar{\zeta}_{\dagger,m_\dagger}) = \Pi_{m_\dagger}[\varsigma;{}_0\bar{\varsigma}_{\dagger,m_\dagger}]\frac{d}{d\varsigma} - \dot{\Pi}_{m_\dagger}[\varsigma;\bar{\varsigma}_{\dagger,m_\dagger}]$$
$$+ \tfrac{1}{2}\Delta\sigma_{1;\dagger'\dagger}\,\Pi_{m_\dagger}[\varsigma;\bar{\varsigma}_{\dagger,m_\dagger}] \quad \text{for } \sigma_{0;\dagger} = \sigma_{0;\dagger'}\,. \qquad (4.3.12^*)$$

Here dot denotes the derivative with respect to $\varsigma$,

$$\Delta\sigma_{r;\dagger\dagger'} \equiv \sigma_{r;\dagger} - \sigma_{r\dagger'}\,, \qquad (4.3.13)$$

and $\bar{\varsigma}_{\dagger,m_\dagger}$ is a full set of zeros of the polynomial $L_{m_\dagger}^{(\lambda_\dagger)}(o_{1;\dagger}\varsigma)$ expressed in terms of $\varsigma$. As a result of the form-invariance of the given RSL equation under basic CDTs, the GS Heine polynomials turn into the GL polynomials for $m_\dagger = 0$. Indeed, applying the HPGs

$${}_0\hat{g}(2\lambda_{\dagger'}, \Delta\sigma_{1;\dagger'\dagger}\, \nu_o; \infty) = \varsigma\frac{d}{d\varsigma} + \lambda_{\dagger'} - \tfrac{1}{2}\Delta\sigma_{1;\dagger'\dagger}\varsigma \quad \text{for } \sigma_{0;\dagger} = -\sigma_{0;\dagger'} \qquad (4.3.14)$$

and

$$\nu_o^{-1}\, {}_0\hat{a}(\Delta\sigma_{1;\dagger'\dagger}\nu_o; \bar{\zeta}_{\dagger,m_\dagger}) = \frac{d}{d\varsigma} + \tfrac{1}{2}\Delta\sigma_{1;\dagger'\dagger} \qquad \text{for } \sigma_{0;\dagger} = \sigma_{0;\dagger'}. \qquad (4.3.14^*)$$

to the LG polynomial $L_{m_{\dagger'}}^{(\lambda_{\dagger'})}(o_{1;\dagger'}\varsigma)$ and putting $o_{1;\dagger'}\varsigma = x$ one comes to the conventional formulas for GL polynomials in the conveniently compiled list in [87]. Namely, use of (4.3.14) gives the backward shift relation (E.3) in [87] for $\sigma_{1;\dagger} = -\sigma_{1;\dagger'}$ and the combination of (E.2) and (E.12),

$$x\dot{L}_n^{(\alpha)}(x) + \alpha L_n^{(\alpha)}(x) = nL_{n+1}^{(\alpha-1)}(x), \qquad (4.3.15)$$

for $\sigma_{1;\dagger} = \sigma_{1;\dagger'}$. Similarly, use of (4.3.11*) leads to (E.2) for $\sigma_{1;\dagger} = \sigma_{1;\dagger'}$ and the combination of (E.2) and (E.11),

$$\dot{L}_n^{(\alpha)}(x) - L_n^{(\alpha)}(x) = -L_n^{(\alpha+1)}(x), \qquad (4.3.16)$$

for $\sigma_{1;\dagger} = -\sigma_{1;\dagger'}$.

One can derive the formulas for the energy differences directly from (4.1.11a), (4.1.11b), and (4.1.17) by formally changing $\mu_o$ for $\mu_o = \nu_o / \iota$ and making the limiting transition:

$$\lim_{\iota\to 0}\left[({}_1\varepsilon_{\dagger,m_\dagger} - {}_1\varepsilon_{c,0})/\iota\right] = {}_0\varepsilon_{t,m_\dagger} - {}_0\varepsilon_{c,0}. \qquad (4.3.17)$$

The outlined procedure is nothing but the particular case of the general transition to the confluent hypergeometric equation performed in [12] for the generic *r*-GRef potential. Namely, we just start from the reference PF

$$ {}_1I^{o}[\iota\xi;\lambda_o,\nu_{o\iota}] = \frac{1-\lambda_o^2}{4\iota^2\xi^2} + \frac{1-\nu_{o\iota}^2}{4(1-\iota\xi)^2} \tag{4.3.18} $$

and then make limiting transition

$$ \lim_{\iota\to 0}\left\{\iota^2\, {}_1I^{o}[\iota\xi;\lambda_o,\nu_{o\iota}]\right\} = {}_0I^{o}[\xi;\lambda_o,\nu_o], \tag{4.3.19} $$

where we put

$$ \lim_{\iota\to 0}[\iota^2(\nu_{o\iota}^2-1)] = \nu_o^2 - 1. \tag{4.3.20} $$

Some details of the transition from $X_m$-Jacobi polynomials to their confluent counter-parts can be found in Odake and Sasaki's paper [22].

It is essential that the number of $\mathfrak{a}$-sequences of $X_m$-Jacobi polynomials increases with increase of $\nu_{o1}$ and becomes infinite in the confluent limit. On other hand, the number of $\mathfrak{b}$-sequences of exceptional orthogonal polynomials remains finite at any fixed value of the exponent difference $\lambda_o$. Our results are fully consistent with the LKH theorem that states that the number of GL polynomials with negative index $-\lambda_o$ is equal to $[\lambda_o]$. As demonstrated below, the sets of orthogonal *c*-Heine polynomials constructed by means of CDTs with FFs of type $\mathfrak{b}$ match (up to some scale factors) the *finite* sequence of Case-II exceptional Laguerre polynomials rigorously introduced in [24, 35]. Note that whereas the range of conditionally exact solvability for Case-II REs of the isotonic oscillator monotonically narrows with increase of $m_{\mathfrak{b}}$, contrary to Cases I and II.

Up to some scale factors, $X_m$-Laguerre polynomials are represented by the following sets of orthogonal Heine polynomials:

$$ Hi^{(m+1)}_{m+v}[\varsigma\,|\,{}^1_0\mathcal{G}^{101}_{\mathfrak{a},m};c,v] = \nu_o^{m-1}\, {}_0\hat{a}(2\nu_o;\bar{\xi}_{\mathfrak{a},m})\, L_v^{(\lambda_o)}(\varsigma), \tag{4.3.21a} $$

$$ Hi^{(m+1)}_{m+v}[\varsigma\,|\,{}^1_0\mathcal{G}^{101}_{\mathfrak{b},m};c,v] = \nu_o^{m}\, {}_0\hat{g}(2\lambda_o,0;\bar{\xi}_{\mathfrak{b},m})\, L_v^{(\lambda_o)}(\varsigma). \tag{4.3.21b} $$

There is also a sequence of orthogonal Heine polynomials

$$\mathrm{Hi}^{(2j+1)}_{2j+v}[\varsigma\,|\,{}^{1}_{0}\mathcal{G}^{101}_{\mathsf{d},2j};\mathsf{c},v] = \nu_o^m\ {}_0\hat{g}(-2\lambda_o,-2\nu_o;\bar{\xi}_{\mathsf{d},2j})\ L^{(\lambda_o)}_v(\varsigma) \qquad (4.3.21^*)$$

forming a subset of Grandati's [32] L3 series $(2j<\lambda_o)$ which exists for any value of $\lambda_o>1$. Each polynomial in this sequence satisfies the orthogonality relation

$$\int_0^\infty \frac{\zeta^{\lambda_o}\exp(-\varsigma)}{\Pi^2_{2j}[\varsigma;\bar{\varsigma}_{\mathsf{d},2j}]}\mathrm{Hi}^{(2j+1)}_{2j+v}[\varsigma\,|\,{}^{1}_{0}\mathcal{G}^{101}_{\mathsf{d},2j};\mathsf{c},v]\,d\varsigma = 0 \qquad (4.3.22)$$

similar to (4.1.46).

Coming back to the X-OPSs, note that, according to (7) in [25],

$$\Pi_m[\varsigma;\bar{\varsigma}_{\mathsf{t},m}] = (-\sigma_{1;\mathsf{t}})^m m!\,L^{(\lambda_{\mathsf{t}})}_m[\sigma_{1;\mathsf{t}}\,\varsigma]. \qquad (4.3.23)$$

Substituting explicit expressions (4.3.8*) and (4.3.8) for the HPGs into the right-hand side of (4.3.21a) and (4.3.21b) and excluding the first derivatives of the GL polynomials via the conventional 'forward shift relation' in terms of [87] one finds

$$\mathrm{Hi}^{(m+1)}_{m+v}[\varsigma\,|\,{}^{1}_{0}\mathcal{G}^{(0)}_{\mathsf{a},m};\mathsf{c},v] \qquad (4.3.24a)$$
$$= (-1)^{m+1}.m!\,\{[\,L^{(\lambda_o+1)}_{m-1}(-\varsigma)+L^{(\lambda_o)}_m(-\varsigma)]\,L^{(\lambda_o)}_v(\varsigma)+L^{(\lambda_o)}_m(-\varsigma)L^{(\lambda_o+1)}_{v-1}(\varsigma)\,\}$$

and

$$\mathrm{Hi}^{(m+1)}_{m+v}[\varsigma\,|\,{}^{1}_{0}\mathcal{G}^{101}_{\mathsf{b},m};\mathsf{c},v] = (-1)^m.m!\,\{[\varsigma L^{(-\lambda_o+1)}_{m-1}(\varsigma)+\lambda_o\,L^{(-\lambda_o)}_m(\varsigma)]\,L^{(\lambda_o)}_v(\varsigma)$$
$$-\varsigma\,L^{(-\lambda_o)}_m(\varsigma)L^{(\lambda_o+1)}_{v-1}(\varsigma)\}, \qquad (4.3.24b)$$

respectively. Representing (22.8.6) in [113] as

$$\varsigma L^{(\alpha+1)}_{m-1}(\varsigma)-\alpha L^{(\alpha)}_m(\varsigma) = (\alpha+m)[L^{(\alpha)}_{m-1}(\varsigma)-L^{(\alpha)}_m(\varsigma)] \qquad (4.3.25)$$

and coupling the latter formula with (22.7.30) there, one can directly match (4.3.24a) and (4.3.24b) to Gomez-Ullate, Kamran, and Milson's expressions in [42] for Case-I and Case-II exceptional Laguerre polynomials:

$$\mathrm{Hi}_{m+v}^{(m+1)}[\varsigma\,|\,{}^{1}\mathcal{G}^{(0)}_{\mathsf{a},m};\mathsf{c},v] = (-1)^{m+1}.m!\,L^{I(\lambda_o)}_{m+1,m+v+1}(\varsigma) \quad (4.3.26a)$$

and

$$\mathrm{Hi}_{m+v}^{(m+1)}[\eta\,|\,{}^{1}\mathcal{G}^{(0)}_{\mathsf{b},m};\mathsf{c},v] = (-1)^{m}.m!\,L^{II(\lambda_o)}_{m+1,m+v+1}(\eta). \quad (4.3.26b)$$

Multiplying Heine equation (2.57) for $\iota = 0$ by the scale factor $\nu_o^{m-1}$ and setting $\varepsilon$ to $\varepsilon_{\mathsf{t}',m'}$, we find that Heine polynomials of our interest satisfy the equation:

$$\{\hat{D}_{\mathsf{t},m;\mathsf{t}'} + \tilde{C}_m[\varsigma;\mathsf{t},m;\mathsf{t}',m']\}\mathrm{Hi}^{(m+1)}_{m+m'}[\varsigma\,|\,{}^{1}\mathcal{G}^{(0)}_{\mathsf{t},m};\mathsf{t}',m'] = 0 \quad (4.3.27)$$

where

$$\hat{D}_{\mathsf{t},m;\mathsf{t}'} \equiv \varsigma\Pi_m[\varsigma;\bar{\varsigma}_{\mathsf{t},m}]\frac{d^2}{d\varsigma^2} + 2\tilde{B}_m[\varsigma;\mathsf{t},m;\sigma_{\mathsf{t}'}]\frac{d}{d\varsigma}. \quad (4.3.28)$$

The polynomial coefficients appearing (4.3.27) and (4.3.28) are defined as follows

$$\tilde{B}_m[\varsigma;\mathsf{t},m;\bar{\sigma}_{\mathsf{t}'}] \equiv \nu_o^{m-1}\,{}_0B_{m+1}[\zeta/\nu_o;\sigma_{0;\mathsf{t}'}\,{}^{1}\lambda_{o;\mathsf{t}},\sigma_{1;\mathsf{t}'}\nu_o;\bar{\zeta}_{\mathsf{t},m}] \quad (4.3.29)$$

$$= (-1)^m m!\left\{\tfrac{1}{2}[-(1+\sigma_{0;\mathsf{t}'}\,{}^{1}\lambda_{o;\mathsf{t}}) + \sigma_{1;\mathsf{t}'}\varsigma]L_m^{(\sigma_{0;\mathsf{t}}\lambda_o)}(\sigma_{1;\mathsf{t}}\varsigma) - \sigma_{1;\mathsf{t}}\varsigma L_{m-1}^{(\sigma_{0;\mathsf{t}}\lambda_o+1)}(\sigma_{1;\mathsf{t}}\varsigma)\right\}$$

and

$$\tilde{C}_m[\varsigma;\mathsf{t},m;\mathsf{t},m'] \equiv \nu_o^{m-1}\,{}_0C^o_m[\varsigma;\varepsilon_{\mathsf{t}',m'}\,|\,{}^{1}\tilde{\mathcal{G}}^{101}_{\mathsf{t},m};++]$$

$$= \tfrac{1}{4}(-1)^m m!\{(\sigma_{1;\mathsf{t}} - \sigma_{1;\mathsf{t}'} + \sigma_{\mathsf{t}\mathsf{t}'}\,{}^{1}\lambda_{o;\mathsf{t}} + \tilde{\varepsilon}_{\mathsf{t}',m'} - \tilde{\varepsilon}_{\mathsf{t},m})L_m^{(\sigma_{0;\mathsf{t}}\lambda_o)}(\sigma_{1;\mathsf{t}}\varsigma)$$

$$-[(\sigma_{0;\mathsf{t}} - \sigma_{0;\mathsf{t}'})^{1}\lambda_{o;\mathsf{t}} - (\sigma_{1;\mathsf{t}} - \sigma_{1;\mathsf{t}'})\varsigma]L_{m-1}^{(\sigma_{0;\mathsf{t}}\lambda_o+1)}(\sigma_{1;\mathsf{t}}\varsigma)\} \quad (4.3.30)$$

where the parameter

$$\sigma_{\mathsf{t}\mathsf{t}'} \equiv \sigma_{0;\mathsf{t}}\sigma_{1;\mathsf{t}} - \sigma_{0;\mathsf{t}'}\sigma_{1;\mathsf{t}'} \quad (4.3.31)$$

Is equal to –2, 0, or +2. In the particular case of X-OPSs polynomial (4.3.30) takes the form

$$\tilde{C}_m[\varsigma; \mathbf{t}_\pm, m; \mathbf{c}, v]$$

$$= \tfrac{1}{4}(-1)^{m+1} m! \left\{ [\pm 1 + 1 + 2\, {}^1\lambda_{o;\mathbf{t}_\pm} + \tilde{\varepsilon}_{\mathbf{t}_\pm, m} - \tilde{\varepsilon}_{\mathbf{c}, v}] L_m^{(\lambda_{\mathbf{t}_\pm})}(\mp\varsigma) \right.$$

$$\left. + [(\pm 1 - 1)\, {}^1\lambda_{o;\mathbf{t}_\pm} + (\pm 1 + 1)\varsigma]\, L_{m-1}^{(\lambda_{\mathbf{t}_\pm} + 1)}(\mp\varsigma) \right\}, \qquad (4.3.32)$$

where $\mathbf{t}_+ = \mathbf{a}$ and $\mathbf{t}_- = \mathbf{b}$.

## 5. Four P-CES Quadrants for Signed Exponent Differences in Gauss-Seed Heun Equation

In general, one can construct 16 P-CES equations with m+1+ $\iota$ finite singular points. Using Takemura's notation [40] in his Conjecture 1 with $m_k = 2$ for k=1,..., M $\equiv$m the RSL equation with m+3 regular singular points (including infinity) can be formally represented as

$$\left[ \hat{L}_{\mathbf{t}, m_\mathbf{t}; \bar{\sigma}} + \frac{C_{m_\mathbf{t}}[z; \varepsilon \,|\, {}^1_1 G^{201}_{\mathbf{t}, m_\mathbf{t}}; \bar{\sigma}]}{4z(z-1)\Pi_{m_\mathbf{t}}[z; \bar{z}_{\mathbf{t}, m_\mathbf{t}}]} \right] F[\xi; \varepsilon \,|\, \mathbf{t}, m_\mathbf{t}; \bar{\sigma}] = 0, \qquad (5.1)$$

where

$${}_1\hat{L}_{\mathbf{t}, m_\mathbf{t}; \bar{\sigma}} \equiv \frac{d^2}{dz^2} + \left\{ \frac{\sigma_0 \lambda_{o;\mathbf{t}} + 1}{z} - \frac{\sigma_1 \mu_{o;\mathbf{t}} + 1}{1 - z} - 2 \sum_{k=1}^{m_\mathbf{t}} \frac{1}{z - z_{\mathbf{t}, m_\mathbf{t}; k}} \right\} \frac{d}{dz} \qquad (5.2)$$

and

$$C_{m_\mathbf{t}}[z; \varepsilon \,|\, {}^1_1 G^{201}_{\mathbf{t}, m_\mathbf{t}}; \bar{\sigma}] = C^0_{m_\mathbf{t}}[\xi \,|\, {}^1_\iota G^{K01}_{\mathbf{t}, m_\mathbf{t}}; \bar{\sigma}] - {}_1 d\, \varepsilon\, \Pi_{m_\mathbf{t}}[z; \bar{z}_{\mathbf{t}, m_\mathbf{t}}]. \qquad (5.3)$$

Making use of (2.22*) and (2.23*) one can verify that the leading coefficient of polynomial (3.55) is equal to

$$C^0_{m_\mathbf{t}; m_\mathbf{t}} \,|\, {}^1_1 G^{201}_{\mathbf{t}, m_\mathbf{t}}; \bar{\sigma}) = \tfrac{1}{4}(\sigma_0 \lambda_{o;\mathbf{t}} + \sigma_1 \mu_{o;\mathbf{t}} - 2m_\mathbf{t} + 1)^2. \qquad (5.4)$$

so that the leading coefficient of polynomial (5.3) for $\bar{\sigma} = -\bar{\sigma}_{\dagger}$ vanishes at the energy $\varepsilon_{\bar{\sigma}_{\dagger}, m_{\dagger}}$, as expected. Heine equation (5.1) may have a polynomial solution of order n at energy $\varepsilon(\bar{\sigma}, n)$ iff n is a nonnegative integer root of the quadratic equation

$$. \ n^2 + (\sigma_0 \lambda_{o;\dagger} + \sigma_1 \mu_{o;\dagger} - 2m_{\dagger} + 1)n + C^0_{m_{\dagger};m_{\dagger}} \mid {}^1_1\mathcal{G}^{201}_{\dagger, m_{\dagger}}; \bar{\sigma}) = \varepsilon(\bar{\sigma}, n) \,. \qquad (5.5)$$

Substituting (5.4) into (5.5) then brings us back to necessary condition (3.18) in Section 3.

In addition to purely polynomial solutions (4.1.20a) – (4.1.20d), Fuschian equation (5.1) has three infinite series of ‘quasi-algebraic’ solutions which can be possibly used to construct the partner integral equations by extending Lambe and Ward’s arguments [48] to Takemura’s equations (5.1)-(5.2).

In Quesne’s linear case for ι=1 [6, 13] both DTs in the pair $\dagger_{-}$ = a, b or in the pair lead to the same PF beam with the reference PF:

$$I^o[z \mid {}^1_1\mathcal{G}^{202}_{\dagger_{\pm}, 1}] = {}^1_1 I^o_{\pm}[z \mid \lambda_{o;\dagger_{\pm}}, \mu_{o;\dagger_{\pm}}], \qquad (5.6)$$

where

$${}^1_1 I^o_{\pm}[z \mid \lambda, \mu] = {}_1 I^o[z \mid \lambda, \mu] - \frac{2}{[z - z_{\pm}(\lambda, \mu)]^2} - \frac{2z-1}{z(1-z)[z - z_{\pm}(\lambda, \mu)]} \qquad (5.6^*)$$

$$\text{for } \lambda \neq \mp \mu,.$$

with the common singular point

$$z_{\pm}(\lambda, \mu) \equiv \frac{\lambda}{\lambda \pm \mu} \quad \text{for } \lambda, \ \mu > 0 \ (\lambda \neq \mp \mu), \qquad (5.7)$$

for each of two pairs.

In the particular case of equations with 4 regular singular points, including infinity (ι=m=1), 16 second-order differential equations (5.1) thus turn into a single Heun equation

$$\left[\hat{L}(\lambda, \mu) + \frac{{}_1 d\,\varepsilon}{4z(1-z)}\right] Hf[z; {}_1 d\,\varepsilon; \lambda, \mu] = 0 \qquad (5.8)$$

where

$$\hat{L}(\lambda,\mu) \equiv \frac{d^2}{dz^2} + \left\{ \frac{1+\lambda}{z} + \frac{1+\mu}{z-1} - \frac{2(\lambda-\mu)}{(\lambda-\mu)z-\lambda} \right\} \frac{d}{dz} + \frac{\tilde{C}_1[z;\lambda,\mu]}{z(z-1)[(\lambda-\mu)z-\lambda]}. \quad (5.9)$$

In the limit $\mu \to \lambda \neq 0$ the polynomial coefficient

$$\tilde{C}_1[z;\lambda,\mu] \equiv \tfrac{1}{4}\,{}_1d\,\varepsilon_{\mathbf{c};0}(\lambda,\mu)[(\lambda-\mu)z-\lambda] - [\lambda(z-1)+\mu z](\lambda-\mu), \quad (5.10)$$

where

$${}_1d\,\varepsilon_{\mathbf{c};0}(\lambda,\mu) \equiv O_0^0(\lambda,\mu) + 4\,{}_1C_0^0(\lambda,\mu), \quad (5.11)$$

is reduced to a constant, the singularity at the point $z = \lambda/(\lambda-\mu)$ disappears and solutions of Heun equation (5.8) turn into the hypergeometric functions. If $\lambda$ and $\mu$ have the same sign, then $(\lambda-\mu)/\lambda<1$ and Heun equation has an outer singular point. In particular, if both parameters $\lambda$ and $\mu$ are positive, Heun equation (5.8) is conditionally exactly quantized via $X_1$-Jacobi polynomials under the boundary conditions imposed at 0 and 1. If $\lambda$ and $\mu$ lie in the second and fourth quadrants, then the third singular point is located between 0 and 1 and Heun equation (5.8) for $|\mu| > |\lambda|$ is conditionally exactly quantized on the negative semi-axis, with the appropriate polynomial solutions represented by CD transforms of Romanovsky-Jacobi polynomials.

It has been proven by Gomez-Ullate, Kamran, and Milson [1, 2] that Heun equations (5.8) with the outer singular point is solvable via $X_1$-Jacobi polynomials provided that

$$\lambda,\ \mu > -1,\ \lambda \neq \mu,\ \text{and}\ \lambda\mu > 0. \quad (5.12)$$

Similarly (as discussed in subsection 4.2 above), the given Heun equation is quantized via a finite set of orthogonal polynomials [20] for $\lambda\,\mu<0$ ($|\lambda|| \geq 1$, $|\mu| \geq 2$) under the boundary condition that the eigenfunctions in question are squarely integrable on the negative semi-axis.

In case of the outer singular point the eight infinite polynomial sequences defined via (4.1.20a) - (4.1.20d) for $m_\dagger = 1$ turn into the following four (not necessarily distinct) sequences of Heun polynomials:

$$\mathrm{Hi}_{m+1}^{(3)}[z;\lambda,-\mu;+-]=\frac{m!}{(2+\mu-m)(\lambda+\mu+m)(\lambda-\mu)}\mathrm{Hp}_{m+1}[z;\lambda,\mu;+-], \qquad (5.13a)$$

$$\mathrm{Hi}_{m+1}^{(3)}[z;-\lambda,\mu;-+]=\frac{m!}{(\lambda+m)(\lambda+\mu+m)(\lambda-\mu)}\mathrm{Hp}_{m+1}[z;\lambda,\mu;-+], \qquad (5.13b)$$

$$\mathrm{Hi}_{m}^{(3)}[z;\lambda,\mu;++]=\frac{1}{(m-1)(\lambda+\mu+m-2)(\mu-\lambda)}\mathrm{Hp}_{m}[z;\lambda,\mu;++], \qquad (5.13c)$$

$$\mathrm{Hi}_{m+2}^{(3)}[z;\lambda,\mu;--]=\frac{1}{(\lambda+\mu-m+1)_2(\mu-\lambda)}\mathrm{Hp}_{m+2}[z;\lambda,\mu;--]. \qquad (5.13d)$$

It is convenient to define Jacobi polynomials with arbitrary (generally complex [122]) indexes as a is a polynomial in each of three variables y, $\lambda$, and $\mu$:

$$P_m^{(\mu,\lambda)}(y)=\frac{1}{2^m}\sum_{k=0}^{m}\frac{1}{k!(m-k)!}(\mu+m)_{m-k}(\lambda+m)_k(y-1)^k(y+1)^{m-k} \qquad (5.14)$$

so that and therefore all the recurrence relations between these polynomials as well as between their coefficients can be automatically extended to the field of complex numbers. The same is obviously true for the polynomials

$$\mathrm{Hp}_{m+1}[z;\lambda,\mu;+-]\equiv(\lambda-\mu)\ \hat{a}\Big(-\mu-1;\frac{\lambda}{\lambda-\mu}\Big)P_m^{(\mu+1,\lambda-1)}(2z-1), \qquad (5.15a)$$

$$\mathrm{Hp}_{m+1}[z;\lambda,\mu;-+]\equiv(\lambda-\mu)\ \hat{b}\Big(-\lambda-1;\frac{\lambda}{\lambda-\mu}\Big)P_m^{(\mu-1,-\lambda-1)}(2z-1), \qquad (5.15b)$$

$$\mathrm{Hp}_{m}[z;\lambda,\mu;++]\equiv(\mu-\lambda)\ \hat{c}\Big(\frac{\lambda}{\mu-\lambda}\Big)P_m^{(\mu-1,\lambda-1)}(2z-1), \qquad (5.15c)$$

$$\mathrm{Hp}_{m+2}[z;\lambda,\mu;--]\equiv(\mu-\lambda)\hat{d}\Big(\lambda+1,-\mu-1;\frac{\lambda}{\mu-\lambda}\Big)P_m^{(\mu+1,\lambda+1)}(2z-1). \qquad (5.15d)$$

Therefore, despite the fact that the derivation of (5.13a) –(5.13d) was made under constraints (3.22a) and (3.22b) Heun equation (5.8) is conditionally exactly solvable via Heun polynomials (5.15a) –(5.15d) for any values $\lambda$ and $\mu$ assuming that the normalizing factors in the right-hand side of (5.13a) –(5.13d) remain finite.

For positive values of signed exponent differences λ and μ both sequences a and b of Heun polynomials turn into $X_1$-Jacobi polynomials. By representing (3.18) as

$$_1d[\varepsilon_k(\lambda,\mu;++)-\varepsilon_0(\lambda,\mu;++)] = k(\lambda+\mu+k) \tag{5.16}$$

for μ >λ ≥1, one can directly relate the standard parameters α, β, γ, and q in Takemura's equation (6.6) for the Heun polynomial of order n = k+1 to the parameters λ and μ via the formulas

$$\alpha = -k-1, \beta = k+\lambda+\mu, \quad \gamma = \lambda+1, \tag{5.17}$$

$$q = t\{d[\varepsilon_{c;0}(\lambda,\mu)-\varepsilon_{c,k}(\lambda,\mu)]+h-g\}, \tag{5.17*}$$

where

$$t = \frac{g+\frac{1}{2}}{g-h} = -|\lambda|/|\mu| \tag{5.18}$$

and

$$g+h = \lambda+\mu-1. \tag{5.18′}$$

Note also that parse sequence (5.15c) starts from a constant and misses a polynomial of the first order, in agreement with the general case of Heine polynomials of type $c\dagger$. Heun equation (5.8) thus provides a very interesting illustrative example of a variety of polynomial sequences associated with Heine equation (5.1). It certainly deserves a more serious attention which is beyond the scope of this paper.

## 6. Concluding Comments and Further Developments

As demonstrated in the paper, the recently developed theory of exceptional orthogonal polynomials can be treated as an integral part of a more general approach dealing with CDTs of the RSL equations solvable via hypergeometric and confluent hypergeometric functions. It is shown that the resultant Heine equations become PES in Gomez-Ullate, Kamran, and Milson's terms [24, 35] if characteristic exponents of the appropriate Bose invariants are energy-independent at both end points. (In following Junker and Roy [8] we refer to these equations as 'P-CES' to stress that they are exactly solvable only under some constraints imposed on exponent differences and outer singular points of the appropriate Bose invariants.) In both cases of regular-at-infinity (K=2, ι=1) and confluent (K=1, ι=0) Bose invariants there are 16 infinite sets (n=0, 1,...) of GS Heine polynomials. $Hi^{(m+1)}_{m+n}[\varsigma \mid {}^{1}_{\iota}G^{K01}_{\dagger,m};\dagger',n]$ for each m >1. Two of these sixteen infinite sets are formed by $X_m$-Jacobi and $X_m$-Laguerre polynomials.

As demonstrated in subsection 4.2 there are three sequences of isospectral REs of the h-PT potential which are quantized via CDTs of Romanovsky-Jacobi polynomials. Two of them have infinitely many members: one starts from the BQR potential [108] restricted to the validity range of SUSY quantum mechanics ($\lambda_o > 1$) whereas the first potential in other sequence is generated using the Jacobi polynomial of a higher order equal to the number of bound energy states in the h-PT potential. There is also a finite sequence of isospectral RPs starting again from the BQR potential [108] with the only restriction: $\lambda_o \neq 0$ or 1.

As clarified in Section 4, an analysis of eigenfunctions of the Schrödinger equation with REs of the DPT potential and isotonic oscillator makes it possible to construct infinite series of orthogonal GS Heine polynomials only within the validity range of SUSY quantum mechanics. As an illustrative example, we pointed to the fact that the *r*-XQ potential within the range $0 < \lambda_{o;\dagger}, \mu_{o;\dagger} < 1$ has quantized solutions expressible via X-Jacobi polynomials though this parameter range cannot be reached via DTs with regular FFs.

In Part II we will study an extension of our analysis to multi-step Darboux transforms of the DPT potential and isotonic oscillator conditionally exactly quantized via multi-index infinite sequences of orthogonal X-Jacobi and X-Laguerre [43-50] polynomials.

Both here and in Part II we focus solely on orthogonal sets of Heine polynomials associated with energy-independent characteristic exponents. However, it is nothing but a very particular case of the general formalism [11, 17-19] using CDTs with AEH FFs to generate exactly-solvable and CES rational potentials. One of most important future developments would be our analysis of rational potentials *exactly quantized* via Heun and *c*-Heun polynomials. In our future studies we would like also to analyze in more details a possible relation between our prescription for constructing multi-index (generally nonorthogonal) Heine polynomials and recent works of Sasaki, Takemura, and Ho [101, 102] dealing with second-order differential equations solvable via multi-index polynomials.

In this paper we discussed only CDTs of RSL equations with real singular points. As a brand new direction we plan to study more thoroughly REs of the so-called 'Scarf II' [123] potential quantized via a finite set of orthogonal Romanovsky polynomials [79] (see [124] for details). The remarkable feature of this shape-invariant reduction of the Milson potential [54] is that the appropriate RSL equation has *energy-independent* characteristic exponents at both complex conjugated singular points. As a result two-step 'juxtaposed' DTs [125, 126] of bound eigenfunctions result in sparse sequences of orthogonal Heine polynomials. As shown in [127], a finite sequence of bound eigenfunctions for the Milson potential is accompanied by a sequence of AEH solutions of type **d** formed by Routh polynomials [128]. It starts from the basic solution which can be used as the FF for the CDT to generate the RE of the Milson potential quantized via Heine polynomials. In the limiting case of singular points on the imaginary axis this RE collapses into the 'Scarf II' potential. A nontrivial question which has to be explored in more detail is whether there are any Routh polynomials with no zeros on the real axis.

**Acknowledgements**

I express my gratitude to M. Kirchbach for brining my attention to a series of recent works on REs of shape-invariant potentials.

## Appendix A

### Breakup of SUSY Quantum Mechanics by Darboux Transformations of CB potentials Using Irregular-at-Origin FFs

In this paper we are only interested in the radial Schrödinger equation with the second-order pole at the origin, namely, we assume that

$$\lim_{r\to 0}[r^2V(r)] = (s - 1/2)^2 - 1/4. \qquad \text{(A1)}$$

The potential has an infinitely deep well (instead of the infinitely high barrier) if the CB parameter s is smaller than 1 and the standard approach (see, e.g., [114, 115]) is to simply choose s > 1. However, as pointed by Negro, Nieto, and Rosas-Ortiz [129] for the DPT potential, the potential has a continuous spectrum for s < 3/2 if the boundary condition is defined via the requirement that any regular solution must be quadratically integrable at the given end point. One can then follow Kemble's arguments [117] for SL problems with singular end points to prove existence of the discrete spectrum, orthogonality of the appropriate eigenfunctions as well as a decrease of spacing of the nodes with an increase of the energy -- the well-known results for non-singular SL problems [116].

From purely mathematical point of view), one comes to the accurately defined SL problem by defining a regular solution $\psi_{\mathfrak{a}}(r)$ at the origin as the one with a larger characteristic exponent: x)

$$\lim_{r\to 0}[r^{1/2}\psi_{\mathfrak{a}}(r)] = 0. \tag{A2}$$

assuming that s > ½. (The exponent difference becomes equal to zero if s = ½ so that the conventional approach is not applicable anymore.) However, the author is not aware of the proof which would extend Kemble's arguments to discrete energy levels embedded in the continuous energy spectrum.

As pointed out in Introduction, successive DTs with irregular FFs decrease the parameter s by 1 until the latter becomes smaller than 3/2. After that any DT with an irregular FF does not affect the pole coefficient. As the CB parameter s becomes smaller than 3/2, DTs with FFs irregular at the origin converts bound eigenfunctions into solutions of type $\mathfrak{b}$ so that the appropriate 'SUSY partners' have completely different discrete energy spectra, compared with the original potential. It should be stressed in this connection that both Andrianov et al. [71, 72] and Sukumar [73] explicitly restrict an

---

x)Contrary to the statement made in [101], we doubt that only the 'most regular' solution should be interpreted as the physical one.

analysis of SUSY quantum mechanics to potentials on the line. This important point is usually ignored in more recent presentations of SUSY quantum mechanics (see [123, 130], for example).

**Appendix B**

**Supersymmetry of Sturm-Liouville Problem Written in Canonical Form**

In [11] we have analyzed in detail the covariant factorization of the ST equation

$$\left\{\frac{d^2}{d\xi^2} + I^o[\xi\,|\,Q^o] + \varepsilon\wp[\xi]\right\}\Phi[\xi;\varepsilon\,|\,Q^o] = 0 \qquad \text{(B1)}$$

written in the canonical form using the so-called 'generalized' [56-61] Darboux transformations

$$\hat{L}_\tau = \wp^{-1/2}[\xi]\,\hat{d}\{\varphi_\tau[\xi\,|\,Q^o]\} \qquad \text{(B2)}$$

where $\phi_\tau[\xi\,|\,Q^o]$ is a solution of SL equation (B1) at the energy $\varepsilon = \varepsilon_\tau$:

$$\begin{aligned} I^o[\xi\,|\,Q^o,\tau] &= -\ddot{\phi}_\tau[\xi\,|\,Q^o]/\phi_\tau[\xi\,|\,Q^o] - \varepsilon_\tau\,\wp[\xi] \\ &= -l\dot{d}\{\phi_\tau[\xi\,|\,Q^o]\} - ld^2\{\phi_\tau[\xi\,|\,Q^o]\} - \varepsilon_\tau\,\wp[\xi] \end{aligned} \qquad \text{(B3)}$$

and

$$\hat{d}\{\varphi_\tau[\xi\,|\,Q^o]\} \equiv \frac{d}{d\xi} - ld\{\varphi_\tau[\xi\,|\,Q^o]\}, \qquad \text{(B4)}$$

with the symbol *ld* standing for the logarithmic derivative. Differential operators (B4) were originally introduced by Rudyak and Zakhariev [55] in the scattering theory and then studied more cautiously in the papers of Schnizer and Leib [56-58] and Suzko [59-61] and in the same context. In our original study [11] of transformation properties of bound eigenfunctions under action of these operators we overlooked a more recent work of Gomez-Ullate, Kamran, and Milson [34] who developed a more general factorization technique for the Fuschian equation with polynomial coefficients, including an additional term with the first derivative absent in (B1). To stress that we only deal with the specific case of this technique in all the problems of our interest, we refer to the mentioned operations as Canonical Darboux Transformations (CDTs).

SL equation (B1) can be thus factorized as

$$({}^{*}\hat{L}_{\tau}\,\hat{L}_{\tau}+\varepsilon_{\tau}-\varepsilon)\,\Phi[\xi;\varepsilon\,|\,Q^{o}]=0 \qquad \text{(B5)}$$

and a similar factorization is valid for the partner equation

$$(\hat{L}_{\tau}\,{}^{*}\hat{L}_{\tau}+\varepsilon_{\tau}-\varepsilon)\,{}^{*}\Phi[\xi;\varepsilon\,|\,Q^{o};\tau]=0 \qquad \text{(B5*)}$$

where

$${}^{*}\hat{L}_{\tau}=\wp^{-1/2}[\xi]\,\hat{d}\{{}^{*}\phi_{\tau}\,|\,Q^{o}\} \qquad \text{(B6)}$$

and ${}^{*}\Phi[\xi;\varepsilon\,|\,Q^{o};\tau]$ is the canonical Darboux transform of $\Phi[\xi;\varepsilon\,|\,Q^{o}]$:

$${}^{*}\Phi[\xi;\varepsilon\,|\,Q^{o};\tau]=\hat{L}_{\tau}\Phi[\xi;\varepsilon\,|\,Q^{o}]=\frac{W\{\Phi[\xi;\varepsilon\,|\,Q^{o}],\phi_{\tau}[\xi\,|\,Q^{o}]\}}{\wp^{1/2}[\xi]\phi_{\tau}[\xi\,|\,Q^{o}]}, \qquad \text{(B7)}$$

which satisfies the partner SL equation:

$$\left\{\frac{d^{2}}{d\xi^{2}}+{}^{*}I^{o}[\xi\,|\,Q^{o};\tau]+\varepsilon\wp[\xi]\right\}{}^{*}\Phi[\xi;\varepsilon\,|\,Q^{o};\tau]=0\,. \qquad \text{(B8)}$$

A pivotal feature of CDTs is that factorization functions $\phi_{\tau}[\xi\,|\,Q^{o}]$ and ${}^{*}\phi_{\tau}[\xi\,|\,Q^{o}]$ for operator (B2) and its reverse (B6) are related via Suzko's reciprocal formula [59-61]:

$${}^{*}\phi_{\tau}[\xi\,|\,Q^{o}]=\wp^{-1/2}[\xi]/\phi_{\tau}[\xi\,|\,Q^{o}] \qquad \text{(B9)}$$

which plays a pivotal role in our analysis of characteristic exponents of singular points in the partner RSL equation.

To emphasize that we deal with some intrinsic feature of the S-L problem, a majority of the results in [11] were obtained with no reference to the Liouville transformation [89-94] using the change of variable

$$x=\int d\xi\,\wp^{1/2}[\xi]\ , \qquad \text{(B10)}$$

which converts Sturm-Liouville problem (B1) into the Schrödinger equation:

$$\left\{\frac{d^2}{dx^2} + V[\xi(x)\,|\,Q^o] + \varepsilon\right\}\Psi[\xi(x);\varepsilon\,|\,Q^o] = 0 \tag{B11}$$

with the potential

$$V[\xi(x)\,|\,Q^o] = -\wp^{-1}[\xi(x)]\, I^o[\xi;|\,Q^o] - \tfrac{1}{2}\{\xi, x\}, \tag{B12}$$

where the symbol { ξ ,x} denotes the so-called Schwartz derivative (see, i.g. [131]). Here we take just the opposite approach and use the dualism between two equations to take advantage of the results well-known in the theory of Darboux transformations of the Schrödinger equation. The mentioned dualism is based on Bose's observation [50] that solutions of equations (B1) and (B11) are related in a simple manner:

$$\Psi[\xi;\varepsilon\,|\,Q^o] = \wp^{1/4}[\xi]\Phi[\xi;\varepsilon\,|\,Q^o]\,. \tag{B13}$$

A similar relation can be written for FFs

$$\psi_\tau[\xi] = \wp^{1/4}[\xi]\phi_\tau[\xi], \tag{B14}$$

$$*\psi_\tau[\xi] = \wp^{1/4}[\xi]\,*\phi_\tau[\xi] \tag{B14*}$$

which directly leads to factorization (B5) of SL equation (B1)..

Making use of the conventional formula] for the potential difference

$${}^1V[\xi(x)\,|\,Q^o;\tau] - V[\xi(x)\,|\,Q^o] = -2\frac{d^2}{dx^2}\,ln\left|\psi_\tau[\xi(x)\}\right| \tag{B15}$$

to represent the difference between the final and initial reference PFs as

$${}^1I^o[\xi;Q^o;\tau] - I^o[\xi;Q^o] = \ 2\,\wp^{1/2}[\xi]\frac{d}{dz}\left\{\wp^{-1/2}[\xi]\,ld\psi_\tau[\xi]\right\}. \tag{B16}$$

We refer the reader to Appendix A in Part II for an extension of (B16) to multi-step CDTs.

**Appendix C**

**Subsets of Jacobi Polynomials With Zeros Outside Quantization Interval**

The purpose of this Appendix is to derive explicit conditions for the polynomial $P_n^{(\lambda,\mu)}(y)$ not to have zeros within the intervals $-1< y <+1$, $-\infty < y < -1$, $+1 < y < +\infty$, using formulas (6.72.4)-(6.72.8) in [10] for the number of its zeros, $N_1$, $N_2$, and $N_3$, respectively. An analysis of the cited formulas shows that

$$N_1 = 0 \quad \text{iff} \quad u_1(\lambda,\mu;m) \le 1 \text{ and } (-)^m(\lambda+1)_m(\mu+1)_m > 0, \tag{C1}$$

$$N_2 = 0 \quad \text{iff} \quad u_2(\lambda,\mu;m) \le 1 \text{ and } (\lambda+\mu+m+1)_m(\mu+1)_m > 0, \tag{C2}$$

$$N_3 = 0 \quad \text{iff} \quad u_3(\lambda,\mu;m) \le 1 \text{ and } (\lambda+\mu+m+1)_m(\lambda+1)_m > 0, \tag{C3}$$

where $(\lambda)_m$ is the rising factorial,

$$u_1(\lambda,\mu;m) = \tfrac{1}{2}\left(|2m+\lambda+\mu+1| - |\lambda| - |\mu| + 1\right), \tag{C4}$$

and

$$u_2(\lambda,\mu;m) = u_3(\mu,\lambda;m) = \tfrac{1}{2}\left(-|2m+\lambda+\mu+1| + |\lambda| - |\mu| + 1\right). \tag{C5}$$

We say that the nodeless Jacobi polynomials with indexes satisfying conditions (C1), (C2), (C3) belongs to Classes J-I, J-II, and J-III, respectively. The classical orthogonal Jacobi polynomials obviously belong to the intersection of Classes J-II and J-III. 'Romanovsky-Jacobi' [84, 86] polynomials [83] orthogonal over the interval $(-\infty, -1)$ for

$$2m - |\lambda| + \mu + 1 < 0 \text{ and } -\lambda > \mu > 1 \tag{C6}$$

belong to the intersection of Classes J-I and J-III, as expected. In fact, it directly follows from (C4) that the constraint

$$u_k(\lambda,\mu;m) \le 1 \tag{C7}$$

for k=1 holds if either

$$-1 \le 2m+\lambda+\mu+1 < |\lambda| + |\mu|, \tag{C8a}$$

or

$$-|\lambda|-|\mu|-1 \le 2m+\lambda+\mu+1<0\,. \tag{C8b}$$

For $-\lambda>\mu>0$ condition (C8b) takes the form:

$$-2|\mu|-1 \le 2m+1<-\lambda-\mu\,. \tag{C9}$$

Representing the first of inequalities (C6) as

$$m+\lambda<-m-\beta-1<0 \tag{C10}$$

and taking into account that the second constraint in (C1) trivially holds for $\mu>-1$ and $\lambda+m<0$, we confirm that the Romanovsky-Jacobi polynomials do belong to Class J-II. Similarly since both quantities $m+\lambda+\mu+1$ and $\lambda+m$ are negative, the second constraint in (C3) must also hold whereas

$$u_3(\lambda,\mu;m)<1/2 \quad \text{if } |\lambda|>|\mu|, \tag{C11}$$

which completes the proof.

Condition (C7) with k=1 can be alternatively represented as

$$\left(2m+\lambda+|\lambda|+\mu+|\mu|+2\right)\left(2m+\lambda+\mu-|\lambda|-|\mu|\right)\le 0\,. \tag{C12}$$

which holds iff

$$m \le 1/2\left(|\lambda|-\lambda+|\mu|-\mu\right). \tag{C13}$$

We thus need to consider separately three cases:

$$0<m<-\lambda \quad \text{for } \lambda\le -1,\ \mu\ge 0, \tag{C14a}$$

$$0<m<-\mu \quad \text{for } \lambda\ge 0,\ \mu\le -1; \tag{C14b}$$

and

$$0<m<-\lambda-\mu \quad \text{for } \lambda,\ \mu<0. \tag{C14*}$$

It is essential that the second constraint in (C1) holds in both cases (C14a) and (C14b). This is also true if either

$$0<\lambda+m<0<-\mu<1 \tag{C14′}$$

or

$$0 < \mu + m < 0 < -\lambda < 1 \,. \qquad \text{(C14'')}$$

By setting $\lambda = -\alpha - 1$, $\mu = \beta - 1$ and taking into account that $\lambda + m < 0 < -\mu$ for $-1 < \mu < 0$ we can directly verify that (C14a), (C14b), and (C14′) cover the necessary and sufficient conditions formulated in Proposition (4.5) in [35] for $\lambda < 0$, $\mu > -2$:

$$0 < \lambda + m < 1, \;\; -2 < \mu < -1 \;\; (0 < m < 1 - \lambda < -\lambda - \mu), \qquad \text{(C15a)}$$

$$\lambda + m < 0 \,, \;\; \mu > -1 \,. \qquad \text{(C15b)}$$

-- the cases (A) and (B) in subsection 5.2 in [42].

For $\lambda, \mu < -1$ condition (C14*) leads to nodeless AEH solutions of type **d** which, as expected, lie below the ground energy level. Taking into account that the second constraint in (C1) holds for even m = 2j if

i) $2j + \lambda,\ 2j + \mu < 0$ ; (C16)

ii) $1 < [-\lambda]\ ,\ [-\mu] < 2j$ (C16*)

provided that λ and μ are either both odd or both even;

iii) $1 < -\lambda < 2j < -\mu$ for even $[-\kappa]$ (C16′)

or $1 < -\mu < 2j < -\lambda$ for even $[-\mu]$. (C16″)

Combining (C16) with (C8b) written as

$$|\lambda| + |\mu| > 2j + 1 \qquad \text{(C17)}$$

we conclude that there is a regular sequence of AEH solutions of type **d** formed by (−1,+1)-nodeless Jacobi polynomials of even order

$$2 \le m = 2j < \min\{|\lambda|, |\mu|\} \,. \qquad \text{(C18)}$$

For negative λ and μ within the range

$$m \le |\lambda + \mu| \le 2m \qquad \text{(C19)}$$

there are also (−1,+1)-nodeless Jacobi polynomials of either even or odd order depending on evenness of the sum [−λ]+[−μ], similarly to irregular AEH solutions discovered by Grandati [31] for the isotonic oscillator using the LKH theorem.